\documentclass[a4paper,11pt]{article}
\pdfoutput=1 

\usepackage{jheppub} 

\usepackage[T1]{fontenc} 

\usepackage{xcolor}

\title{Nonlinear dynamics of flux compactification}

\author[a,b]{Maxence Corman}
\author[a]{William E. East}
\author[a,c]{Matthew C. Johnson}

\affiliation[a]{
   Perimeter Institute for Theoretical Physics,
   Waterloo, Ontario N2L 2Y5, Canada.
}
\affiliation[b]{
   Department of Physics and Astronomy,
   University of Waterloo,
   Waterloo,
   Ontario N2L 3G1,
   Canada
}

\affiliation[c]{
   Department of Physics and Astronomy,
   York Univeristy,
   Toronto,
   Ontario M3J 1P3,
   Canada
}

\emailAdd{mcorman@perimeterinstitute.ca}
\emailAdd{weast@perimeterinstitute.ca}
\emailAdd{mjohnson@perimeterinstitute.ca}

\abstract{
We study the nonlinear evolution of unstable flux compactifications, applying
numerical relativity techniques to solve the Einstein equations in $D$
dimensions coupled to a $q$-form field and positive cosmological constant.  We
show that initially homogeneous flux compactifications are unstable to
dynamically forming warped compactifications.  In some cases, we find that the
warping process can serve as a toy-model of slow-roll inflation, while in other
instances, we find solutions that eventually evolve to a singular state. 
Analogous to dynamical black hole horizons, we use the geometric properties of marginally
trapped surfaces to characterize the lower dimensional vacua in the
inhomogeneous and dynamical settings we consider.  We find that
lower-dimensional vacua with a lower expansion rate are dynamically favoured,
and in some cases find spacetimes that undergo a period of accelerated
expansion followed by contraction. 
}

\begin{document}
\maketitle
\flushbottom


\section{Introduction}
\label{sec:intro}

The standard cosmological model invokes accelerated expansion of the Universe
both at early times, in an inflationary era, and at late times, in the current
epoch of dark energy domination. Determining the physical mechanism(s)
responsible for the accelerated expansion of the Universe is among the most
important challenges in modern cosmology. One proposed framework for tackling
this is string/M-theory, where the mechanisms responsible for dark energy
and inflation would ideally just be one feature of a complete description of
gravity and the standard model of particle physics. 

A major complication in developing these phenomenological connections are the
extra spatial dimensions invoked to make string theory a consistent quantum
theory of gravity.  There are two dominant paradigms for explaining why we
cannot experimentally probe extra spatial dimensions: they are small 
(compactification~\cite{Kaluza:1921tu,Klein:1926tv}) or the
standard model degrees of freedom are constrained to move in only four
dimensions (the braneworld
scenario~\cite{ArkaniHamed:1998rs,Randall:1999ee,Randall:1999vf}). The 
specific choice of compactification or realization of the braneworld scenario has implications for
phenomenology, dictating the particle content and vacuum structure, as well as
the types and strengths of interactions, in the effectively four-dimensional
theory that results. The proliferation of four-dimensional theories (known as
the string theory landscape~\cite{Susskind:2003kw}) intertwines string theory
with cosmology in many fundamental ways. In this paper, our main point of
contact will be the evolution of the size and shape of a compactification,
which (in this picture) are part of our cosmological history, and can provide
the physical mechanism for inflation and dark energy. 

What dynamics might be associated with extra dimensions? In the simplest
scenario, the Universe remains effectively four-dimensional and small
deformations of the extra dimensions correspond to a set of fields known as
Kaluza Klein (KK) modes. Even this is highly non-trivial, requiring the
addition of various sources of energy momentum (such as $q$-form gauge fields,
branes, etc.) to stabilize the size and shape of the compactification, and
verifying that the resulting four-dimensional effective theory has the desired
properties. Beyond studying linear perturbations of such static stable configurations, very little is
known about the dynamics associated with extra dimensions. This is not
surprising given the difficulties in solving Einstein's equations in four
dimensions, let alone ten. Nevertheless, a better understanding is necessary to
fully understand cosmology in theories with extra dimensions, and in particular 
address questions such as: How was the Universe we observe selected from the
many possibilities? What features of our Universe are accidental, and which are
inevitable (e.g. fixed by special initial conditions or symmetries)? Why are
there only three large spatial dimensions?

To make progress in this direction, we focus on a simple model that retains
many of the important features of the low-energy limit of string theory:
Einstein-Maxwell theory in $D$-dimensions with a positive cosmological constant
and a $q$-form gauge field. Freund and Rubin \cite{Freund:1980} showed that this
theory admits solutions in which the extra dimensions are compactified on a
sphere, stabilized against collapse by the positive curvature of the
compactification and a homogeneous configuration of the gauge field over the
sphere. If a positive bulk cosmological constant is included
\cite{Bousso:2003}, it is possible to find solutions in $D=p+q$ dimensions that
are a product space of $p$-dimensional anti-de Sitter, Minkowski or de Sitter
space and a $q$-dimensional sphere. The size of the compact sphere and the
magnitude of the four dimensional cosmological constant are adjusted with the
number of units of flux of the $q$-form gauge field wrapping the $q$-sphere.
This simple model figures prominently in the AdS/CFT
correspondence~\cite{Maldacena:1997re}, serves as a simple example of flux
compactifications in string theory~\cite{Douglas:2006es,Denef:2007pq}, and has
been employed to study the cosmological constant
problem~\cite{Denef:2007pq,Carroll:2009dn,Brown:2013fba,Asensio:2012pg}, flux
tunneling~\cite{Aguirre:2009tp,BlancoPillado:2009di,Brown:2010bc,BlancoPillado:2010et},
and dimension-changing
transitions~\cite{Contaldi:2004hr,Krishnan:2005,Carroll:2009dn,BlancoPillado:2009mi},
among other phenomena. Another interesting feature of the Einstein-Maxwell
model is that in addition to spherical compactifications, it also admits stable
solutions where the compact space is inhomogeneous, or
``warped"~\cite{Kinoshita:2007,Kinoshita:2009hh,Lim:2012,Dahlen:2014}. In string
constructions, warped extra dimensions are essential in models that address the
hierarchy between the gravitational and electroweak
scales~\cite{Giddings:2001yu}, dark energy~\cite{Kachru:2003aw}, and cosmic
inflation (e.g.~\cite{Kachru:2003sx}). A complete understanding of the
dynamical generation of such structure is an important missing component of the
cosmology of these models. 

In the Einstein-Maxwell model, the linear stability and mass spectrum of
the Freund-Rubin solutions were studied in
refs.~\cite{DeWolfe:2001nz,Kinoshita:2007,Kinoshita:2009hh,Lim:2012,Brown:2013,Hinterbichler:2014}.
Their analysis showed that the stability of the solution to small perturbations
depends on the relative value of the flux density or Hubble parameter compared
to the cosmological constant as well as on the dimension of the internal
manifold. There are two types of dynamical instabilities:
\begin{itemize}
\item The \emph{total volume instability} can be attributed to homogeneous
perturbations ($\ell=0$ modes) of the internal space and arises whenever the
density of flux lines warping the $q$-sphere is too small, or equivalently, when
the Hubble expansion rate of the external de Sitter space is too large causing
the internal manifold to either grow or shrink. The endpoint of this
instability was found to be either decompactification to empty $D$-dimensional
de Sitter space or flow in towards a different configuration where total flux
integrated over the compact space is the same but the volume is smaller hence
flux density larger \cite{Krishnan:2005}.
\item The \emph{warped instability} arises when $q\geq4$ (in contrast to 
the volume instability which already exists when
$q\geq2$) and is due to inhomogeneous perturbations. 
Mathematically, this instability is due to a mode that couples the metric and flux (with $\ell\geq2$ angular dependence) 
and in turn deforms the internal space. One expects that if some configuration is unstable
for a given total flux, then this may signal the presence of another more
stable configuration with the same flux. Indeed
refs.~\cite{Kinoshita:2007,Lim:2012,Dahlen:2014} numerically constructed stationary
warped solutions and ref.~\cite{Kinoshita:2009hh} studied their perturbative
stability. But their connection to the inhomogeneous instability
has not been determined, so it is not known whether these are the
endpoint of the instability.
\end{itemize}
Note that when $q\geq5$, all of the
Freund-Rubin solutions are linearly unstable to one or both types of instability.

The goal of this paper is to go beyond studying stationary or homogeneous
solutions, and their linear perturbations, by performing full nonlinear
evolutions of perturbed Freund-Rubin and warped compactifications.  We do this
by applying modern numerical relativity techniques to probe the inhomogeneous
and strong field regime, as has been done for a number of different
cosmological scenarios,
e.g.~\cite{Garfinkle:2008ei,Wainwright:2013lea,East:2015ggf,East:2016anr,Clough:2016ymm},
though here we study inhomogeneities in a compact extra dimension.  We find rich
dynamics, in some cases finding evolution from unstable to stable stationary
warped solutions, though in other cases finding that unstable solutions evolve towards a singular state (even in some cases
overshooting stable stationary solutions). We comment on some features of the cosmology seen 
by four-dimensional observers, and motivate the use of the cosmological apparent horizon as a useful 
measure of the four dimensional Hubble parameter. The solutions we study provide an important proof-of-principle 
that numerical relativity could be a powerful tool for exploring new phenomena in cosmologies with extra 
spatial dimensions.


\section{Flux compactifications in Einstein-Maxwell theory} 
\label{sec:model}
In this paper, we focus on solutions to Einstein-Maxwell theory in $D=p+q$ spacetime dimensions with a $D$-dimensional 
cosmological constant $\Lambda_D$ and a $q$-form flux that wraps $q$ compact dimensions, leaving $p$ uncompactified dimensions. 
The starting point for the theory is then the following $D=p+q$-dimensional action
\begin{equation}\label{action}
   S= \int d^px d^qy \sqrt{-g} \bigg [\frac{1}{2} {}^{(D)}R-\Lambda_D-\frac{1}{2q!}\mathbf{F}_q^2 \bigg] 
\end{equation}
where we use units with $M_D=c=1$, where $M_D\equiv \left(8\pi G_D\right)^{-1/(D-2)}$ is the
$D$-dimensional Planck mass,  ${}^{(D)}R$ is the $D$-dimensional scalar curvature, 
and $\mathbf{F}_q=F_{M_1...M_q}$ is a $q$-form. Note this choice of units is not conventional, but it leaves us the freedom to fix $\Lambda_D$. 

The Einstein equations which follow from the action \eqref{action} are
\begin{equation}\label{eq:einstein}
    G_{MN}={}^{(D)}R_{MN}-\frac{1}{2} {}^{(D)}R g_{MN}=T_{MN} 
\end{equation}
where the stress-energy tensor is
\begin{equation}\label{eq:T}
   T_{MN}=\frac{1}{(q-1)!}F_{MP_2...P_q}F_{N}^{P_2...P_q}-\frac{1}{2q!}\mathbf{F}_q^2g_{MN}-\Lambda_Dg_{MN}
\end{equation}
with $\mathbf{F}_q^2=F_{M_1...M_q}F^{M_1...M_q}$. The equations governing the $q$-form in the absence of sources are 
\begin{equation}\label{eq:ME}
    \nabla_{[N}F_{MP_2...P_q]}=\nabla^MF_{MP_2...P_q}=0 \ .
\end{equation}
Throughout the paper we will use $M$, $N$, \ldots to denote indices that run over the $D-$dimensions, $\overline{m}$, $\overline{n}$, \ldots for $(D-1)$-dimensional spatial indices, $\mu$, $\nu$, \ldots for $p=4$-dimensional spacetime indices, and $\alpha,\beta,\ldots$ for $q$-dimensional spatial indices. 

The simplest flux compactifications of Einstein Maxwell theory are the Freund-Rubin solutions~\cite{Freund:1980}: product spaces 
$\mathcal{M}_p \times S_q$, where $\mathcal{M}_p$ is a maximally symmetric p-dimensional spacetime and $S_q$ is a q-dimensional sphere.
In this paper, we investigate solutions that are warped along a single internal
direction, the polar angle $\theta$. That is, we study solutions such that the
$p$-dimensional external space is homogeneous in the uncompactified spatial dimensions 
with a warp factor depending on $\theta$, and the $q$-dimensional compact space has
the topology of a sphere with $q-1$ azimuthal symmetries. With these symmetries, the 
metric takes the form:
\begin{equation}\label{metricW}
    \begin{array}{lcl}
        ds^2 &=&-(\alpha^2-\beta_{\theta}\beta^{\theta})dt^2+\gamma_{xx}(\theta,t)d {\vec{x}^2}_{p-1}+2\gamma_{\theta\theta}(\theta,t)  \beta^\theta dt d\theta \\
        && +\gamma_{\theta\theta}(\theta,t) d\theta^2 +\gamma_{\phi_1\phi_1}(\theta,t) d\Omega^2_{q-1}
    \end{array}
\end{equation}
where $d\Omega^2_{q-1}=d\phi_1^2 +\sin^2\phi_1 d\Omega^2_{q-1}$, $\alpha(\theta,t)$ is the lapse and $\beta^{\theta}(\theta,t)$ is by symmetry the only non-zero component of the shift vector.

The $q$-form flux is time-dependent and non-uniformly distributed in the $\theta$-direction,
\begin{equation}
\label{FluxAns}
\begin{array}{lcl}
\mathbf{F}_{q}&=&  Q_B(\theta,t) N(\theta,\phi_1,\dots,\phi_{q-1}) d\theta \wedge \dots \wedge d \phi_{q-1} \vspace{1em}  \\
&&-\alpha  Q_E(\theta,t)  N(\theta,\phi_1,\dots,\phi_{q-1}) dt \wedge d\phi_1 \wedge \dots \wedge d\phi_{q-1}
\end{array}
\end{equation}
where $N(\theta,\phi_1,...,\phi_{q-1})= \sin^{q-1} \theta \sin^{q-2}
\phi_1...\sin \phi_{q-2}$ and $Q_B(\theta,t)$ and $Q_E(\theta,t)$ represent
the magnetic and electric flux strengths, respectively.

In the remainder of this section, we review a variety of features of flux compactifications in Einstein-Maxwell theory. 
In section~\ref{sec:characterizing_solutions}, we define several quantities that will be useful in describing solutions. In 
section~\ref{dimredmain}, we outline how to describe the cosmology of the non-compact space. In section~\ref{sec:FR_branch}, we
review the Freund-Rubin solutions and their stability. Finally, in section~\ref{warped}, we review the warped compactifications 
of refs.~\cite{Kinoshita:2007,Kinoshita:2009hh}. The reader interested in going directly to the results can proceed to section~\ref{sec:results}.


\subsection{Characterizing the solutions}\label{sec:characterizing_solutions}
We now define a few quantities which are helpful in describing the solutions presented below. The compact space is characterized by the volume of the internal $q$-sphere 
\begin{equation}
    \textbf{Vol}_{S^q} \equiv \int \sqrt{\gamma_q} d^q y= \int \sqrt{\gamma_q} d\theta \wedge d \phi_1 \wedge ... \wedge d \phi_{q-1} \ ,
\end{equation}
the total number of flux units, which is a conserved quantity 
obtained by integrating the flux density over the internal $q$-sphere,
\begin{equation}\label{fluxunit}
n \equiv  \int_{S_q} \mathbf{F}_q \ , 
\end{equation}
and the aspect ratio
\begin{equation}
\epsilon = \frac{\int_0^{\pi} \sqrt{\gamma_{\theta\theta}(\theta,t)} d\theta}{ \pi \sqrt{{\widetilde{\gamma}}_{\phi_1\phi_1}(\pi/2,t)}} \ ,
\end{equation}
defined such that spherical solutions have $\epsilon=1$, oblate solutions have $\epsilon <1$ and prolate solutions have $\epsilon >1$.

As a visualisation tool, we also plot the internal metric as an embedding in $q+1$ Euclidean dimensions. The internal metric $ds^2=\gamma_{\theta\theta} d\theta^2 + \widetilde{\gamma}_{\phi_1\phi_1} \sin^2 \theta d{\Omega_{q-1}}^2$ is the induced metric on the surface
\begin{equation}
\begin{array}{lcl}
x_1&=&\int^\theta_{\pi/2} d\theta' \sqrt{\gamma_{\theta'\theta'}(\theta',t)-\left[\partial_{\theta'}\left(\widetilde{\gamma}_{\phi_1\phi_1}(\theta',t)^{1/2} \sin \theta'\right)\right]^2} \vspace{1em} \nonumber \\
x_2 &=& \widetilde{\gamma}_{\phi_1\phi_1}(\theta,t)^{1/2} \sin \theta \cos \phi_1 \vspace{1em} \nonumber \\
x_3 &=& \widetilde{\gamma}_{\phi_1\phi_1}(\theta,t)^{1/2} \sin \theta \sin \phi_1 \cos \phi_2 \vspace{1em} \nonumber \\
\vdots \vspace{1em} \nonumber \\
x_{q} &=& \widetilde{\gamma}_{\phi_1\phi_1}(\theta,t)^{1/2} \sin \theta \hdots \sin \phi_{q-2} \cos \phi_{q-1} \vspace{1em} \nonumber \\
x_{q+1} &=& \widetilde{\gamma}_{\phi_1\phi_1}(\theta,t)^{1/2} \sin \theta \hdots \sin \phi_{q-2} \sin \phi_{q-1} \ .
\end{array}
\end{equation}

\subsection{The lower dimensional cosmology}\label{dimredmain}

If one hopes to make contact with the observable Universe, it is necessary to
determine the effective four-dimensional cosmology sourced by evolution of the
compact extra dimensions. The standard approach is via the procedure of
``dimensional reduction", where one integrates the action over the compact extra
dimensions and identifies a four-dimensional gravitational sector and a set of
moduli fields associated with properties of the compactification, such as the
total volume (see e.g. refs.~\cite{DeWolfe:2002nn,Giddings:2001yu} for an approach
most relevant to the present context). This approach has several limitations.
Perhaps most importantly, because one must identify a set of coordinates to
integrate over, dimensional reduction is intrinsically gauge dependent. Furthermore,
gauge dependence arises when identifying the four-dimensional gravitational sector
and moduli fields; it is typically feasible to do so only in special coordinate
systems where the symmetries of the spacetime are manifest. Without prior
knowledge of the ``right" coordinate system, it is typically only possible to
study small perturbations (see e.g. refs.~\cite{Giddings:2005ff,Frey:2008xw}). In the
context of numerical relativity, one does not have complete freedom to dictate
the coordinate system most convenient for dimensional reduction: in general, it
is necessary to specify the gauge dynamics in a way that leads to well-posed
evolution, while avoiding coordinate
singularities. Another challenge is that in the typical approach to dimensional
reduction, the goal is to find a set of equations of motion for the
four-dimensional variables, while our starting point is the solution itself.
Given a solution and not the four-dimensional equations of motion, it may not
be possible to unambiguously identify the appropriate four-dimensional
variables. These subtleties motivate an alternative approach based on the
geometrical properties of the solutions themselves, which we now outline. Note, to make contact with the observable universe we assume $p=4$.  

To motivate our approach, let us recall some properties of the standard FLRW solution in four dimensions:
\begin{equation}
ds^2 = -\alpha^2(t)dt^2 + a^2(t) (dx^2+dy^2+dz^2) \ .
\end{equation}
The extrinsic curvature of spatial slices is $K_{ii} = - a (da/d\tau)$, with $d/d\tau\equiv (1/\alpha) d/dt$, for $i=x$, $y$ and $z$, and the trace is
\begin{equation}
K = 3 {K^x}_x  = - \frac{d \ln {\rm Vol}_3}{d\tau} = -\frac{3}{a}\frac{da}{d\tau} \equiv - 3 H.
\end{equation}
where $H$ is the Hubble parameter and ${\rm Vol}_3 = \sqrt{\gamma} =a^3$ is the
normalized volume enclosed by a congruence of comoving geodesics. Note that these
equalities are contingent on the time slicing chosen here, which preserves
the homogeneity of the FLRW solution.
In this cosmological slicing, the trace of the extrinsic curvature
(or equivalently the expansion of comoving timelike geodesics) determines the Hubble
parameter. Another useful geometrical quantity is the area of the cosmological
apparent horizon. We define the cosmological apparent horizon 
as a surface where the null
expansion vanishes. (This is analogous to how apparent horizons can be used to
define black hole horizons on a specific timeslice.) In an expanding FLRW universe, the
coordinate radius of the cosmological apparent horizon is simply the comoving
Hubble radius $r_H = (aH)^{-1}$, yielding an area:
\begin{equation}
\mathcal{A}_H = 4 \pi a^2 r_H^2 = 4 \pi H^{-2} 
\end{equation}
Therefore, we see that both the extrinsic curvature and the area of the cosmological apparent horizon can be used as alternative definitions of the Hubble parameter:
\begin{equation}
H 
= - \frac{K}{3} = \sqrt{\frac{4\pi}{\mathcal{A}_H}} \ ,
\end{equation}
where again the equivalence with the usual definition of the Hubble parameter is contingent on choosing a cosmological slicing.

How does this picture generalize to the present context, where we have compact
extra dimensions? The trace of the intrinsic curvature in this case depends on
the position in the compact space and contains terms associated with the expansion
of the volume in the compact space:
\begin{equation}
K(\theta, t) = 3 {K^x}_x + {K^\theta}_\theta + (q-1) {K^{\phi_1}}_{\phi_1} = - \left( \frac{d \ln {\rm Vol}_3}{d\tau} + \frac{d \ln {\rm Vol}_q}{d\tau} \right)
\end{equation}
where for a general slicing, $d/d\tau \equiv (1/\alpha)(\partial_t-\mathcal{L}_{\beta})$,
where the last term is the Lie derivative with respect to the shift vector.
The observers associated with a general time slicing will not necessarily follow geodesics in the full D-dimensional spacetime, and restricting to
geodesic slicing can be problematic due to the appearance of coordinate
singularities. This aside, there are other subtleties associated with finding
an effective four-dimensional Hubble parameter from the extrinsic curvature. If
we were to use the trace of the extrinsic curvature, note that this includes
expansion of both the compact and non-compact space. Should one simply use
$-{K^x}_x$, which characterizes the expansion in the non-compact dimensions, or
some combination of the expansion in the compact and non-compact space? In
addition, the expansion is not homogeneous in the extra dimensions, so one must
define the correct measure of integration over the compact space to obtain the
expansion seen by an ``average" cosmological observer. 

Some insight to these
questions can be gained by investigating the properties of the cosmological apparent
horizon, which as we outlined above, can be used to define the Hubble parameter
in a four-dimensional FLRW Universe. For surfaces of constant time and (uncompactified) radius
$r \equiv \sqrt{x^2+y^2 +z^2}$ with unit inward (outward) normal $s^{\bar{m}}$ the inward (outward) null expansion
\begin{eqnarray}
\Theta_{\mp} = D_{\bar{m}} s^{\bar{m}} + K_{\bar{m}\bar{n}} s^{\bar{m}} s^{\bar{n}} -K
\end{eqnarray}
vanishes on the surface 
\begin{eqnarray}\label{eq:horizonradius}
r_H &=& \frac{\pm 2}{\sqrt{\gamma_{xx}}} \frac{1}{({K^x}_x -K)} \ .
\end{eqnarray}
A marginally inner trapped surface with $\Theta_-=0$ and $\Theta_+>0$ is a generalization
of the de Sitter horizon, while the marginally outer trapped surface with $\Theta_+=0$ and $\Theta_-<0$
that occurs for contracting spacetimes is more similar to that of a black hole apparent horizon\footnote{Though we note that the usual definition of apparent horizon in the context of dynamical black hole spacetimes
typically includes an extra condition specifying that the surface be outermost, or be at the boundary between
a trapped and untrapped region, that excludes the cosmological setting we study here.}.

The area of the cosmological apparent horizon is obtained by integrating over the compact space
\begin{eqnarray}\label{eq:horizonarea}
\mathcal{A}_H (t) &=& \int d\theta d\phi_1 \cdots d \phi_{q-1} \sqrt{\gamma_q} \ 4 \pi r_H^2 \gamma_{xx} \\
&=& \int d\theta d\phi_1 \cdots d \phi_{q-1} \sqrt{\gamma_q} \ 4 \pi \left( \frac{2}{{K^x}_x -K}\right)^2  \ .
\end{eqnarray}
Note that this is a $q+2$ dimensional area with units of $L^{q+2}$ where $L$
is some length scale. One can also use this area as a measure of entropy:
\begin{equation}\label{eq:entropy}
\mathcal{S} \equiv 2\pi \mathcal{A}_H \ ,
\end{equation}
where we recall that in our units $M_D\equiv (8\pi G_D)^{-1/(D-2)}=1$.
The connection between the area of the apparent horizon and
gravitational entropy is related to the thermodynamic
interpretation of Einstein's
equations~\cite{Jacobson:1995ab,Padmanabhan:2002sha} and
has been considered for black holes (e.g.
ref.~\cite{Hayward:1998ee}) and cosmological spacetimes
(e.g.
refs.~\cite{Frolov:2002va,Cai:2005ra,GalvezGhersi:2011tx}).
In ref.~\cite{Kinoshita:2009hh}, it was shown that for a
subset of the solutions we consider below, the entropy as
defined above is a useful indicator of stability. In
particular, for solutions at fixed conserved flux, the
stable solution has the highest entropy. Note that since this analysis is entirely classical, one could simply use the area of the cosmological horizon as a measure of stability. As for a purely
four-dimensional FLRW Universe, a Hubble parameter can be
defined by
\begin{equation}\label{eq:hubbledef}
\frac{H}{M_4} \equiv \pm \sqrt{\frac{4\pi}{\mathcal{A}_H}} 
\end{equation}
where we take the positive (negative) sign when the inward (outward) null expansion
vanishes.
In our results below where we wish to examine the effective four-dimensional
cosmology, we will use this definition of the Hubble parameter. 
Finally, we define the four-dimensional Planck mass as
\begin{equation}\label{eq:M4definition}
M_4^2 \equiv \int d\theta d\phi_1 \cdots d \phi_{q-1} \sqrt{\gamma_q (t=0,\theta)} \ ,
\end{equation}
where $\gamma_q$ refers to the background solution. 

It is useful to examine the Hamiltonian constraint equation in order to make a more direct connection with the effective four-dimensional theory. This is given by 
\begin{equation}
K^2 - K_{\bar{m} \bar{n}} K^{\bar{m} \bar{n}} = 2 \rho - {}^{(D-1)}R
\end{equation}
where $\rho = n^N n^M T_{MN}$ and ${}^{(D-1)}R$ is the intrinsic curvature on spatial slices. The extrinsic curvature term decomposes as follows
\begin{eqnarray}\label{eq:extrinsicfactorization}
K_{\bar{m} \bar{n}} K^{\bar{m} \bar{n}} -K^2 &=& -6 \left( \frac{{K^x}_x -K}{2} \right)^2 +\frac{1}{2} ({K^\theta}_\theta)^2 + \frac{(q+3)(q-1)}{4} ({K^\phi}_\phi)^2 \\
&+& \left({K^\theta}_\theta+\frac{q-1}{2} {K^\phi}_\phi\right)^2 \ . \nonumber
\end{eqnarray}
Note that choosing to isolate the factor of $({K^x}_x -K)/2$, which appeared in the expression for the cosmological apparent horizon, nicely splits the extrinsic curvature term into negative definite and positive definite components. Re-arranging the Hamiltonian constraint equation we obtain:
\begin{equation}\label{eq:Friedmann_D}
\left( \frac{{K^x}_x -K}{2} \right)^2 = \frac{1}{3 M_4^2} \rho_{\rm eff} (\theta, t)
\end{equation}
where we have defined
\begin{eqnarray}
\rho_{\rm eff} (\theta, t)/M_4^2 &\equiv& \rho  -\frac{1}{2} {}^{(D-1)}R +\frac{1}{4} ({K^\theta}_\theta)^2 + \frac{(q+3)(q-1)}{8} ({K^\phi}_\phi)^2 \\
&+& \frac{1}{2}\left({K^\theta}_\theta+\frac{q-1}{2} {K^\phi}_\phi\right)^2 \ . \nonumber
\end{eqnarray}
Equation~\eqref{eq:Friedmann_D} has the form of the Friedmann equation. The expression for the apparent horizon area eq.~\eqref{eq:horizonarea} can be used to define the measure of integration over the Hamiltonian constraint equation to give a four-dimensional Friedmann equation. In particular,
\begin{equation}
H(t)^2 =  \frac{1}{3 M_4^2} \langle \rho_{\rm eff} (t)\rangle = \frac{4\pi M_4^2}{\mathcal{A}_H(t)} \ , 
\end{equation}
where $H$ is defined as in eq.~\eqref{eq:hubbledef} and
\begin{equation}
\langle \rho_{\rm eff} (t)\rangle  \equiv M_4^2 \left[ \int d\theta d\phi_1 \cdots d \phi_{q-1} \sqrt{\gamma_q} \ (\rho_{\rm eff} (\theta, t))^{-1} \right]^{-1} \ .
\end{equation}
Note that with these definitions, the square of the Hubble parameter is inversely proportional to the entropy, so a stability criterion based on maximizing the entropy (or synonymously, the area) is equivalent to one that minimizes this definition for the Hubble parameter. 

For completeness, and because it will be useful in characterizing the properties of the solutions presented below, we sketch the standard procedure of dimensional reduction; further details can be found in appendix{~\ref{dimred}}. We begin with the D-dimensional action in ADM form:
\begin{eqnarray}\label{adm_action}
S &=& \frac{1}{2 } \int d^4 x d^qy \sqrt{-g} \left[ K_{\bar{m} \bar{n}} K^{\bar{m} \bar{n}} -K^2 + {}^{(D-1)}R -2 \Lambda_D-\frac{1}{q!} \mathbf{F}_q^2 \right] \ . 
\end{eqnarray}
The goal is to find an effective action for the four-dimensional metric variables and moduli fields, which can be identified with integrals of combinations of metric functions over the compact space (e.g. the volume). Schematically, for spacetimes that are homogeneous in the three large dimensions, the various terms in the action contribute as follows:
\begin{itemize}
\item $K_{\bar{m} \bar{n}} K^{\bar{m} \bar{n}} -K^2$: The extrinsic curvature term contains time derivatives of the metric functions, and therefore contains the 4-D Ricci scalar and kinetic terms for moduli fields. 
\item $ {}^{(D-1)}R$: The Ricci scalar on spatial slices contains spatial derivatives of the metric functions on the compact space. With our assumption that the metric is independent of the three large dimensions, there are no contributions to the 4-D Ricci scalar. This term therefore contributes only to the potential for moduli fields.
\item $2\Lambda_D + \mathbf{F}_q^2/q!$: The cosmological constant and flux terms contribute to the potential for moduli fields. 
\end{itemize}
Here, we focus on the extrinsic curvature term; additional details for specific examples can be found in appendix{~\ref{dimred}}. Factoring the extrinsic curvature term as in Eq.~\eqref{eq:extrinsicfactorization}, we have
\begin{eqnarray}
S = \frac{1}{2} \int d^4 x d^qy \sqrt{\gamma_q} \alpha \gamma_{xx}^{3/2} \left[ - 6 \left( \frac{{K^x}_x -K}{2} \right)^2 + \ldots \right] 
\end{eqnarray}
Comparing this to the action for four dimensional FLRW solutions, one can try to equate:
\begin{equation}
\sqrt{-g(t)} \ M_4^2 H(t)^2 = \int d^qy \sqrt{\gamma_q} \alpha \gamma_{xx}^{3/2} \left( \frac{{K^x}_x -K}{2} \right)^2
\end{equation}
For a convenient metric ansatz, one can explicitly identify $\sqrt{-g(t)}$,  $M_4^2$ and $H(t)^2$; we outline several examples in appendix{~\ref{dimred}}. A nice feature of the decomposition of the extrinsic curvature we have chosen is that it contains the combination of metric functions that yield a dimensionally reduced action in the four dimensional Einstein frame (e.g. the conformal frame where $M_4$ is constant). For solutions with warping there are some subtleties in finding a unique four-dimensional metric determinant and Hubble parameter which we discuss in appendix{~\ref{dimred}}. In the more general cases we consider below, where we do not have complete freedom to specify a gauge where the metric functions take a convenient form, it is not possible to unambiguously identify the four dimensional Hubble parameter. We therefore utilize the geometrical definition of the Hubble parameter based on the area of the apparent horizon in Eq.~\ref{eq:hubbledef}.

\subsection{Freund-Rubin branch}\label{sec:FR_branch}

\begin{figure}[tbp]
\centering
\includegraphics[width=1.\textwidth]{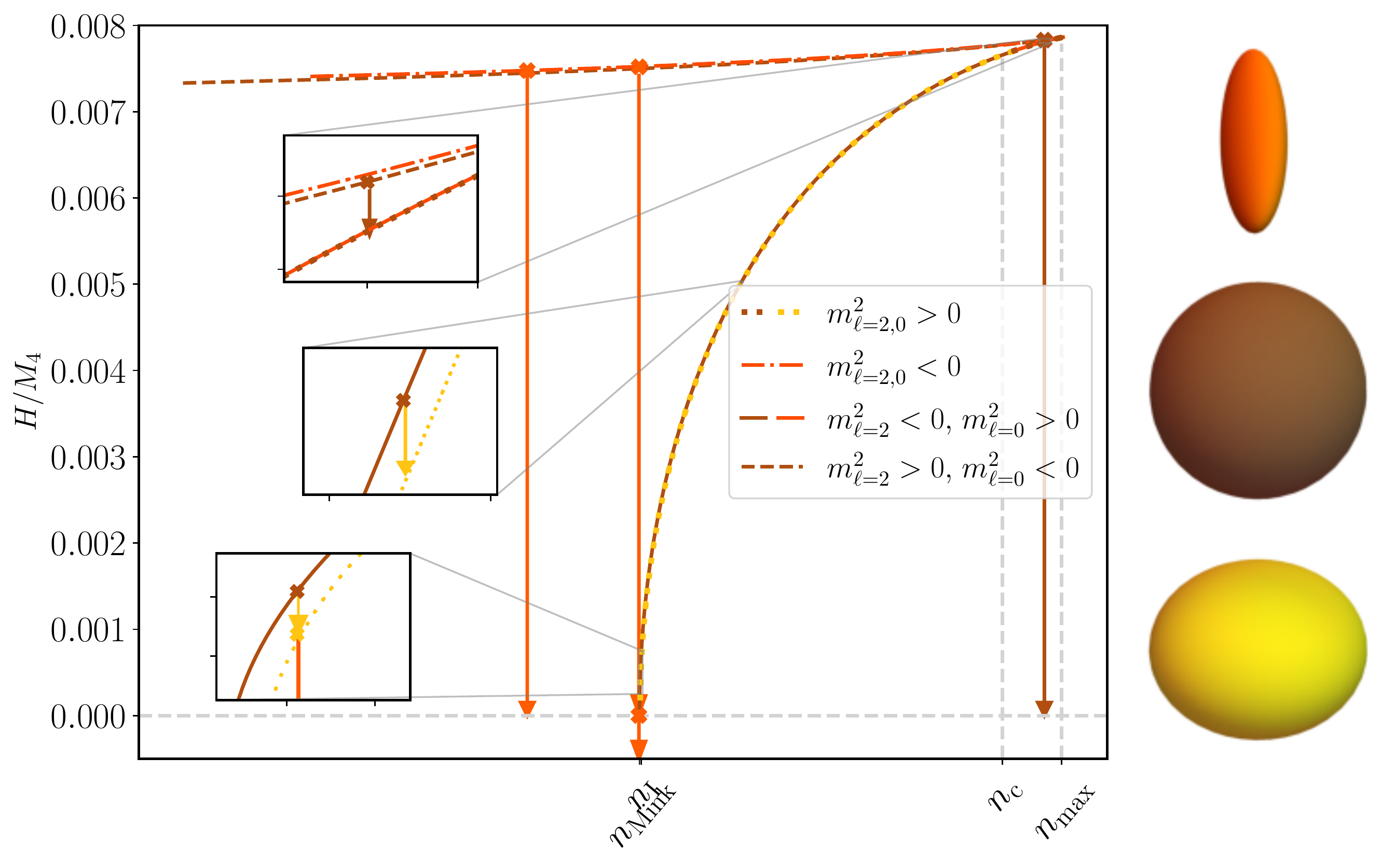}
\caption{\label{fig:cartoon}  A cartoon of the Freund-Rubin and ellipsoidal
solutions in the $(H/M_4,n)$ plane for $\Lambda_D=1$ and $q=4$. For each 
value of the conserved flux number \eqref{fluxunit}, there are two solutions: a
symmetric solution where the compact space is spherical with an aspect ratio
$\epsilon = 1$ (indicated in brown), and a warped solution where the internal
manifold is oblate with $\epsilon <1$ (yellow) or prolate with $\epsilon>1$
(orange). We find three critical values of $n$. First, for $n_{\mathrm{Mink}} <
n < n_{\mathrm{max}}$ there are two Freund-Rubin and two warped solutions: On the Freund-Rubin
branch there is a small and a large volume branch perturbatively stable or
unstable to the volume instability ($m^2_{l=0}>0$ or $m^2_{l=0}<0$)
respectively. At $n=n_{\mathrm{max}}$ the two branches merge and annihilate. On
the warped branch there is one solution on the large Hubble warped branch,
perturbatively unstable to the warped instability ($m^2_{l=2}<0$) and a
solution on the small Hubble warped branch.  At $n=n_c$ the small Hubble warped
branch intersects the small volume Freund-Rubin branch and the two branches are marginally
stable to the warped instability ($m^2_{l=2}=0$). Whenever the ellipsoidal
solution has $\epsilon >1$ it is also perturbatively unstable. Arrows indicate
the specific nonlinear solutions we discuss in section~\ref{sec:results}. They all
point towards a solution with smaller effective Hubble rate and higher entropy (area).
For a small range $n_{\mathrm{Mink}}\leq n<n_I$, solutions tend to a state where $H/M_4 <0$, the cosmological
implications of which are discussed in section~\ref{sec:unstable}
}
\end{figure}

In this paper, we consider the nonlinear evolution of perturbations to 
two classes of stationary solutions of the theory described above.
Namely, we consider the homogeneous Freund-Rubin solutions and warped solutions with a $\theta$-dependence. 
In the symmetric Freund-Rubin solution, a $q$-form flux uniformly wraps the extra dimensions into a $q$-sphere,
\begin{equation}
    \mathbf{F}_q=\rho_B \textbf{vol}_{S^q}
\end{equation}
where $\rho_B$ is the magnetic flux density and $\textbf{vol}_{S^q}=\epsilon$ is 
the volume element on the internal $q$-sphere.  The direct product condition
guarantees that the $p$ extended dimensions form an Einstein space. Restricting
to the trivial case of a maximally symmetric extended de Sitter spacetime, 
\begin{equation}
    ds^2=-dt^2+ e^{2Ht} d {\vec{x}^2}_{p-1} + L^2 d\Omega_q^2
\end{equation}
where $L$ is the radius of $q$-sphere, $H$ is the Hubble parameter \eqref{eq:hubbledef} and in the particular case where $p=4$, $d {\vec{x}^2}_{p-1}=dx^2+dy^2+dz^2$ is the usual 3-Cartesian element. 

The Maxwell equations are trivially satisfied, while the Einstein equations \eqref{eq:einstein} enforce algebraic relations between the parameters $\{ \rho_b,H,L\}$
\begin{equation}\label{FR1}
{\Lambda_D}= \frac{(p-1)^2}{2} H^{2} + \frac{(q-1)^2}{2} L^{-2}
\end{equation}
\begin{equation}\label{FR2}
\rho_B^2= -(p-1) H^{2} + (q-1)L^{-2}
\end{equation}
such that if we fix units with $\Lambda_D=1$, we are left with one free parameter describing the Freund-Rubin solutions.
This parameter can be taken to be the total number of flux units \eqref{fluxunit}
\begin{equation}
n \equiv  \int_{S_q} \mathbf{F}_q = {\rho_b} \mathbf{Vol}_{S_q} \ ,
\end{equation}
where the latter equality is specific to the Freund-Rubin solution.  From
\eqref{FR1} and \eqref{FR2}, we can see that there can be more than one
solution for a given value of $n$.  Figure{~\ref{fig:cartoon}} shows these
different solutions in $(H/M_4,n)$ space. Focusing on the spherical solutions
with aspect ratio $\epsilon=1$, the figure indicates that below some
value $n_{\text{max}}$, there exists two solutions, a small and a large volume
branch. As we will see below, the former is stable to the total-volume
instability ($\ell=0$), but may be unstable to the warped instability
($\ell \geq 2$), while the latter is unstable to the total-volume
instability, with the end point being decompactification or flow towards
the small volume solution. For $n \geq n_{\text{max}}$, there is no
solution.  

\subsubsection{The effective potential}\label{effV} 
In order to give some intuition for the stability of the flux
compactification solutions, we can go back to the dimensional
reduction procedure of section{~\ref{dimredmain}}, and considering the source terms
in eq.~\eqref{adm_action}, think of the radius $L$ of the sphere as a
four-dimensional radion field, living in an effective potential given by
\begin{equation}\label{eq:Veff}
\frac{V(L)}{M^4_4} = \frac{1}{2} \left(\frac{L_0}{L} \right)^q  \left( - \frac{q(q-1)}{L^2} + 2 \Lambda_D +\frac{1}{M_4^4}  \frac{n^2}{L^{2q}} \right) 
\end{equation}
The details of the derivation can be found in appendix{~\ref{dimred_fr}}. 
From left to right, the three terms represent the spatial curvature, the
higher dimensional vacuum energy, and the energy density of the flux,
respectively. The flux term is
repulsive, and tends to push the sphere to larger radius, but the curvature
of the compact space is attractive, such that the interaction of these two
terms can form a minimum of the potential where the radius of the
$q$-sphere can be stabilized, yielding a four-dimensional vacuum. 

\begin{figure}[h]
\centering
\includegraphics[width=.32\textwidth]{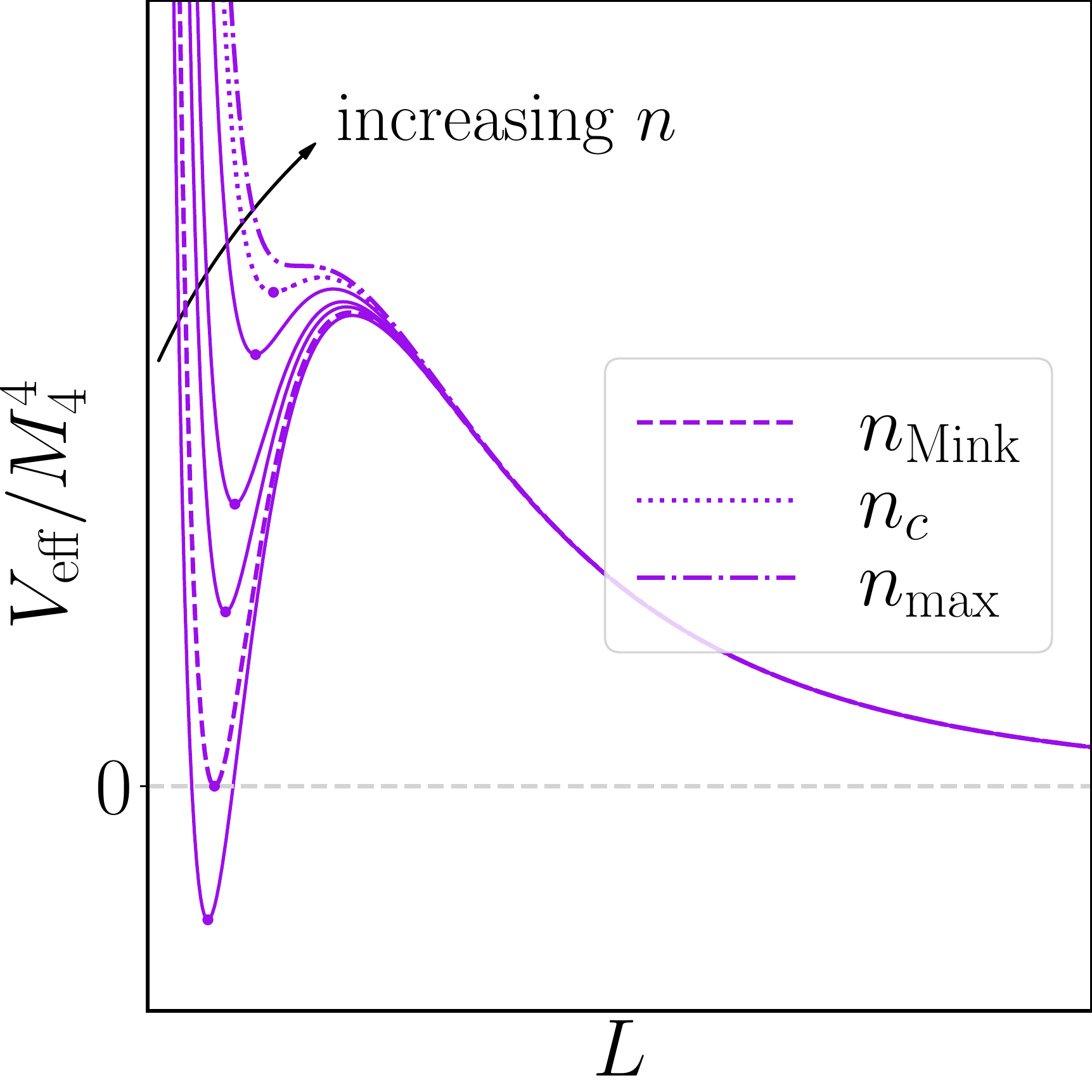}
\includegraphics[width=.32\textwidth]{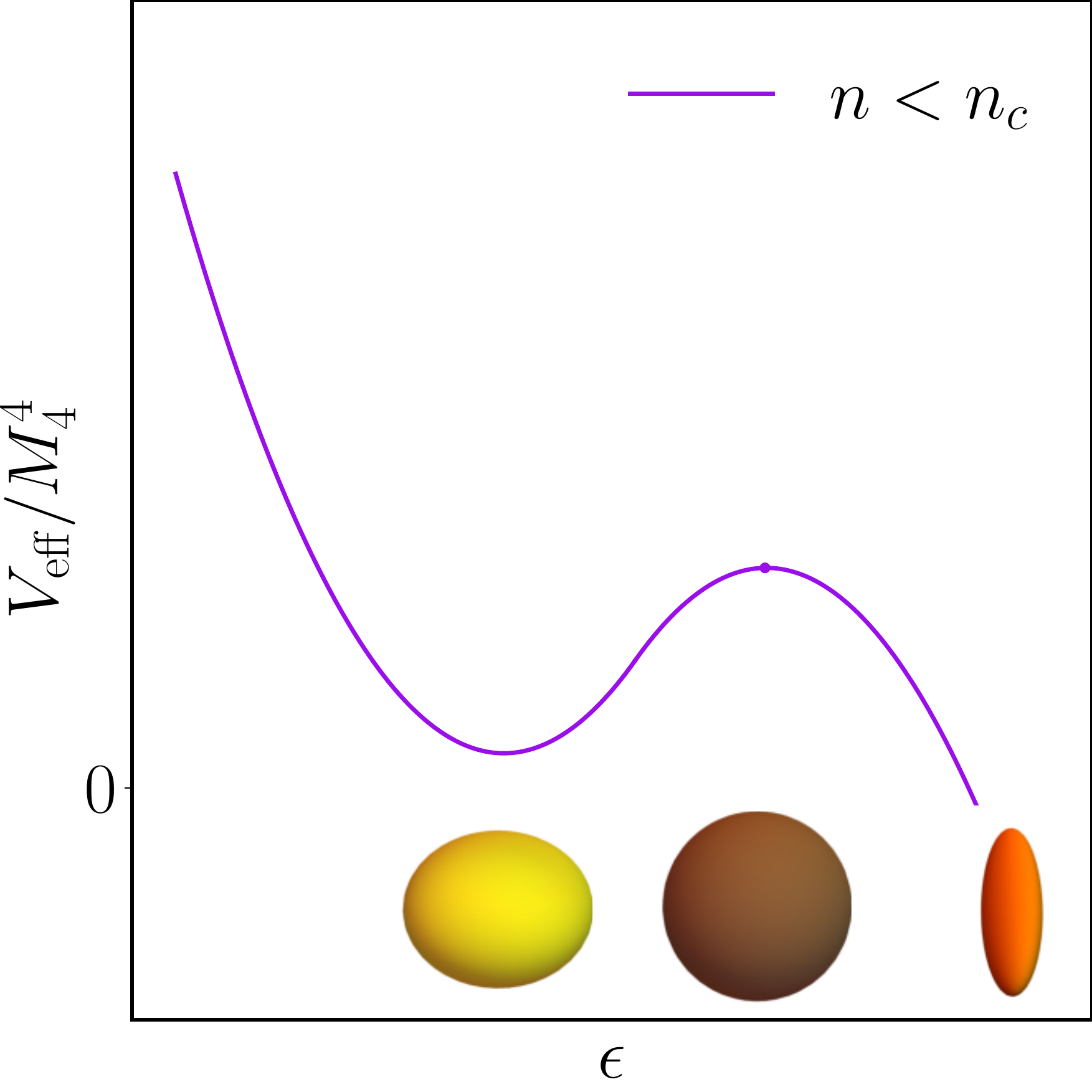}
\includegraphics[width=.32\textwidth]{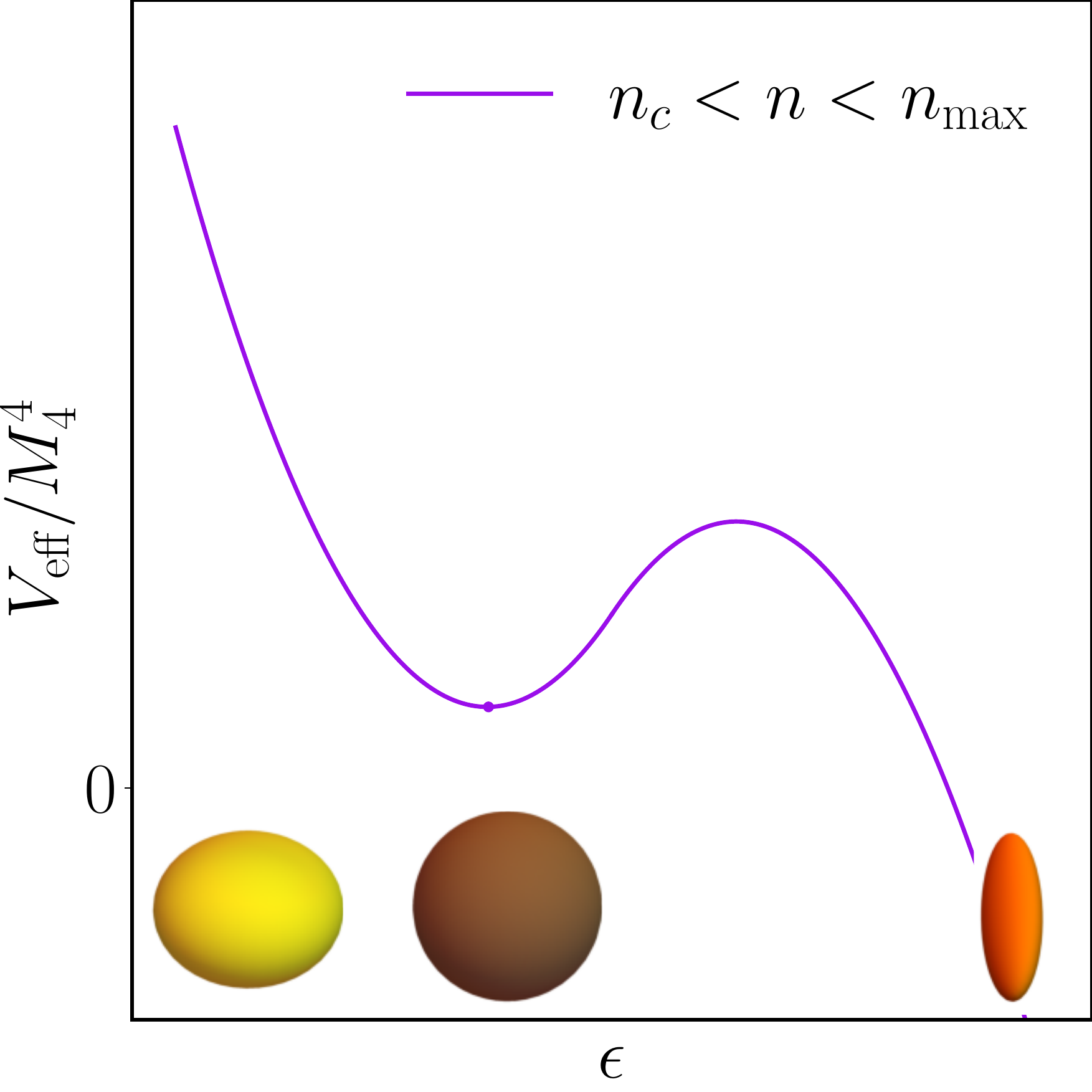}
\caption{\label{fig:Veff} 
Left: The effective radion potential eq.~\eqref{eq:Veff} for $\Lambda_D > 0$
and successively larger values of $n$ from bottom to top, assuming 
the compact space is spherically symmetric. The extrema correspond to the Freund-Rubin
solutions. For small $n$, the effective potential has a maximum
($m^2_{\ell=0} < 0$ and always de Sitter) and minimum ($m^2_{\ell=0} > 0$
and de Sitter or anti-de Sitter). At $n=n_{\mathrm{max}}$ the solutions disappear.
Right: Schematic of the effective potential for a fixed radius (minimum
of $V_{\mathrm{eff}}(L)$), but changing ellipticity. We find that the
effective potential tends to $+\infty/-\infty$ as the internal manifold
becomes increasingly oblate or prolate, respectively. The Freund-Rubin
solution is at a maximum ($m^2_{\ell=2} < 0$) when the corresponding
warped solution is oblate, and a minimum ($m^2_{\ell=2} > 0$) when the warped
solution is prolate. Some solutions escape the potential well of oblate
solution to roll in the prolate direction.
}
\end{figure}

Each allowed value of $n$, $p$ and $q$ defines a set of allowed radion
potentials or landscape of lower dimensional theories. The potential for
fixed $q=p=4$ and $\Lambda_D>0$ is sketched in figure{~\ref{fig:Veff}} for a number
of values of $n$. As we saw in the previous subsection, the number of
extrema depends on the value of $n$. For small enough $n$, the effective
potential has a minimum and a maximum corresponding to the small and large
volume branches, respectively. The extrema merge at $n = n_{\text{max}}=81 \pi^2 /\left(\sqrt{2} \Lambda_8^{3/2}\right)$
and above this value there is no solution.  Note that for small enough $n$ the
four-dimensional vacua are negative, but as $n$ increases, they eventually
become positive, which is important for cosmological
solutions.  To derive this effective potential, we assumed the shape of the
compact space is fixed. However, we will see below that minima of the
effective potential in figure{~\ref{fig:Veff}} can be unstable maxima in
other directions of the field-space that correspond to shape mode
fluctuations.

\subsubsection{Stability}
\label{stabFR}
We now briefly review linear perturbations around Freund-Rubin solutions,
restricting to scalar-type perturbations with respect to, not only the
$p$-dimensional external de Sitter space, but also the $SO(q)$ symmetry of
the background internal space. The full perturbative spectrum was studied in
refs.~\cite{Brown:2013,Hinterbichler:2014}, and we defer the reader to those references 
for a more complete analysis.  We write the perturbed metric as
\begin{equation}\label{pert}
\delta g_{\mu\nu} = -\frac{1}{p-2} g_{\mu\nu} \bar{h} Y_\ell (\theta) \ , \qquad  \delta g_{\alpha\beta}= \frac{1}{q} g_{\alpha\beta} \bar{h} Y_\ell (\theta)
\end{equation}
which tells us that the $q$-sphere is deformed with the shape of a $m=0$ spherical harmonic $Y_\ell(\theta)$ and some amplitude $\bar{h}$.

The perturbed field strength is
\begin{equation}\label{Fmn}
\delta F_{\alpha_1...\alpha_{q}} = -\bar{a} \rho_B {\epsilon}_{\alpha_1...\alpha_{q}} \lambda_\ell Y_\ell(\theta) \ , \qquad \delta F_{\beta\beta_2...\beta_{q}} = \nabla_\beta \bar{a} \rho_B {\epsilon}^{\alpha}_{\beta_2...\beta_{q}} \nabla_{\alpha} Y_\ell(\theta)
\end{equation}
where $\lambda_\ell=\ell(\ell+q-1)/ L^2 > 0$ is the eigenvalue of the
spherical harmonic, $\Box_y Y_\ell(\theta)= -\lambda_\ell Y_\ell(\theta)$
(recalling that $y$ refers to the $q$-dimensional coordinates) 
and $\bar{a}$ is a dimensionless function. Note that the equations require that $\bar{h}$ and
$\bar{a}$ shift in opposite directions ($\text{sign} \
\bar{a}=-\text{sign} (\bar{h})$), which physically means that whenever the
internal radius gets larger, the flux density also gets larger
($\text{sign} \ \delta F_{\alpha_1...\alpha_q}=-\text{sign} \ \delta
g_{\alpha\beta}$). Linearizing the Einstein-Maxwell system, we obtain a set
of ordinary, coupled, second-order differential equations for the
fluctuations, the spectrum of which can be found by diagonalization. We
find two channels of instabilities, the first due to the homogeneous mode,
the so-called volume-instability, and the second due to the inhomogeneous mode,
the so-called warped instability.

We first consider homogeneous ($\ell=0$) fluctuations in the
total volume of the internal manifold. The equation of motion is
\begin{equation}
\Box_x \bar{h}(x)= \frac{1}{L^2}\left(-2(q-1)+\frac{q(p-1)}{p+q-2}{\rho_B}^2 L^2 \right)\bar{h}(x)
\end{equation}
(recalling that $x$ refers to the p-dimensional coordinates),
which implies that the mode has positive mass when
\begin{equation}\label{vol}
{{\rho_B}^2L^2} > \frac{2(q-1)(p+q-2)}{q(p-1)}
\end{equation}
or alternatively, using eqs.~\eqref{FR1}--\eqref{FR2}, when 
\begin{equation}
{H^2} \leq \frac{2\Lambda_D  (p-2)}{(p-1)^2(p+q-2)}, \qquad \text{or} \quad {\rho_B^2} \geq \frac{2\Lambda_D }{(p-1)(q-1)} \ .
\end{equation}
This implies that if the density of the flux lines wrapping the extra
dimensions is too small, or the Hubble parameter of the external space is too
large, then there can be an instability where the total volume of the internal
manifold uniformly grows or shrinks, but the shape of the compactified sphere is
fixed. Stable de Sitter solutions are on the small-volume branch, while 
unstable ones are on the large volume branch and correspond to a maximum of the
effective potential.

Now looking at the coupled scalar sector, which will be the main focus of this
paper, then when $q\geq4$, perturbations with polar number $\ell \geq 2$ can be unstable.
Mathematically, this instability arises
from the coupling of the metric and flux perturbations, their equations of motion being
\begin{gather}
 \Box_x \begin{pmatrix} \tilde{h} \\ \tilde{a} \end{pmatrix}
 = 
  \begin{bmatrix}\frac{1}{L^2} \begin{pmatrix}
   -q\frac{q-1}{p+q-2}{\rho_B}^2L^2 & 0  \\
   0 & 0 
   \end{pmatrix} +M \end{bmatrix} \begin{pmatrix} \tilde{h} \\ \tilde{a} \end{pmatrix}
\end{gather}
where $M$ is a $2 \times 2$ matrix given by 

\begin{gather}
 M
 = \frac{1}{L^2}
  \begin{pmatrix} -L^2\lambda-2(q-1) + q{\rho_B}^2 L^2-2\frac{q-1}{q}L^2 \lambda & \frac{-4}{{\rho_B}^2L^2}\frac{q-1}{q} L^2 \lambda(L^2\lambda+q) \\
  \frac{q-1}{q}{\rho_B}^2L^2 & -L^2\lambda +2\frac{q-1}{q}L^2\lambda
   \end{pmatrix} 
\end{gather}
where $\tilde{h}=\bar{h}-2\lambda_{\ell} a$ and $\tilde{a}=\rho_B \bar{a}$.

The mode will be stable provided the eigenvalues of $M$ are positive, which for $\ell \geq 2$ implies
\begin{equation}\label{lumpy}
{\rho_B}^2L^2< \frac{\ell(\ell+q-1)-2q+2}{2(q-2)}\frac{p+q-2}{p-1} \ ,
\end{equation}
or equivalently when 
\begin{equation}\label{tachyonl2}
{H^2} \geq \frac{2\Lambda_D  \left((p-1)q^2-(3p-1)q+2\right)}{q(q-3)(p-1)^2(p+q-2)}, \qquad \text{or} \quad {\rho_B^2} \leq \frac{4\Lambda_D }{q(q-3)(p-1)} \ .
\end{equation}
Taking $p=4$, one finds that for $q=2$ or $q=3$, de Sitter vacua are only
unstable to the $\ell=0$ mode. For $q=4$, the only excited mode to develop a
negative mass is $\ell=2$. For $q \geq 5$, all de Sitter solutions are unstable to
$\ell=0$ or $\ell=2$ fluctuations. Note that the case of $q=4$ is interesting because it has
a window of stability in the range of fluxes allowed by eqs.~\eqref{vol} and
\eqref{lumpy}. The warped instability signals the presence of a new branch
of deformed solutions, which we describe next.

\subsection{Warped branch}\label{warped}
\begin{figure}[h]
\centering
\includegraphics[width=.47\textwidth]{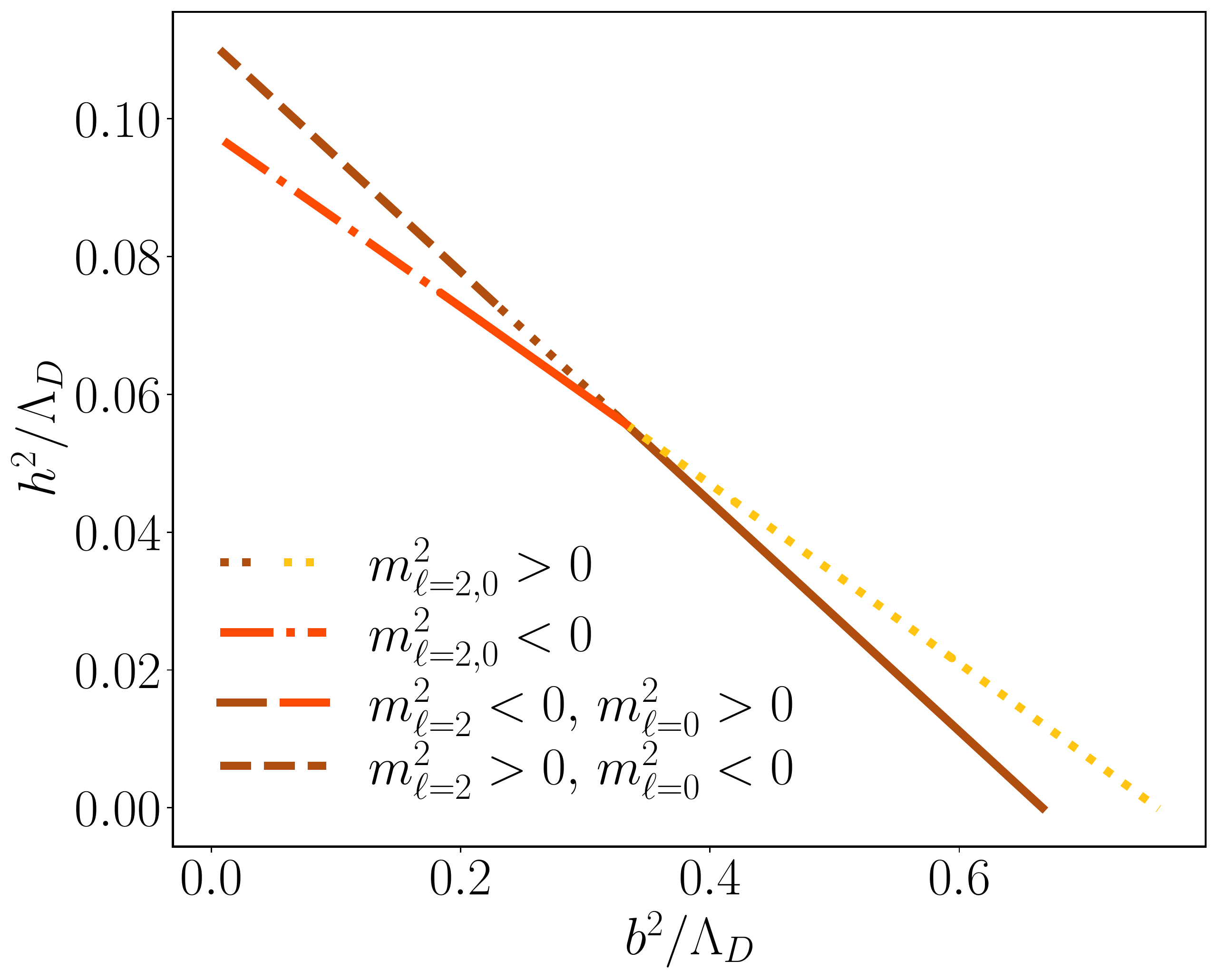}
\includegraphics[width=.47\textwidth]{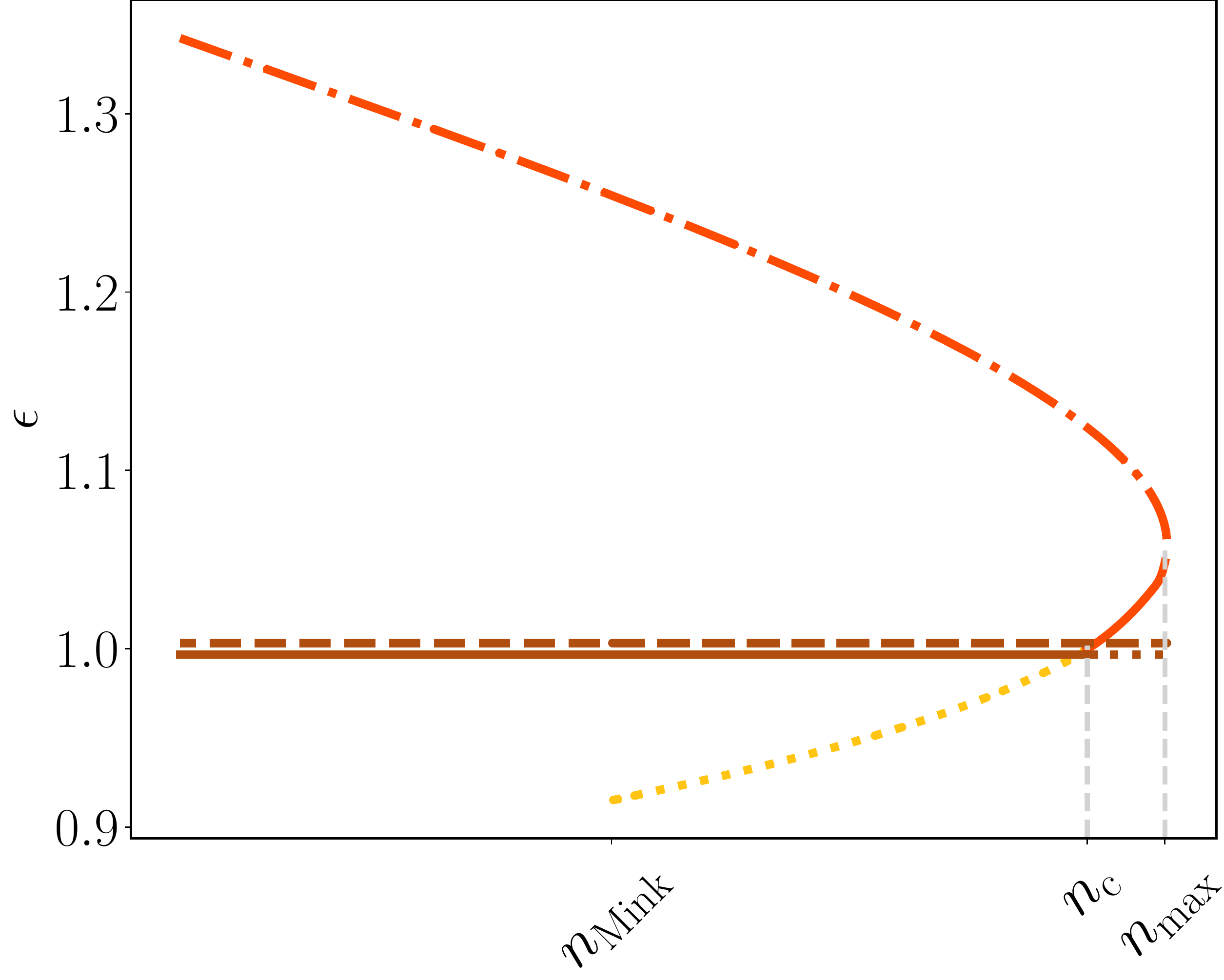}
\caption{\label{fig:warpedspace} Two branches of solutions to
\eqref{eq:warpedeqn} in the $(b^2/\Lambda_D,h^2/\Lambda_D)$ (left) and $(\epsilon,n)$ plane (right).
The brown lines represent the Freund-Rubin branch, while the yellow (oblate) and orange (prolate) lines represent the interpolation of warped solutions reconstructed numerically. The two branches intersect at
$({b_{cr}}^2/\Lambda_D,{h_{cr}}^2/\Lambda_D)=(0.36,0.052)$. The linestyle reflects the perturbatively unstable modes for the two branches. See figure{{~\ref{fig:cartoon}}} for more details.
}
\end{figure}

In the previous section, we described the symmetric Freund-Rubin solutions, and saw
that there is a critical value of $n$ above which inhomogeneous perturbations
develop a tachyonic mass. This suggests that there may be other warped
solutions obeying the Einstein-Maxwell system of equations.
References \cite{Kinoshita:2007,Lim:2012,Dahlen:2014} constructed stationary prolate or
oblate topological spheres numerically, and ref.~\cite{Kinoshita:2009hh} studied
their linear stability. One way to describe such warped solutions is by the
following metric ansatz
\begin{equation}\label{dsKim}
    \begin{array}{lcl}
        ds^2&=&e^{2\phi(\tilde{\theta)}}\left[-dt^2+e^{2ht} d\vec{x}_{p-1}^2\right]+e^{-\frac{2p}{q-2}\phi(\tilde{\theta)}} ( d{\tilde{\theta}}^2 +a(\tilde{\theta})^2 d\Omega^2_{q-1})
    \end{array}
\end{equation}
and flux
\begin{equation}\label{FqW}
F_q(\tilde{\theta})= b \ a(\tilde{\theta})^{q-1} e^{ -\frac{2p(q-1)\phi(\tilde{\theta)}}{(q-2)}}\sin^{-(q-1)}(\tilde{\theta}) \ N(\tilde{\theta},\phi_1,...,\phi_{q-1}) \ d\tilde{\theta} \wedge ... \wedge d\phi_{q-1} 
\end{equation}
where the internal coordinate $\tilde{\theta}$ lies in the finite interval
$\tilde{\theta}_{-} < \tilde{\theta} < \tilde{\theta}_{+}$, with
$\tilde{\theta}_{-/+}$ designating the two poles \cite{Kinoshita:2007},
and where $b$ and $h$ are constants such that $b=\rho_B$ and $h=H$ whenever one
recovers the Freund-Rubin solution with $\phi(\tilde{\theta})=0$ and
$a(\tilde{\theta})=L$. 
Note that eq.~\eqref{dsKim} can be put in the form of
eq.~\eqref{metricW}, provided one performs the following coordinate
transformation
\begin{equation}\label{transf}
\theta \rightarrow \frac{\tilde{\theta}}{L} + \frac{\pi}{2}
\end{equation}
where $L=2\tilde{\theta}_{+}/\pi$. The inhomogeneous flux, eq.~\eqref{FqW},
automatically satisfies Maxwell's equations and the Bianchi identity. Plugging
in our ansatz, the Einstein equations give us two equations involving second
derivatives of the metric 
\begin{equation}\label{eq:warpedeqn}
\phi''= (p-1)h^2 e^{-\frac{2(D-2)\phi}{q-2}} -(q-1)\frac{{a}'}{{a}}\phi'  +  e^{-\frac{2p\phi}{q-2}}\frac{1}{(D-2)} \left ( -2\Lambda_D + (q-1) b^2   e^{ -2p\phi} \right)
\end{equation}
\begin{equation}
\frac{{a}''}{{a}}=-\phi'^2\frac{p(D-2)}{(q-2)^2} - {a}^{-2}+\frac{{a}'^2}{{a}^2}
\end{equation}
and one equation involving first derivatives 
\begin{eqnarray}
(q-1)(q-2)\frac{{a}'^2}{{a}^2}&=&(q-2)(q-1){a}^{-2} +\frac{p(D-2)}{q-2}\phi'^2 +p(p-1)h^2 e^{-\frac{2(D-2)\phi}{q-2}}-2  e^{\frac{-2p}{q-2}\phi}\Lambda_D \vspace{1em} \nonumber  \\
&&+b^2 e^{\frac{-2p(q-1)}{q-2}\phi} \ ,
\end{eqnarray}
where the prime denotes the derivative with respect to $\tilde{\theta}$.
Using the procedure outlined in ref.~\cite{Kinoshita:2009hh}, we solve these
equations, and hence construct warped solutions. We refer the reader to
ref.~\cite{Kinoshita:2009hh} for more details. Note that we assume that the
internal space is symmetric about the equator since the
linear analysis shows that the first mode to become tachyonic is
quadrupolar ($\ell=2$).  Figure{~\ref{fig:warpedspace}} shows the two
one-parameter families of solutions, namely the trivially warped
Freund-Rubin solutions, and the non-trivially warped solutions, in the
$(b^2/\Lambda_D,h^2/\Lambda_D)$ (left) and $(\epsilon,n)$ (right) planes. This figure shows that the
two branches intersect at a single point
$({b_{cr}}^2/\Lambda_D,{h_{cr}}^2/\Lambda_D)=(0.36,0.052)$ where the only solution is the
trivial one, and the compact space is a perfect sphere. For values of
$b<b_{cr}$, the internal compact space is prolate, while for values
$b>b_{cr}$, it is oblate. This is particularly important as, according to eq.~\eqref{tachyonl2}, this critical point coincides with the point at which
the $\ell=2$ mode of the Freund-Rubin branch becomes massless. In other words, the
warped branch emanates from the marginally stable Freund-Rubin solution, as one
would expect.

\subsubsection{Stability}
The spectrum for scalar perturbations of the warped solutions was studied
in \cite{Kinoshita:2009hh}. Computing the eigenspectrum in a similar way
to the Freund-Rubin solutions, one finds that the marginal stability of
the warped solutions coincides with the marginal stability of the
Freund-Rubin branch. In particular, for the $\ell=2$ mode, when $h^2$
satisfies the first inequality given by eq.~\eqref{tachyonl2}, then the
eigenvalue of the warped branch is positive, while the eigenvalue of the
Freund-Rubin solution becomes negative.  In other words, in the low Hubble
regime, where the Freund-Rubin branch is unstable to inhomogeneous
excitations, the warped branch is perturbatively stable. Conversely, the
warped branch is unstable to inhomogeneous perturbations in the regime
where the Freund-Rubin branch is stable. Additionally, the mass squared of the
warped branch is larger than that of the symmetric branch in the regime
where the latter is unstable, which in turn implies that the warping of the
internal compact space stabilizes the shape mode of the compact space.

Alternatively, one can use a thermodynamic argument. Recall that the
entropy is defined by eq.~\eqref{eq:entropy}, where $H$ is defined by the
cosmological apparent horizon (see appendix{~\ref{dimred}} for an
explicit derivation for the warped metric ansatz). As shown in
\cite{Kinoshita:2009hh}, the thermodynamic stability of these solutions
agrees with their dynamical stability. In other words, when
$n_c<n<n_{\mathrm{max}}$, where $n_c=32 \sqrt{3} \pi^2/\Lambda_8^{3/2}=0.97 n_{\rm max}$, the small volume Freund-Rubin branch has a smaller
Hubble parameter, or larger entropy (area), and hence is thermodynamically
preferred. On the other hand, when $n<n_c$, the warped branch has
smaller Hubble or larger entropy and is thermodynamically preferred.
This is shown in figure{~\ref{fig:cartoon}}.  At the linear level, the
dynamical and thermodynamic stability of the Freund-Rubin solutions
determine the shape and stability of the warped solutions.
Reference~\cite{Dahlen:2014} sketched an effective potential that
neatly encapsulates this behaviour. In the effective theory described
by eq.~\ref{eq:Veff}, only the radius of the solution is treated as a
dynamical radion field. If we now allow the shape of the compact space
to vary as well, we must treat the aspect ratio as a dynamical field,
and extend the effective potential to be a function of $L$ and
$\epsilon$. Minima of the potential in the $L$ direction are now minima
or maxima in the $\epsilon$ direction depending on whether the solution
is stable or unstable to shape fluctuations which in turn depends on
its conserved flux number.  Reference \cite{Dahlen:2014} argued that
this effective potential is captured by a cubic potential schematically
drawn in figure{~\eqref{fig:Veff}}, and that the effective potential
asymptotes to $V \rightarrow + \infty$ in the oblate direction and $V
\rightarrow - \infty$ in the prolate direction.  Intuitively, one would
expect that in the direction of decreasing $\epsilon$, the equatorial
radius is increasing, and flux is concentrating there such that the
solution eventually settles to a minimum. On the other hand, as
$\epsilon$ increases, the equatorial radius will decrease and the flux
concentrates at the poles. Having no flux to support the equator, the
sphere collapses to zero radius and the potential \eqref{eq:Veff} tends
to $V \rightarrow - \infty$. What happens to the solution as it rolls
down the potential is unclear. In the next section, we verify this
general picture nonlinearly and study the endpoint of the solutions.
The evolution and endpoints of the unstable solutions are summarized in
figure{~\ref{fig:cartoon}}.

\section{Numerical implementation}\label{sec:numerics}
We evolve the Einstein equations using the generalized harmonic formulation
\cite{Garfinkle:2001ni,Pretorius:2005}, where the gauge degrees are specified by choosing the
source functions which determine the covariant d'Alembertian of the
coordinates, $H^M=\Box x^M$. 
We choose our evolution variables according to a space-time decomposition of 
the metric.
See appendix{~\ref{eom}} for the evolution variables and equations of
motion. 

To numerically evolve the system, we discretize in time and $\theta$.
To avoid solving the equations directly on the poles we use a shifted grid
\begin{equation}
\theta_j=\frac{j+1/2}{N_{\theta}}\pi, \qquad j=0,\ 1,\dots,\ N_{\theta}-1 \ .
\end{equation}
We expand the evolution variables as a sum of sines or cosines, depending on
the parity of the function around the pole, and use pseudospectral methods to
calculate the spatial derivatives. 
The variables are evolved in time using fourth-order Runge-Kutta time stepping.
High-frequency spectral noise is reduced by applying an exponential filter
\cite{Majda7511}. This filter is applied to the coefficients of every
derivative function and directly to the coefficients of the solution at the end
of each time step. The coordinate freedom is fixed by choosing the source
functions. These are set to be such that the shift is driven to zero and the
lapse remains approximately constant when the solution remains close
to the background solution. This avoids extra dynamics coming solely
from gauge transitions (as opposed to physical instability).
  
During the evolution,
we search for, and in some cases find, trapped regions: 
points where both (i.e. the nominally ``inward" and ``outward") null geodesics moving in the $\theta$ direction 
must have the same sign for the derivative with respect to the
affine parameter $d\theta/d\lambda$. 
In such cases, we excise a causally disconnected region
bounded by such a point where the null geodesics are both ingoing, 
and instead use fourth order finite difference stencils 
to calculate derivatives, and
Kreiss-Oliger dissipation to reduce the high-frequency noise \cite{Kreiss}. In this
way, we continue to evolve the spacetime outside the trapped regions.

We construct initial data describing perturbed Freund-Rubin
or stationary warped solutions. To do this, we take the background metric on the initial 
time slice and add the perturbation given by eq.~\eqref{pert} with some specified amplitude. We then solve for the
initial electric and magnetic forms using the Hamiltonian and momentum constraints (hence our perturbed solutions
still exactly satisfy the constraints). This procedure gives rise to a slightly perturbed flux number, although close enough to 
background value to not affect the properties relevant for assessing stability. See appendix{~\ref{sec:conv}} for results illustrating that
we start with sufficiently small perturbations so as to be in the linear instability regime, as well as numerical convergence.  

In the case where the background solution is a warped solution, we construct the background solution using the procedure
described in section~\ref{warped}. Note that most of the solutions presented below are for $q=4$, where 
only the $\ell=0$ or $\ell=2$ modes can be perturbatively unstable. We therefore only consider $\ell=0$ or $\ell=2$ perturbations, 
leaving the investigation higher modes in solutions with more dimensions for future work.

\section{Results}
\label{sec:results}
We now present our numerical solutions, restricting to $p=4$ to make contact with cosmology.

\subsection{Total volume instability of Freund-Rubin solutions}\label{hgFR}
We begin with a discussion of the total volume instability, which affects Freund-Rubin solutions
on the large volume branch. We obtain 
results similar to ref.~\cite{Krishnan:2005}, which studied the cases where the
dimensionality of the $q$-sphere was two or three. In those cases,
the homogeneous mode is the only one excited, and hence the
inhomogeneous perturbations ($\ell > 0$) can be set to zero. Here, our initial conditions are  
Freund-Rubin solutions on the large volume branch in theories with $q=4$. 
Note that for $q=4$, the small volume branch is vulnerable to the warped instability, 
but the large volume branch is not. This guarantees that, at least initially, time evolution 
does not break the spherical symmetry of the compact space. 

In the absence of the warped instability, we can understand the time evolution entirely from the 
perspective of the four dimensional effective theory (see appendix{~\ref{dimred}} for further details): 
Einstein gravity with a scalar field describing the radius of the compact sphere that evolves in the potential depicted in figure{~\ref{fig:Veff}}. Our 
initial condition lies at the maximum of the effective potential, and the evolution will take the solution 
either to the potential minimum (corresponding to the $dS_p \times S_q$ solution on the small-volume branch) 
or to a solution that decompactifies to $D=p+q$ dimensional de Sitter space. 

We find that the results of the full nonlinear evolution away from the large volume Freund-Rubin 
solutions are as expected from the four dimensional effective theory. A small positive perturbation to the total volume leads 
to decompactification while a small negative perturbation evolves toward the stable small-volume $dS_p \times S_q$ solution. 
For the solutions that decompactify, we confirm that the 
curvature scalar asymptotes to what is expected for $D=p+q$ dimensional de Sitter space with a
cosmological constant $\Lambda_D$, i.e.
\begin{equation}
{}^{(D)}R=\frac{2 D \Lambda_D }{D-2} \ .
\end{equation}
The evolution does not lead to any significant growth away from homogeneity as the solution decompactifies, as 
expected based on the absence of the perturbative warped instability on the large volume branch.

\begin{figure}[h]
\centering
\includegraphics[width=.327\textwidth]{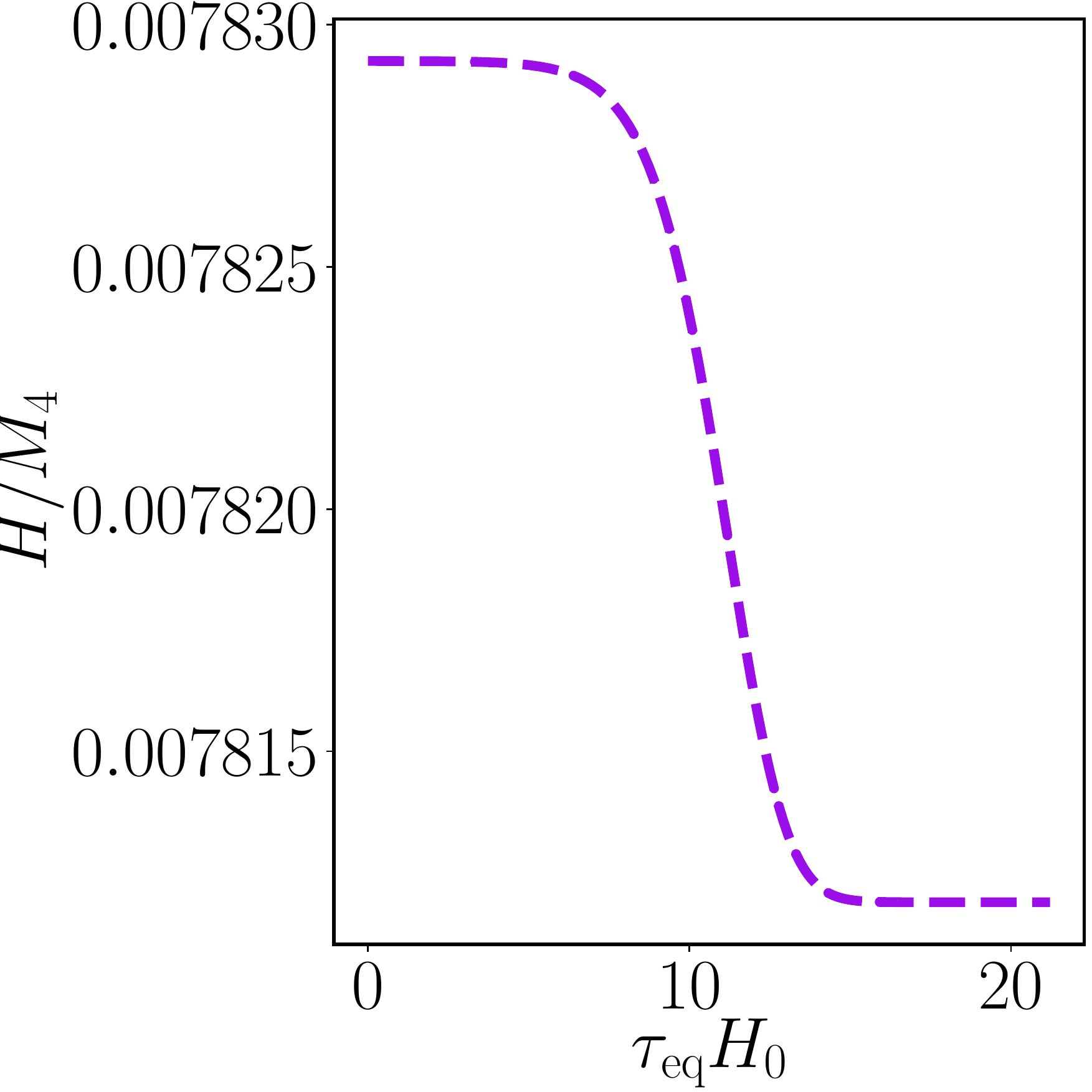}
\includegraphics[width=.327\textwidth]{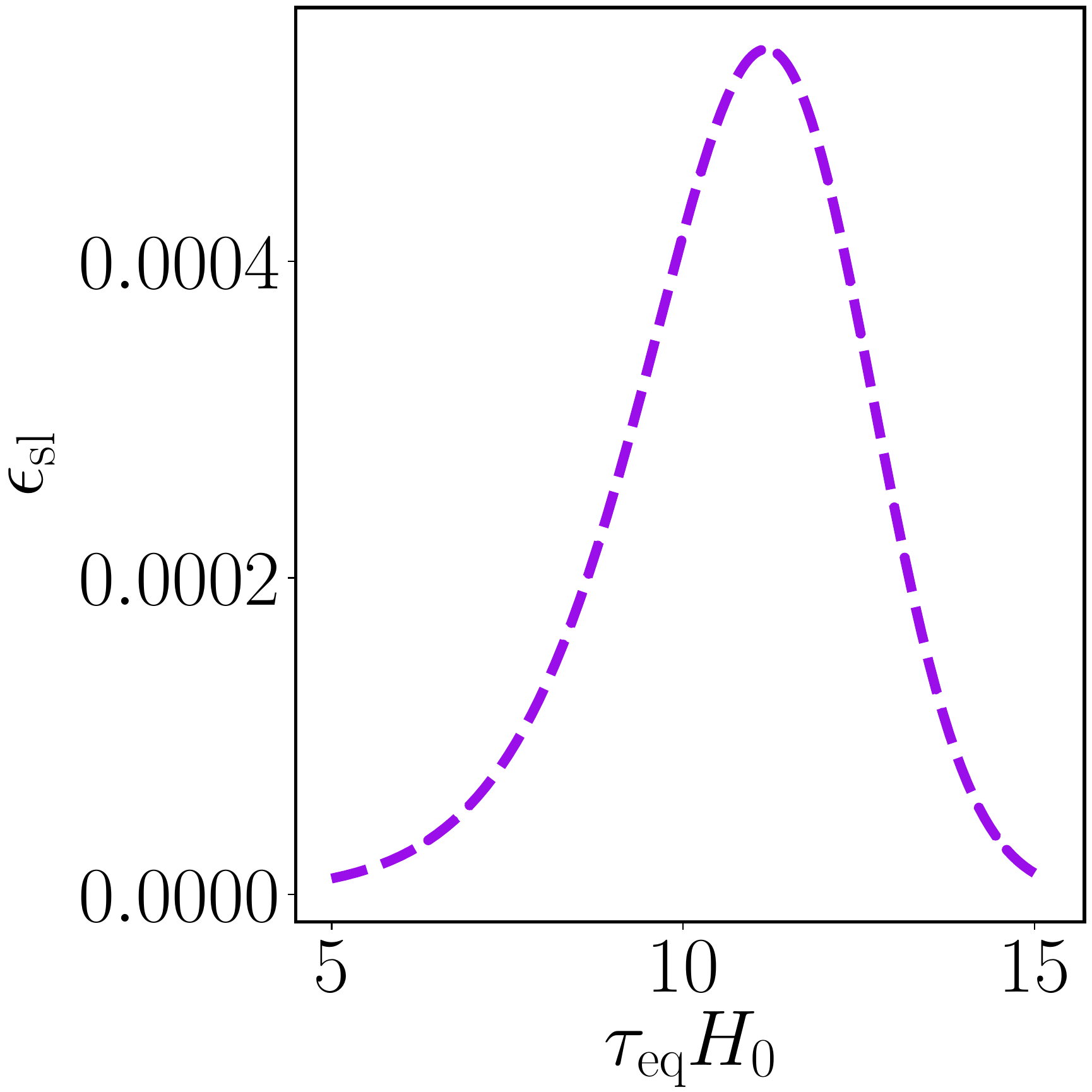}
\includegraphics[width=.327\textwidth]{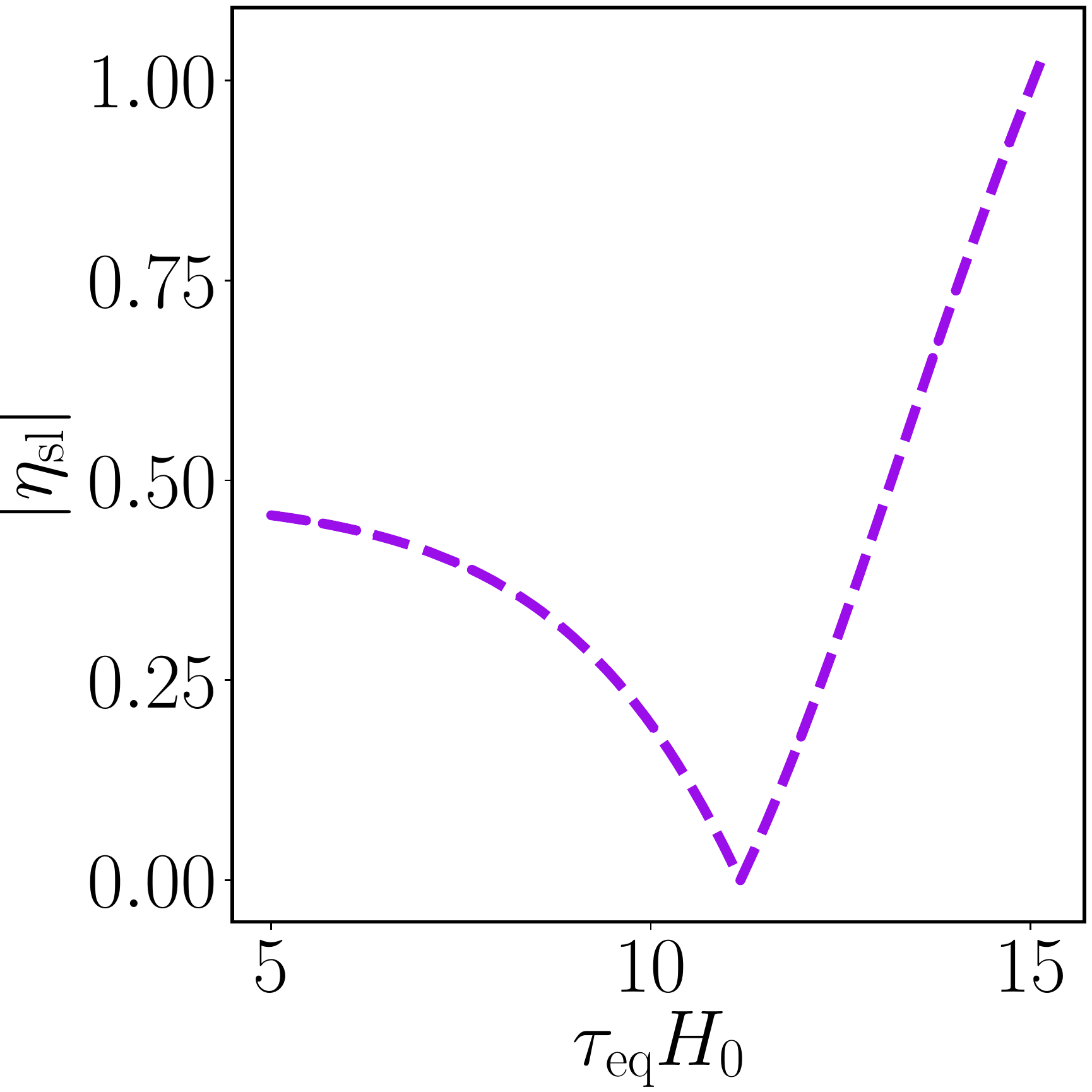}
\caption{\label{fig:q4l0A3.5} Sample solution on the large volume branch
with $p=q=4$, $ \Lambda_D=1, H_0/M_4=0.0078$ and an initial $\ell=0$
perturbation. Left: The effective Hubble rate, $H/M_4$, flows to the
solution with the same value of $n$ on the small volume branch. The slow
roll parameters $\epsilon_{\mathrm{sl}}$ (middle) and $\eta_{\mathrm{sl}}$
(right) during the transition. All plots are shown as a function of proper
time (in particular as measured at $\theta=\pi/2$, though here the
solutions remain homogeneous).}
\end{figure}

For a negative total volume perturbation, the solution eventually settles
to the small volume $dS_p \times S_q$ solution with the same conserved
flux $n$ as the initial condition.   This end state has slightly smaller
radius and a slightly lower Hubble parameter (as computed from
eq.~\eqref{eq:hubbledef}) compared to the initial large volume Freund
Rubin solution. In the left panel of figure{~\ref{fig:q4l0A3.5}}, we show
the Hubble parameter as a function of the proper time at the equator,
defined by $d\tau = \alpha(t,\theta=\pi/2) dt$ (though here the solutions
remain homogeneous and the value of $\theta$ is irrelevant), where it can be seen that the evolution
smoothly connects the large and small volume solutions.
A cosmological observer in four
dimensions would observe a brief period of quasi-de Sitter expansion,
followed by pure de Sitter evolution. In the center and right panels of
figure{~\ref{fig:q4l0A3.5}}, we plot the slow-roll parameters defined
by:
\begin{equation}\label{eq:slow_roll}
\epsilon_{\rm sl} \equiv - \frac{1}{H^2} \frac{dH}{d\tau_{\rm eq}}, \ \ \ \ \eta_{\rm sl} \equiv \epsilon_{\rm sl} - \frac{1}{2 H \epsilon_{sl}} \frac{d\epsilon_{\rm sl}}{d\tau_{\rm eq}} \ .
\end{equation}
Both remain less than one over the duration of $N = H_0 \tau_{\rm eq} \sim 5$ e-folds. This implies that the transition from the large to the small-volume branch describes a short bout of slow-roll inflation. The evolution described here therefore serves as a toy model of inflation as driven by the volume modulus of a compactification. Solutions for other choices of the flux $n$ are qualitatively similar.

\subsection{Warped instability of Freund-Rubin solutions}
We now move on to study perturbations that excite the warped instability of Freund-Rubin 
solutions on the small volume branch. Recall that the small volume solutions are stable to the 
total volume instability (in the four dimensional effective theory they sit at a minimum of the 
effective potential), but when $q \geq 4$ they may be vulnerable to
the warped instability. We focus on the case where an $8$-dimensional spacetime is
compactified down to four dimensions, and the internal manifold has the
topology of a $4$-sphere. This is an interesting scenario because, as we saw in
Sec.~\ref{stabFR}, it features both a window of stability $n_c < n < n_{\rm max}$ in the range of fluxes
allowed by eqs.~\eqref{vol} and \eqref{lumpy}, as well as perturbatively unstable solutions for $n < n_c$. For initial
 conditions, we start from a small volume Freund-Rubin solution with an $\ell=2$ perturbation; this is the 
 only unstable mode in the linear regime. 

\subsubsection{Linearly stable solutions: $n_c < n < n_{\rm max}$}
We first explore the range of fluxes $n_c < n < n_{\rm max}$ where the Freund-Rubin solution is linearly stable
to both homogeneous and inhomogeneous perturbations. Although sufficiently small
perturbations should decay, we can ask what will happen if one adds a
sufficiently large homogeneous ($\ell=0$) or inhomogeneous ($\ell = 2$) perturbation of the form given by eq.~\eqref{pert}. In particular,
appealing to the effective potential picture, it is not hard to imagine that
the solution, originally sitting at the minimum of the potential well, will be
kicked out, provided the perturbation is sufficiently large. 

For large $\ell=0$ perturbations, we expect the solution to reach the maximum of the
effective potential after which it will decompactify. 
Indeed we find that when the size of the perturbation is such that the initial volume of the perturbed solution exceeds its large volume value it will decompactify.

In the effective potential depicted in the right panel of figure{~\ref{fig:Veff}}, one can think of the potential maximum 
as corresponding to the stationary but unstable prolate solution on the warped branch with the same value 
of the conserved flux $n$ as the small-volume Freund-Rubin solution. Adding $\ell =2$ perturbations of 
increasing size, one eventually approaches a configuration close to this unstable prolate solution. Once the size 
of the perturbation exceeds this point, we expect the solution to become increasingly prolate. However, since the effective 
potential for the aspect ratio $\epsilon$ is only qualitative, we do not have a concrete prediction for the end-state. Likewise, with 
no stable warped solution to flow to, thermodynamic arguments are not of much help in determining the end-state. 
Note that we still put in an initial perturbation to the metric of the form given by eq.~\eqref{pert}
(but with nonlinear corrections to the $q$-form through the constraints, as described in section~\label{sec:numerics}), even as we consider
large perturbations beyond the linear regime.

\begin{figure}[h]
\centering
\includegraphics[width=.327\textwidth]{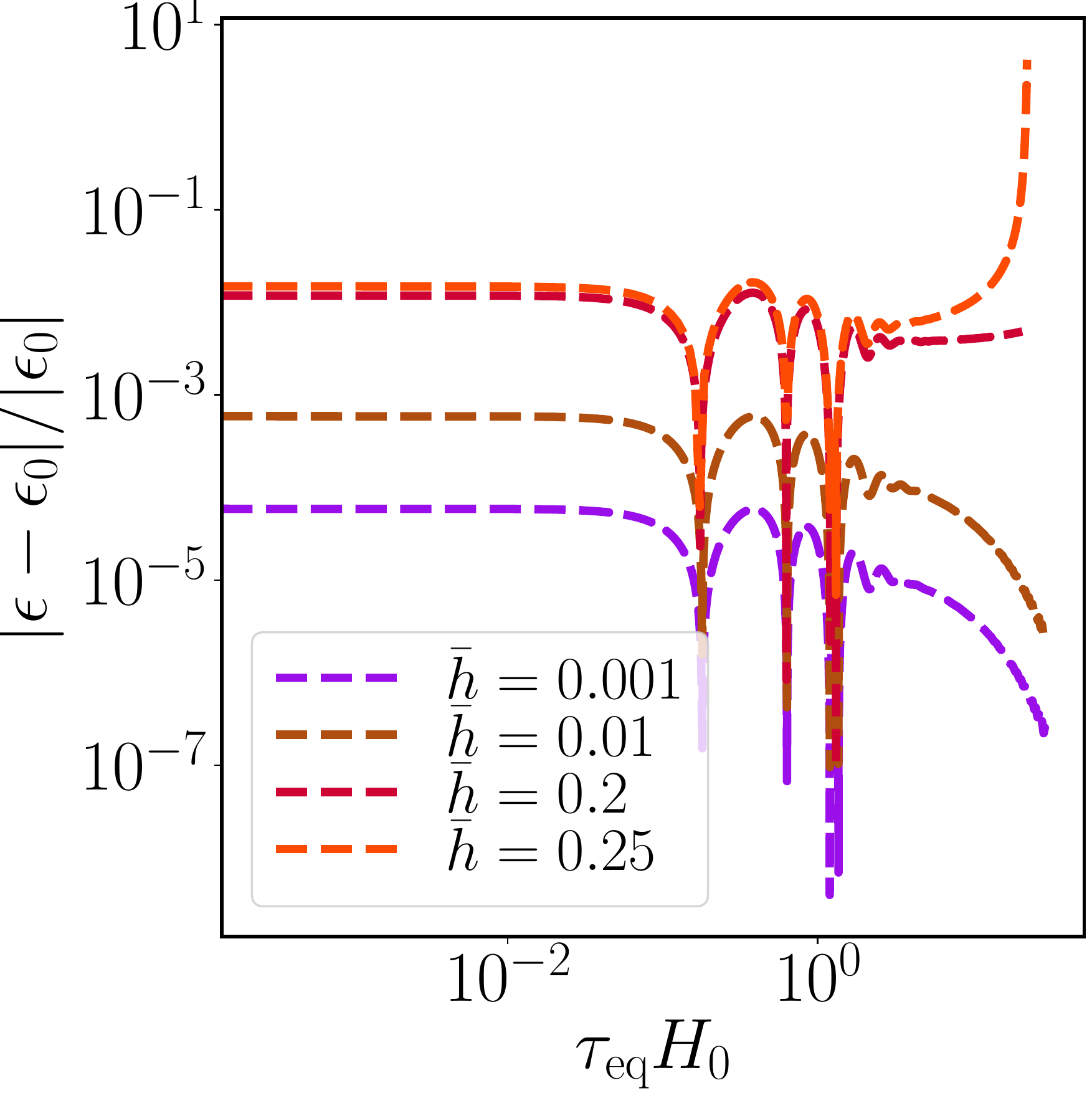}
\includegraphics[width=.327\textwidth]{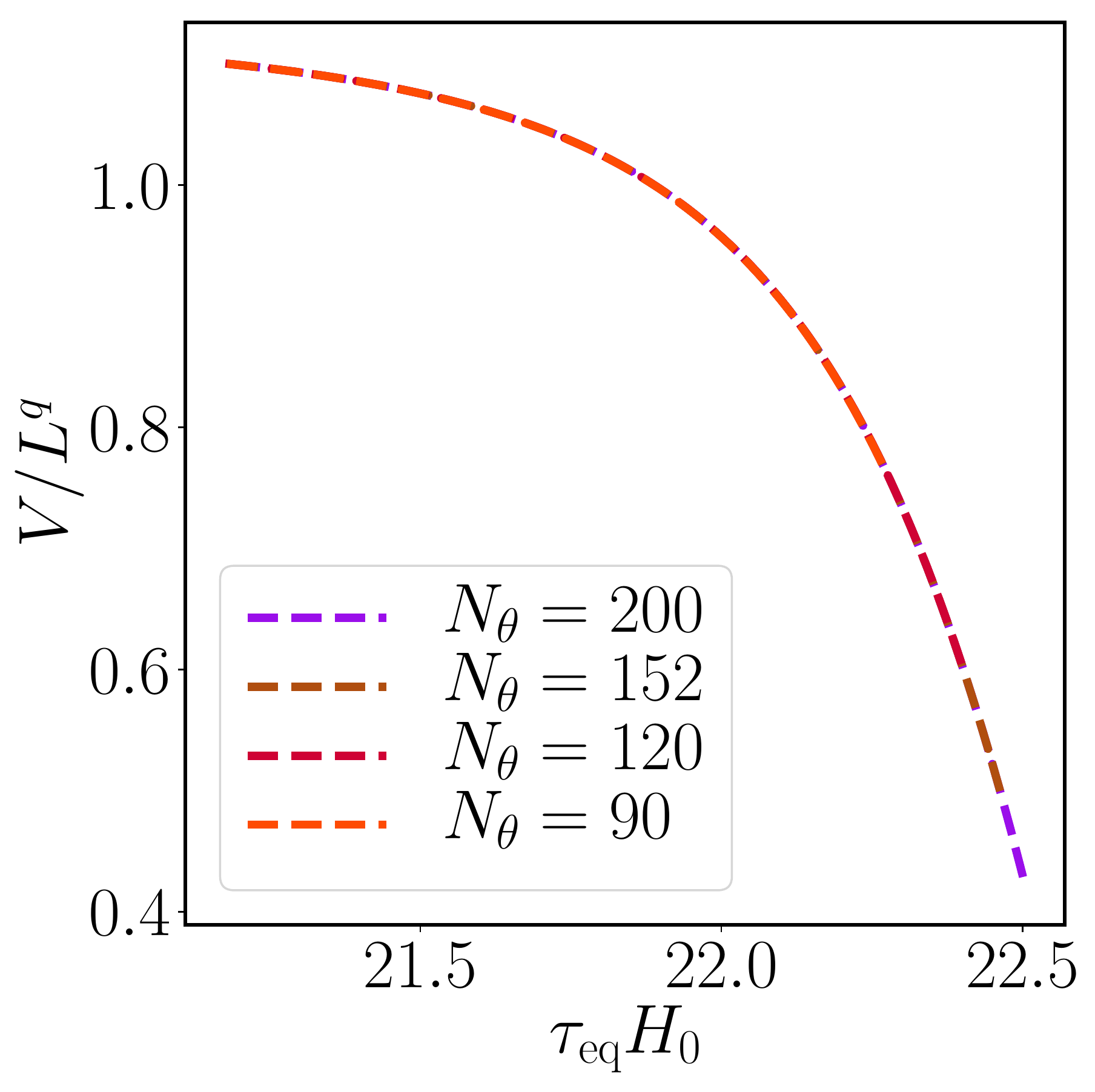}
\includegraphics[width=.327\textwidth]{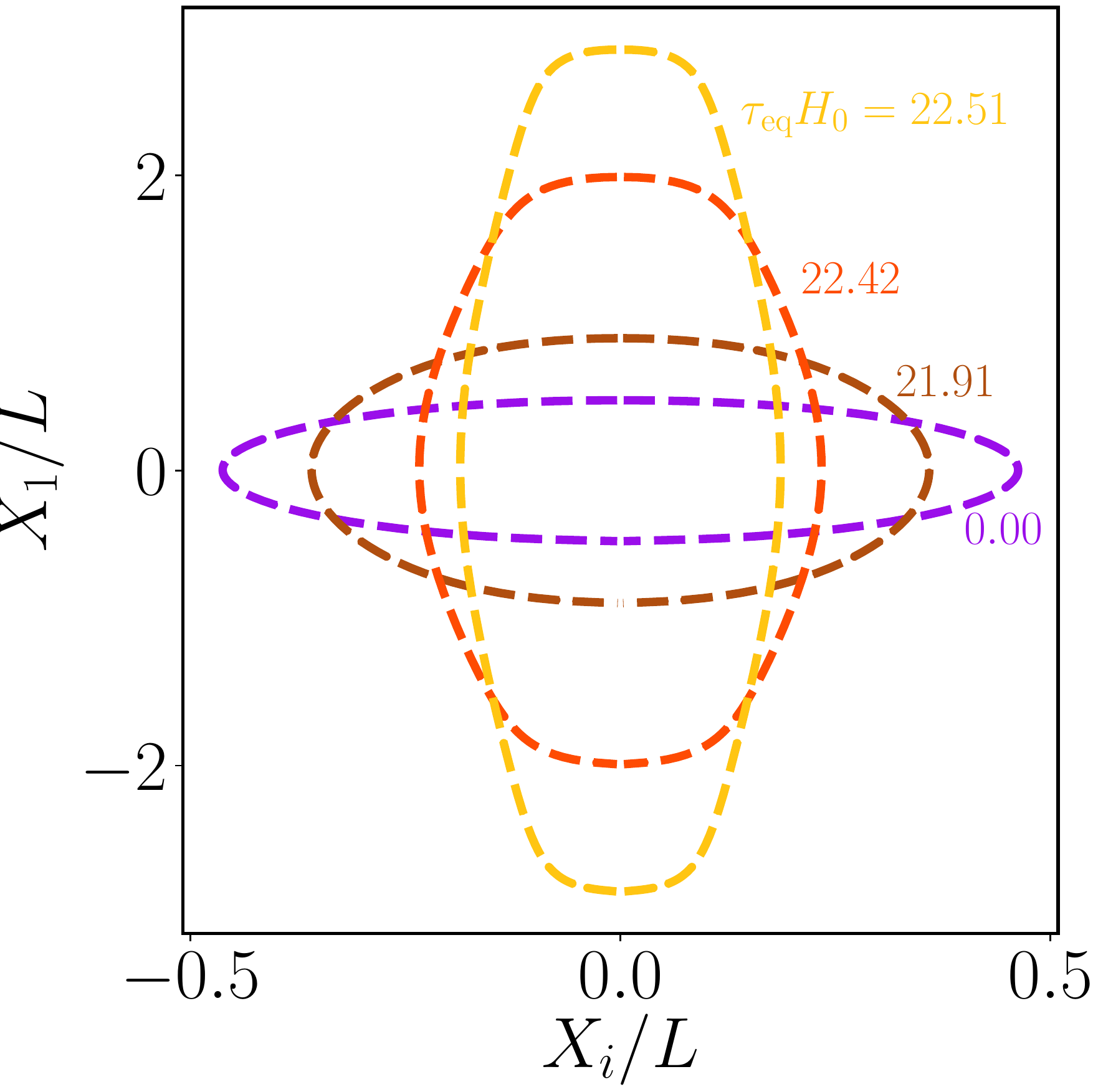}
\caption{\label{fig:q4A4.0}  A sample solution on the small volume branch
with $p=q=4$, $ \Lambda_D=1$, $H_0/M_4=0.0077$ and an $\ell=2$ perturbation. Left: The relative difference between the aspect ratio
of the sample solution and the background Freund-Rubin solution for successively larger perturbations. Middle: The
$q$-dimensional volume of the internal space for successively higher resolutions for initial data with $(\ell,\bar{h})=(2,0.3)$.
Right: Several snapshots of the embedding of internal space for
initial data with $(\ell,\bar{h})=(2,0.3)$. }
\end{figure}

In figure{~\ref{fig:q4A4.0}}, we show the evolution of the aspect ratio for a
stable Freund-Rubin solution when perturbed with successively larger $\ell = 2$
perturbations.  As expected, perturbations given by eq.~\eqref{pert} with
sufficiently small $h$ decay. However, there is a critical initial amplitude
above which the solution evolves to become more and more prolate.  As this
threshold is approached (around $\bar{h} \sim 0.2$), the instability timescale
(after a brief transient  where the aspect ratio undergoes a few damped
oscillations) increases---consistent with the initial condition approaching a
maximum in the effective potential.  Beyond the threshold, as the
compactification becomes increasingly prolate, the compactified (but not uncompactified) 
volume rapidly decreases, as can be seen in figure{~\ref{fig:q4A4.0}}. By adding higher numerical
resolution, we can reach higher aspect ratios and smaller compactified volumes
(which have higher magnitudes of the scalar curvature) before the evolution
breaks down, but at all resolutions we see no evidence that the solution is
asymptoting to some non-singular state.  As we will see below, this behaviour
seems to be generic for solutions where the internal space becomes prolate.
Note that as the prolate solutions evolve to their ultimately singular end, the
four dimensional effective theory, and the effective potential depicted in
figure{~\ref{fig:Veff}} eventually are no longer valid. The effective potential
in the prolate direction is therefore only indicative of the general direction
of evolution.

\subsubsection{Linearly unstable solutions: $n < n_c$}
\label{sec:unstable}
We now discuss the evolution of Freund-Rubin solutions that are linearly unstable to $\ell=2$ perturbations, 
which have a flux less than the critical value $n<n_c$. As illustrated in figure{~\ref{fig:cartoon}}, and outlined in the previous sections, 
at each flux $n_{\rm Mink} < n < n_c$ there exists a linearly unstable Freund-Rubin solution, 
as well as a corresponding linearly stable oblate warped solution with the same flux. 
The warped solution being thermodynamically preferred (e.g. with a higher entropy/lower Hubble parameter), 
these solutions are a natural candidate for the end point of the instability~\cite{Kinoshita:2009hh,Dahlen:2014}. 
This expectation is reflected in the effective potential for the aspect ratio sketched in the middle panel of figure{~\ref{fig:Veff}}. 
The Freund-Rubin solution is at the maximum of the effective potential, and a negative $\ell=2$ perturbation would 
cause the solution to evolve towards the oblate warped solution at the potential minimum. 
Sampling initial conditions with a wide range of fluxes, we find that the endpoint of the $\ell = 2$ Freund-Rubin instability (in the $-\epsilon$ direction)
is in most cases the stable warped solution with corresponding flux. The notable exceptions occur in a window of flux 
between $n_{\rm Mink} < n < n_I= 435.56$, where the end state is instead a crunching prolate solution. We discuss these solutions in more detail below. 
The confirmation of the thermodynamic arguments in previous literature, with interesting exceptions, is one of our primary results. 

\begin{figure}[h]
\centering
\includegraphics[width=.327\textwidth]{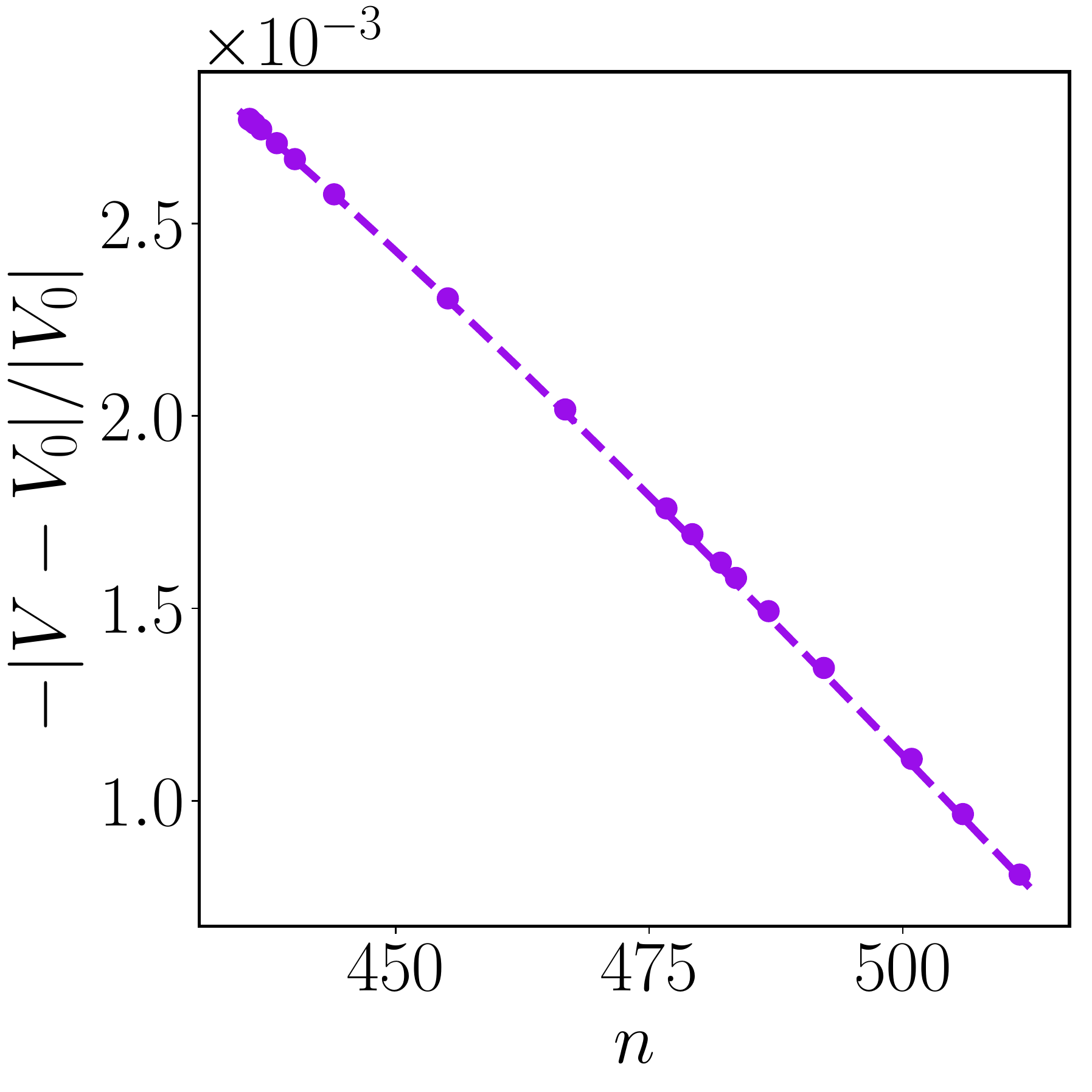}
\includegraphics[width=.327\textwidth]{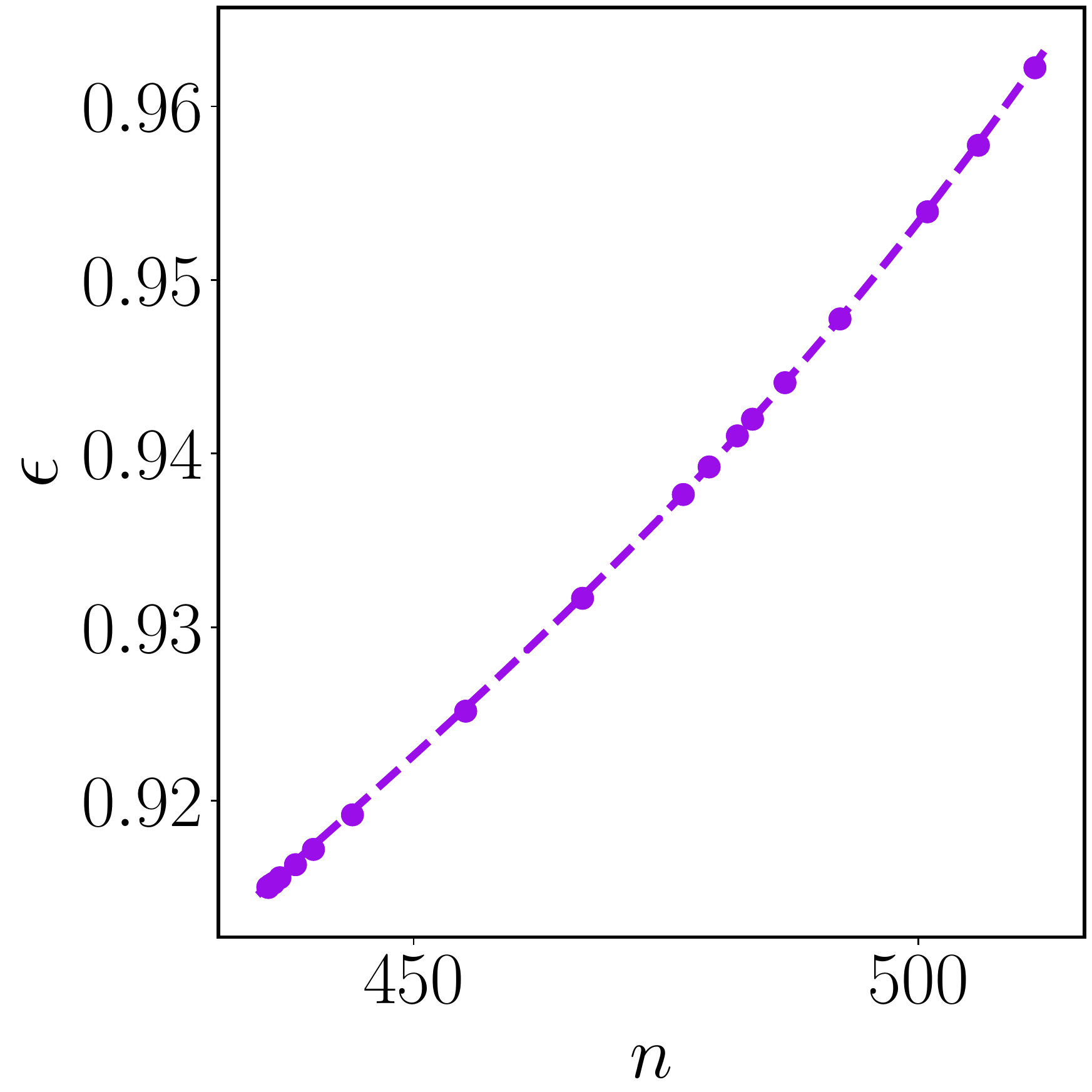}
\includegraphics[width=.327\textwidth]{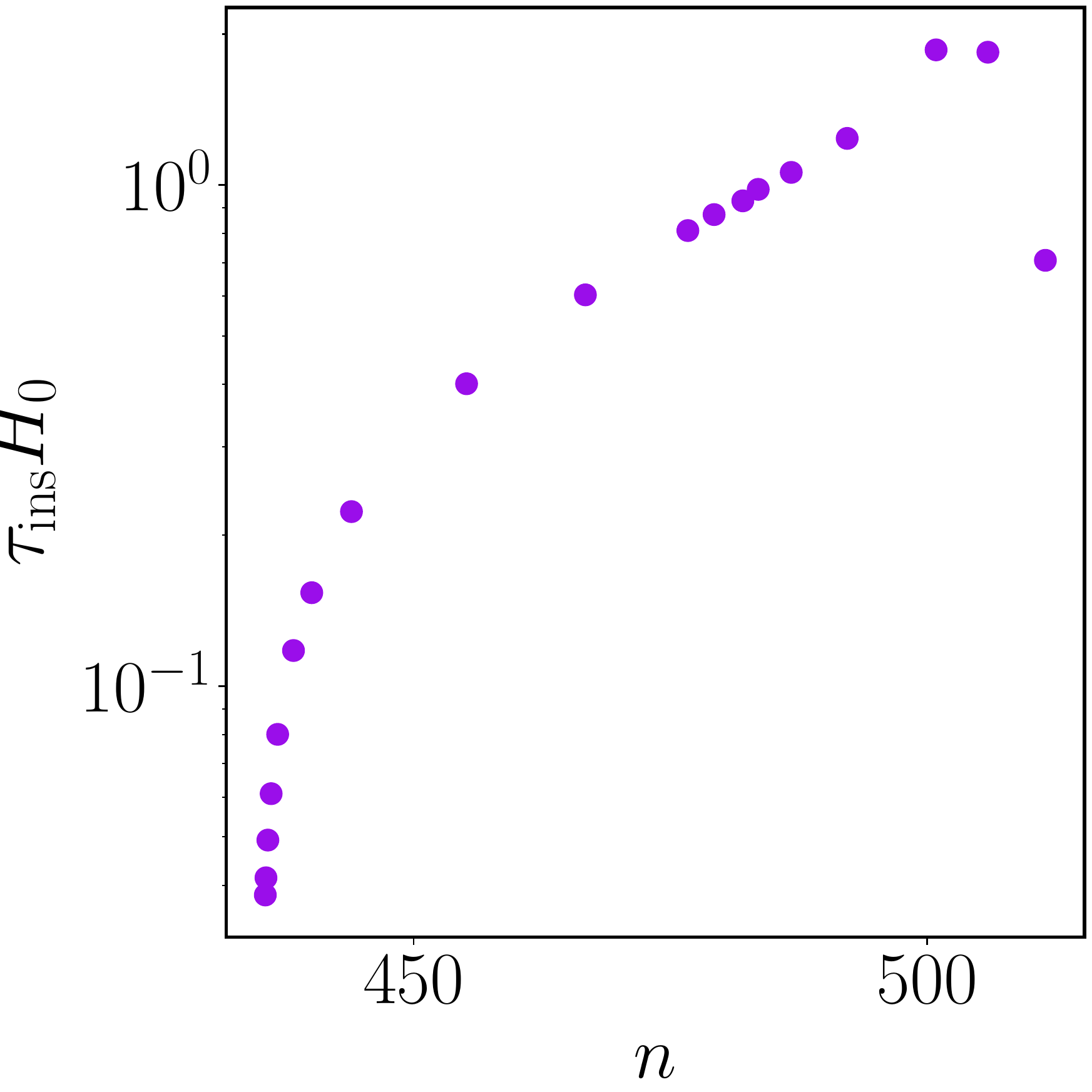}
\caption{\label{fig:OblComp} The volume (left), aspect ratio (middle) and ratio
of instability timescale to e-fold period (right) of warped solutions with
$n_I < n < n_c$ on the small volume branch for the special case $p=q=4$. We
find that in this regime each unstable symmetric solution evolves to an 
ellipsoidal solution with $\epsilon<1$, which has roughly the same internal
volume but lower effective potential. The dashed lines represent the
interpolation of stationary solutions constructed as described in section~\ref{warped}, while the
dots represent the end states of evolving 
a symmetric solution with an $\ell=2$ perturbation. For the aspect ratio, the
maximum of the difference is $\sim10^{-4}$.}
\end{figure}

To explicitly verify that the endpoints of the Freund-Rubin instability are indeed the stable warped solutions, in 
figure{~\ref{fig:OblComp}} we show the volume of the compact space (left) and the aspect ratio (middle) of the numerical solutions at late times (dots) 
compared to the corresponding quantities for the warped solutions (dashed line) from section~\ref{warped}. The agreement
is excellent. Note that the warped solutions have roughly the same internal volume as the symmetric
solution they evolve from. In figure{~\ref{fig:OblComp}}, we also plot the instability timescale measured from the linear regime of the 
numerical evolution, which grows as the flux is increased. We can understand this as follows. The Hubble parameter
for the Freund-Rubin and warped solution match at $n_c$. Therefore, as $n$ increases, the extrema of the effective potential in the 
$\epsilon$ direction merge, and the curvature at the maximum goes to zero---we therefore expect an increasing instability timescale as 
$n \rightarrow n_c$.

\begin{figure}[h]
\centering
\includegraphics[width=.31\textwidth]{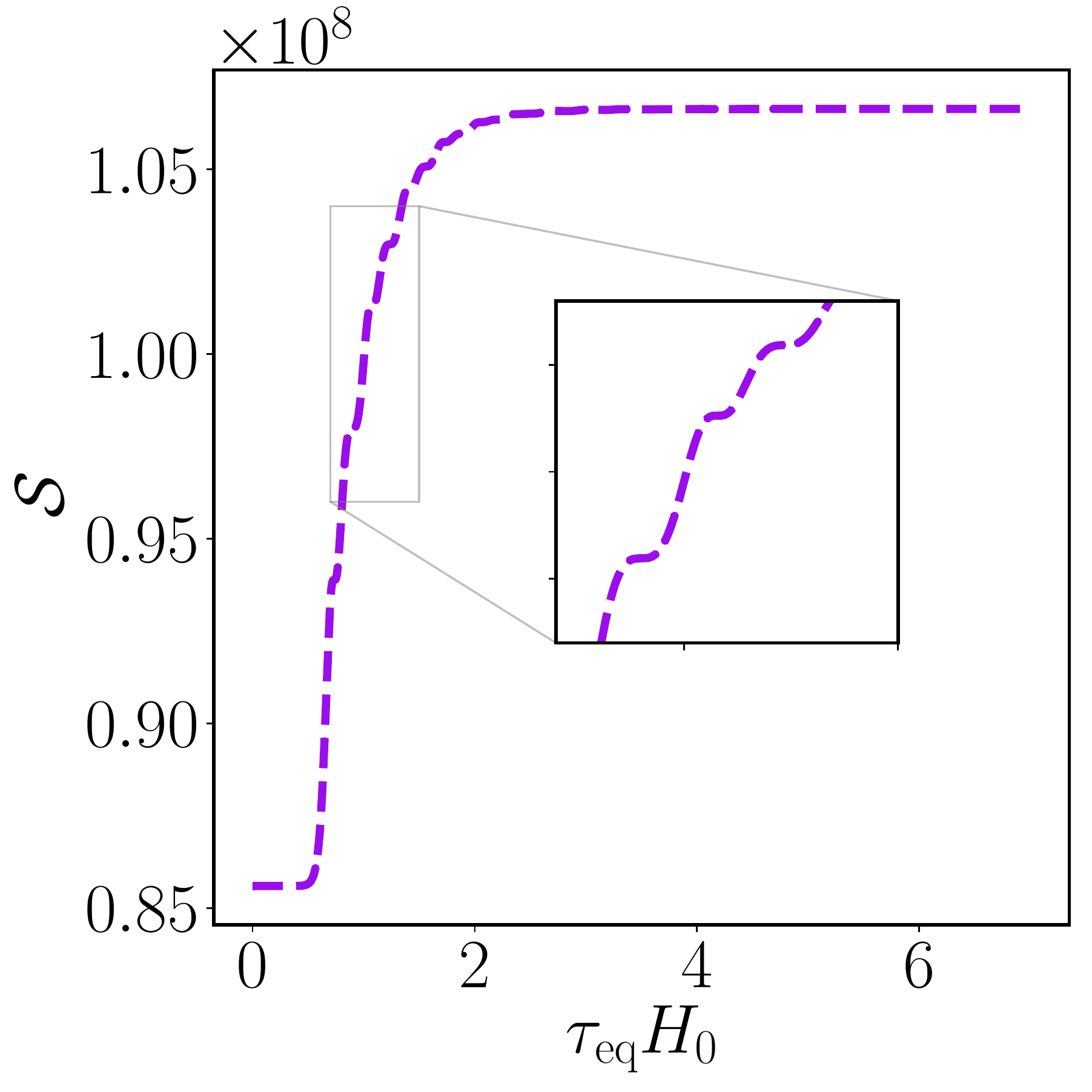}
\includegraphics[width=.31\textwidth]{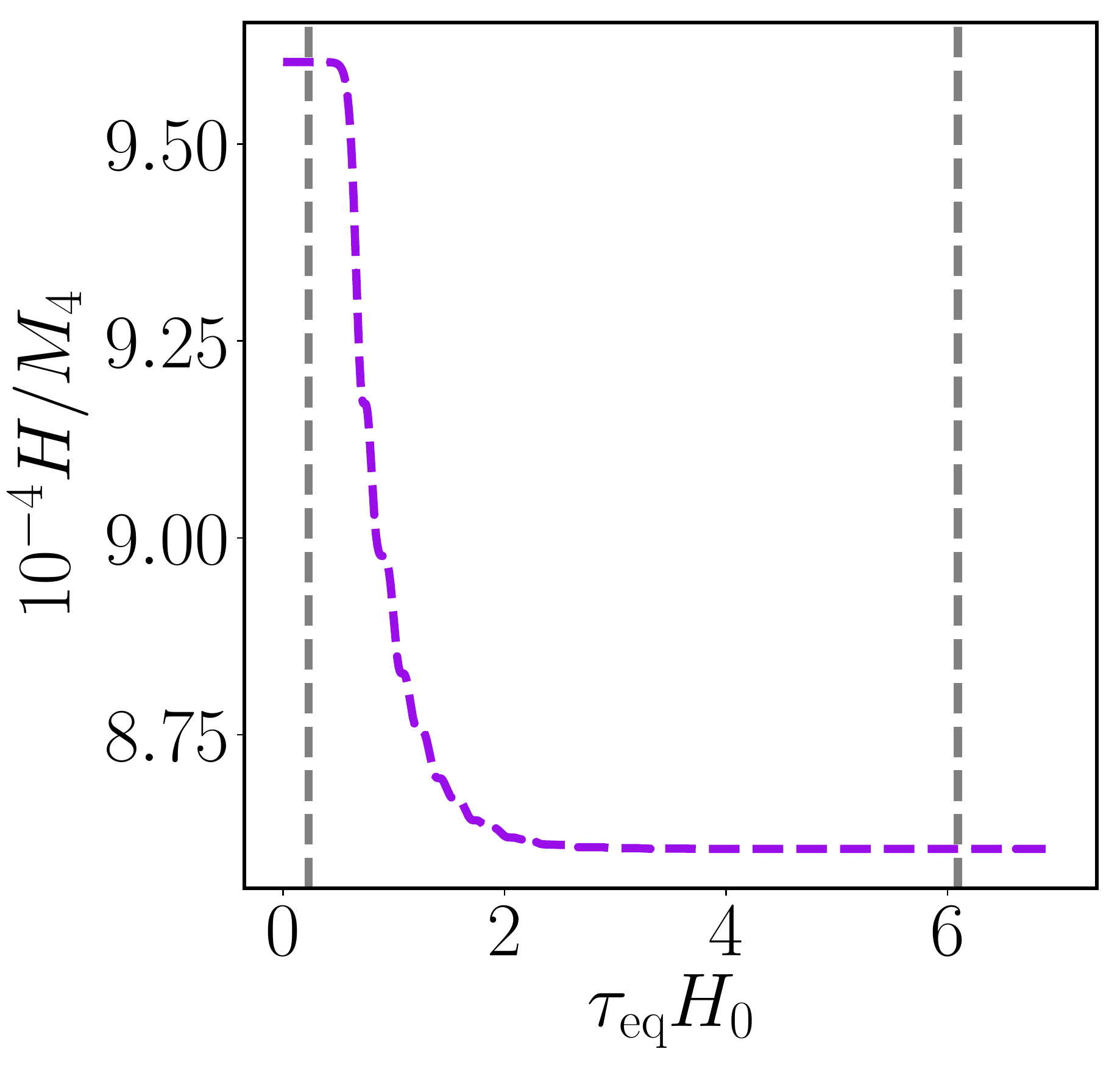}
\includegraphics[width=.36\textwidth]{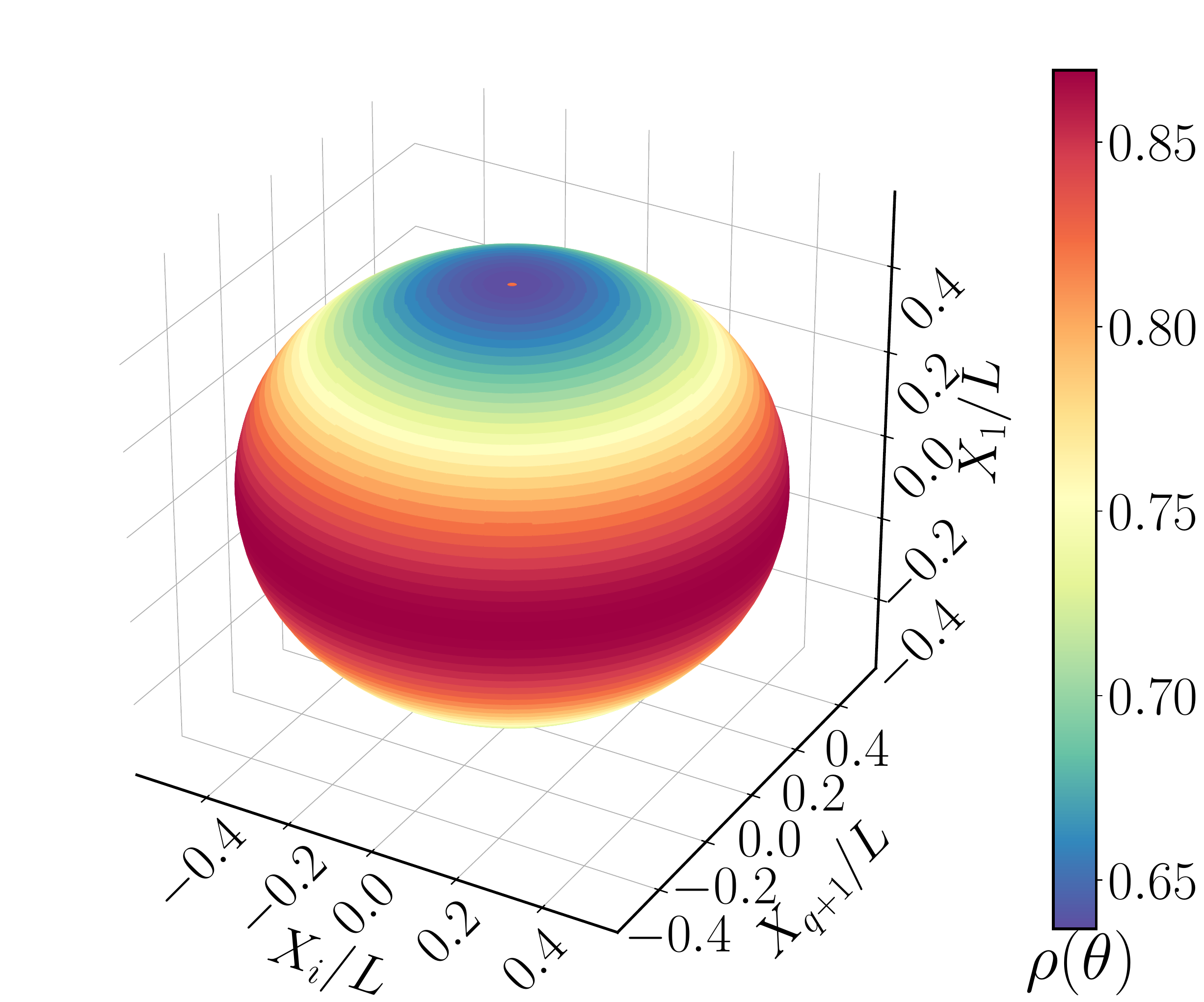}
\caption{\label{fig:q4l2A45} A sample solution on the small volume branch with
$p=q=4$, $\Lambda_D=1$, $H_0/M_4=0.00096$, $n=436.13$ and an initial $\ell=2$
perturbation. Left: The effective entropy which is increasing between the two
stationary solutions. Middle: The effective four-dimensional Hubble rate
eq.~\eqref{eq:hubbledef}, with grey lines indicating the 
approximate initial and final times of the transition period, as defined in the text. Right: The three dimensional projection of the embedding of
internal space at $\tau_\mathrm{eq} H_0=8.6$. The color shows the flux
density.}
\end{figure}

We now focus on the specific example shown in figure{~\ref{fig:q4l2A45}} to illustrate 
the transition between an initially unstable spherical
solution and its endpoint, a stable oblate solution. In the left panel we show the entropy, which 
increases monotonically as expected. There are interesting step-like features in the evolution which persist 
at increasing resolution, and are therefore not likely to be numerical artifacts. In the middle panel 
we show the Hubble parameter, which decreases monotonically over the course of $\sim 5$ e-folds to its asymptotic value. 
We show the embedding of the compact space in the right panel. Note that the flux is
distributed on the ellipsoid the way you would expect it from the linear
analysis. We found in section~\ref{stabFR} that the unstable mode has inversely
correlated flux $\bar{a}$ and shape $\bar{h}$ components, and similarly we find that
whenever the radius gets larger, the flux density does too, such that, for an
oblate solution, the flux is concentrated around the equator. This makes intuitive sense,
as a region of larger radius implies higher curvature, and hence a larger flux
density to support the region against collapse.

\begin{figure}[h]
\centering
\includegraphics[width=.492\textwidth]{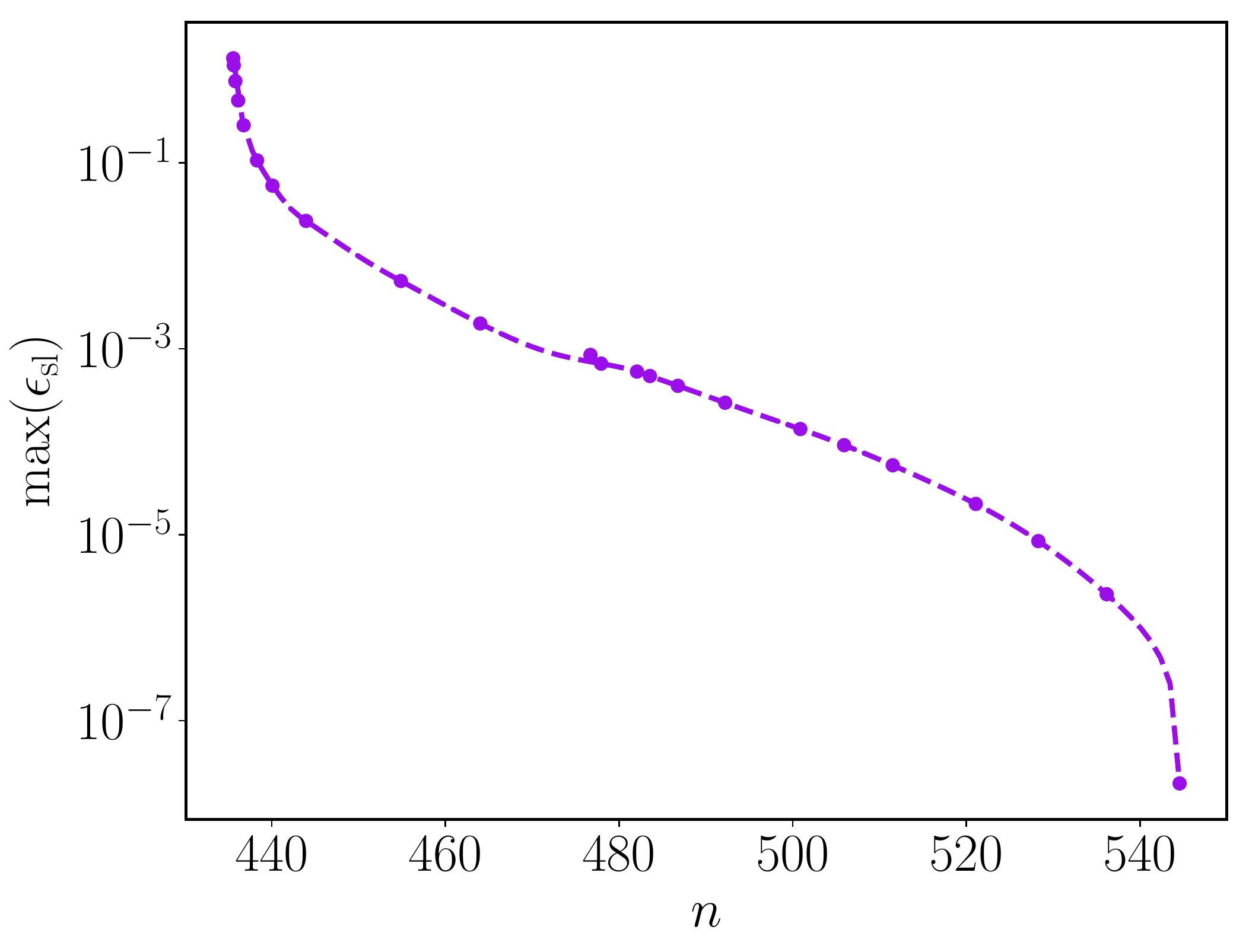}
\includegraphics[width=.492\textwidth]{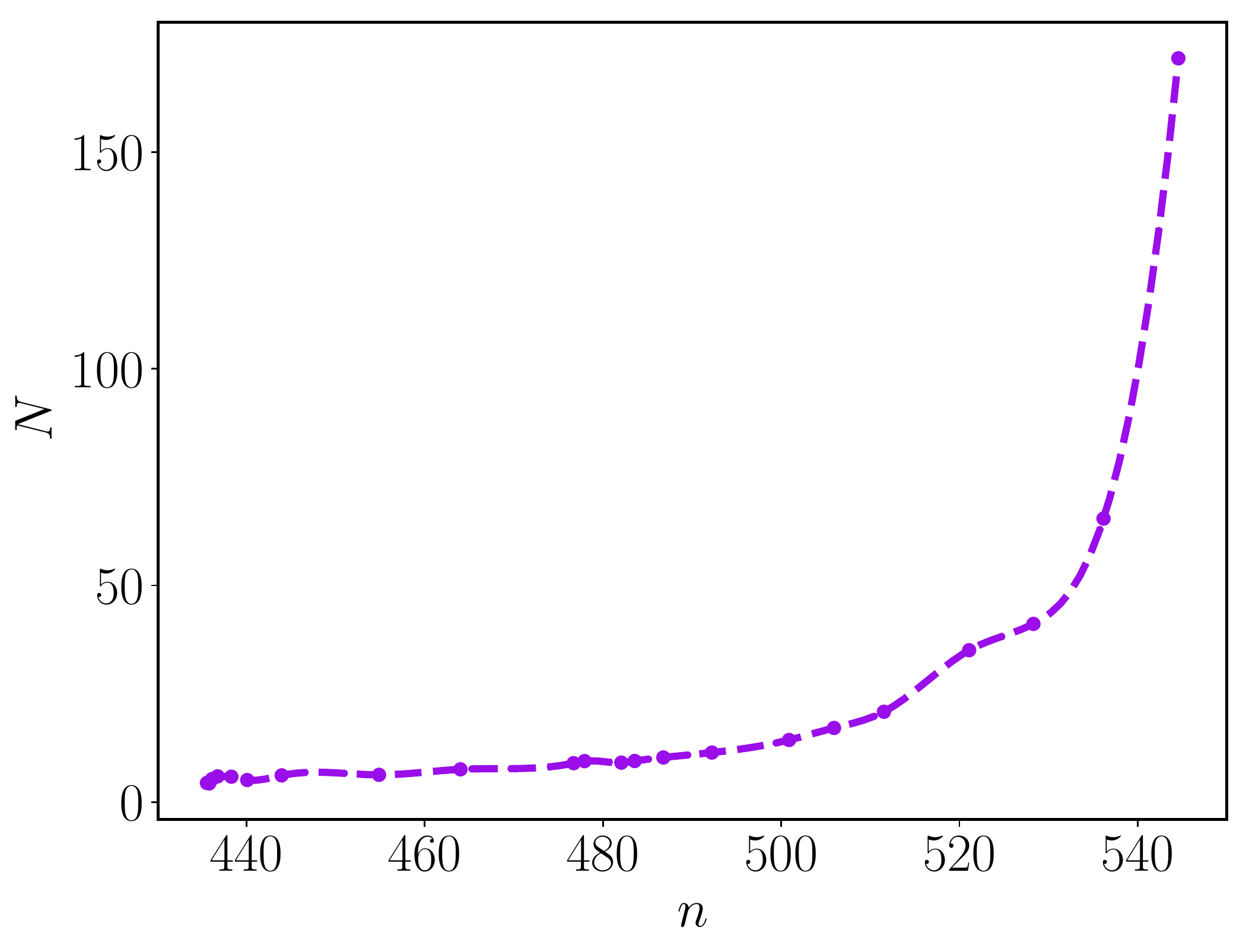}
\caption{\label{fig:inflation}  For unstable Freund-Rubin solutions with $n_I < n< n_c$ we plot the maximum value the slow-roll parameter $\epsilon_{\rm sl}$ takes during the evolution (left) and the number of e-folds of expansion in the four-dimensional effective theory elapsed during the transition to the stable warped end-point (right).}
\end{figure}

For the choice of flux in figure{~\ref{fig:q4l2A45}}, the transition between
the spherical and oblate solution occurs over the span of $\sim 5$ e-folds. As
noted above, the instability timescale increases with $n$. Therefore, an
interesting question is whether the transition period can persist over a larger
number of e-folds. In this case, the four-dimensional effective theory includes
a period of slow-roll leading to an asymptotic regime of pure de Sitter
expansion \footnote{Note that one is usually interested in computing the number
of e-folds from a de Sitter phase to a universe with a small or zero cosmological
constant, rather than to another de Sitter phase with smaller but comparable expansion rate.}. In
figure{~\ref{fig:inflation}}, we show the maximum value that the slow-roll
parameter $\epsilon_{\rm sl}$ (defined in eq.~\ref{eq:slow_roll}) takes during
the evolution (left) as well as the elapsed number of e-folds during the
transition. To be more precise, we define the number of e-folds between the
initial ($i$) Freund-Rubin and final ($f$) stationary warped solution as
$N=\int^f_i H \ d\tau_{\rm eq}$, where the time at which the slow-roll period
starts (ends) is defined as when the Hubble factor differs by $10^{-4}$ relative
to its initial (final) value (indicated by grey dashed lines in
figure{~\ref{fig:q4l2A45}}). We see that it is possible to get $\sim 10$--$100$
e-folds of slow-roll inflation as $n \rightarrow n_c$. We conclude that the
evolution of unstable Freund-Rubin solutions provides a viable toy model for
slow-roll inflation in flux compactifications. One interesting application of
these solutions is to use the full higher dimensional picture to explicitly
compute the effect of extra dimensions on the spectrum of linear scalar and
tensor perturbations. This would make contact with phenomenology and
cosmological observables such as the cosmic microwave background. We defer this
and other possible explorations to future work.  

\begin{figure}[h]
\centering
\includegraphics[width=.328\textwidth]{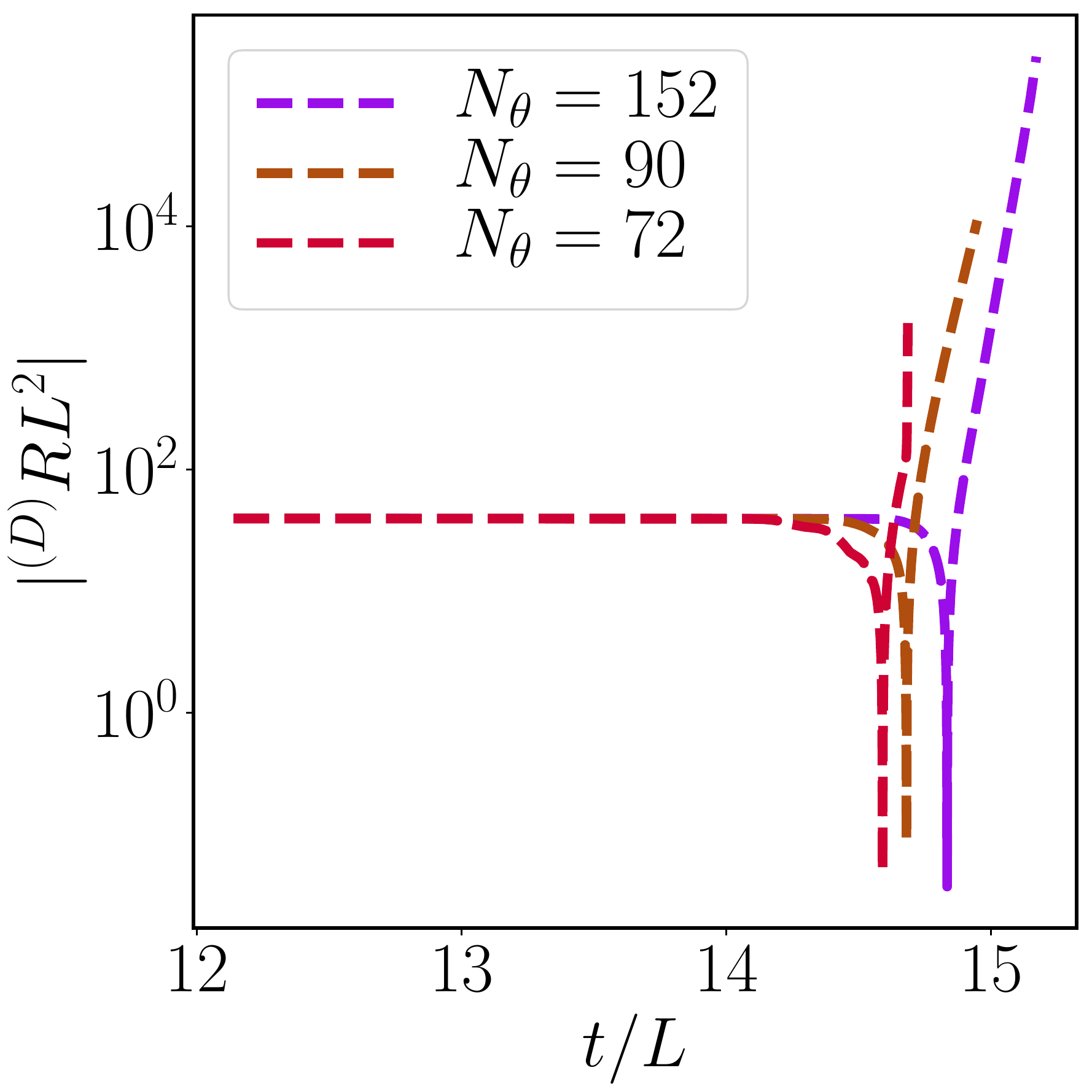}
\includegraphics[width=.328\textwidth]{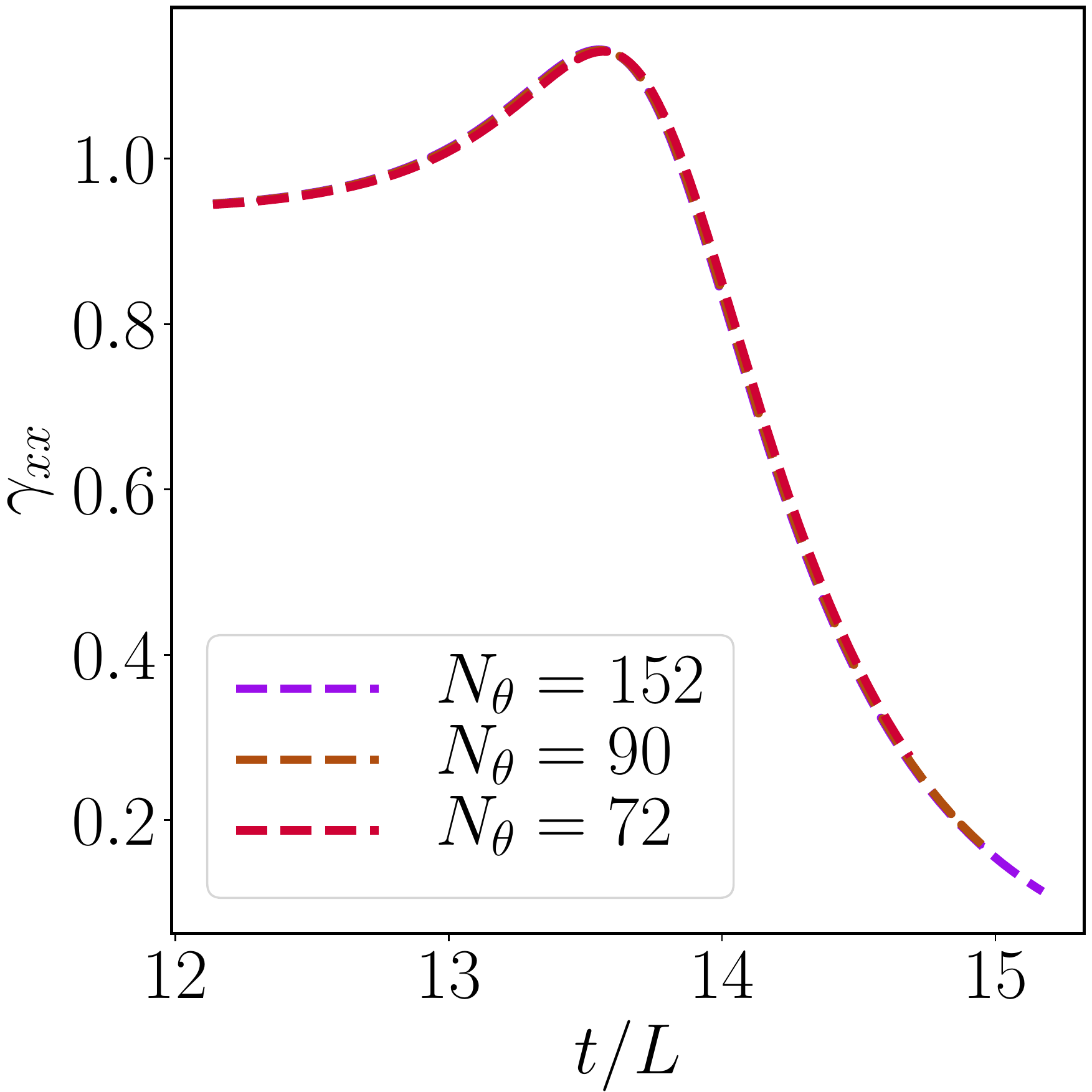}
\includegraphics[width=.328\textwidth]{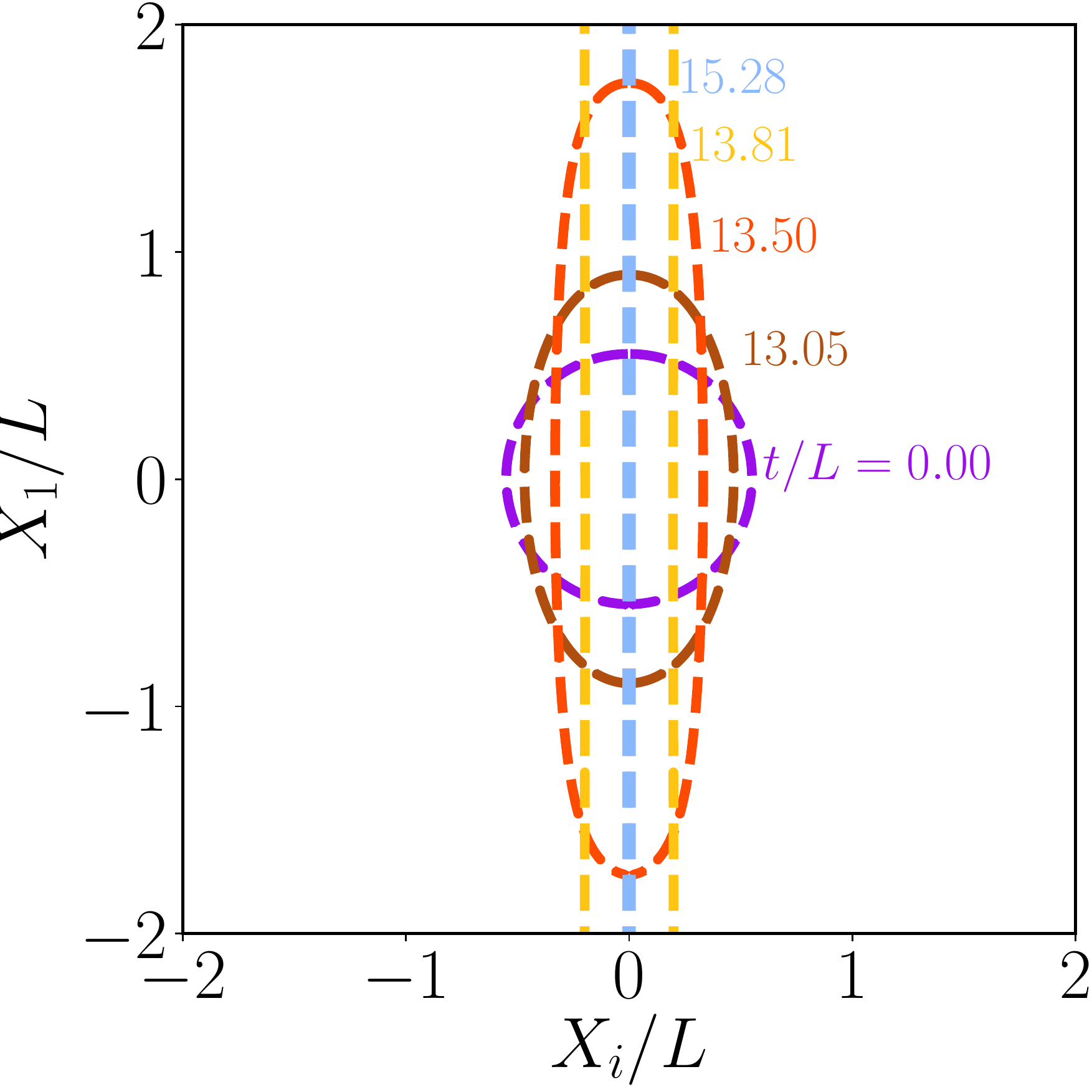}
\caption{\label{fig:football} End state of solution on small volume Freund-Rubin branch
with $p=q=4$, $\Lambda_D=1$, $n=n_{\text{Mink}}$ and an initial $\ell=2$
perturbation for different resolutions. The $D$-dimensional
curvature scalar which becomes negative at the end (left), scale factor
at the equator (middle) and embedding of internal space (right). }
\end{figure}

So far, the linear analysis has been very good at predicting what
happens in the full nonlinear case. But this is not always true. In the range
of flux $n_{\rm mink} < n<n_I= 435.56$ we find that while the solution does transition to the
corresponding oblate solution briefly, but does not settle there. Rather, it oscillates 
between the oblate and nearly spherical solution before running away
to prolate values. In figure{~\ref{fig:football}}, we illustrate this behaviour for a limiting case
where the external spacetime is Minkowski. There it can be seen that after a brief period
 where the internal space is oblate, the
solution transitions to being more and more elongated, with the flux concentrating around the
poles. In this case (in contrast to the prolate solutions
discussed above), we find regions where the characteristics are ingoing, which
allows us to excise a region around the poles and continue the evolution.  
We find that the spacetime curvature blows up, and the scale factor and
equatorial circumference both tend to zero, consistent with a crunch. 

Note that for the Minkowski spacetime, there are no oblate solutions with the same value of $n$, so it is not surprising that the solution goes prolate. 
However other solutions with $n_{\mathrm{mink}}< n <n_I$ have a solution on the oblate branch, and yet show the same behaviour as the Minkowski solution.
We can understand this as follows. First, recall from figure{~\ref{fig:OblComp}} that as $n$ is
decreased to approach $n_I$, the instability timescale decreases.  Hence, in the language of the effective
potential picture, the velocity approaching the minimum of potential will be
larger, and there will be a greater tendency to overshoot, and, due to nonlinear
effects, eventually roll back up the potential towards larger $\epsilon$.
The smaller the flux, the shorter the timescale of instability, and we observe that solutions undergo fewer oscillations 
about the oblate solution as the flux is decreased. The Hubble parameter decreases monotonically, passing through zero 
as the solutions roll back up the potential to increasing $\epsilon$. As $H$ goes through zero, the 
expanding marginally inner trapped cosmological horizon becomes infinite and disappears, and
a marginally outer trapped horizon appears and begins contracting. Again, this is consistent with a crunch.

\begin{figure}[h]
\centering
\includegraphics[width=.327\textwidth]{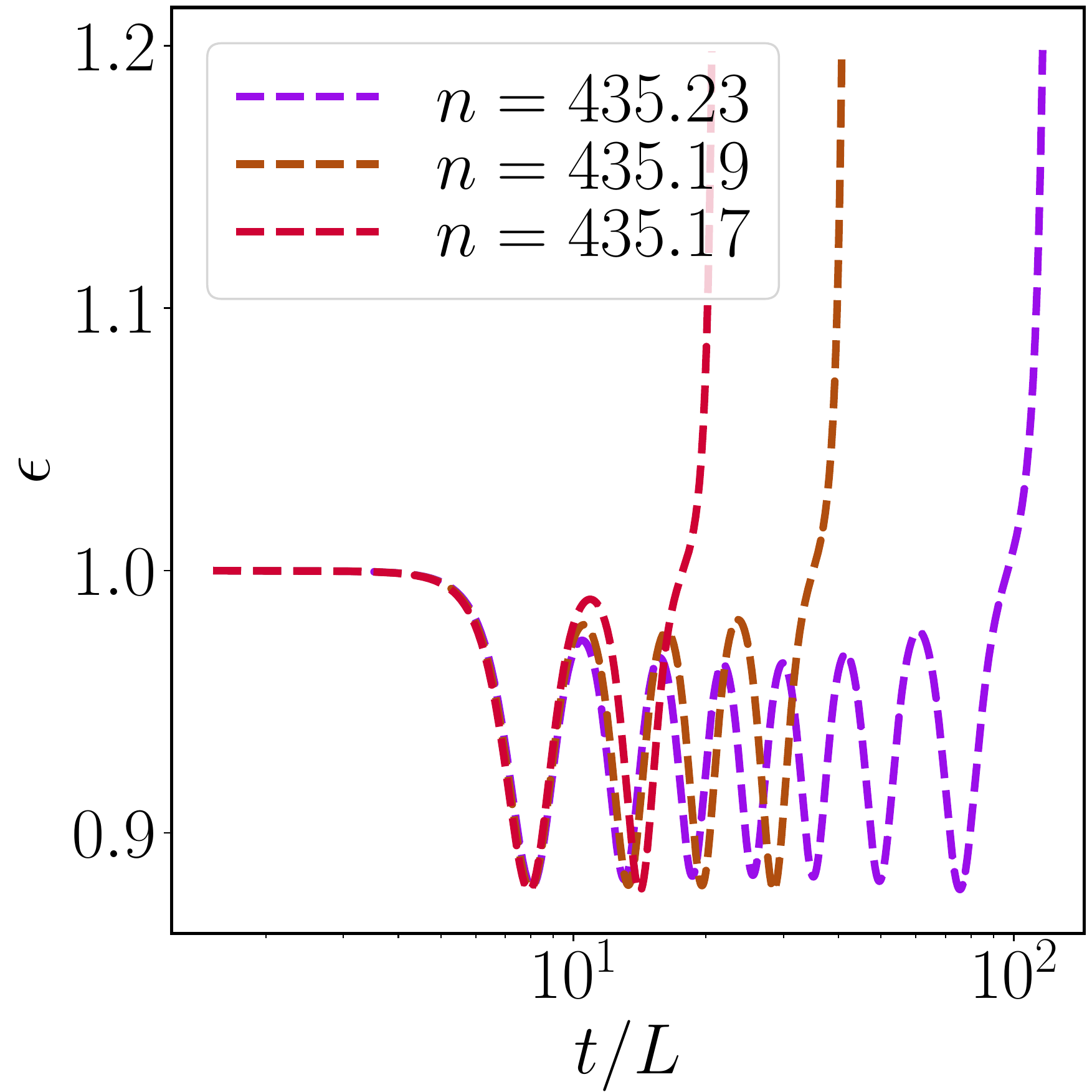}
\includegraphics[width=.327\textwidth]{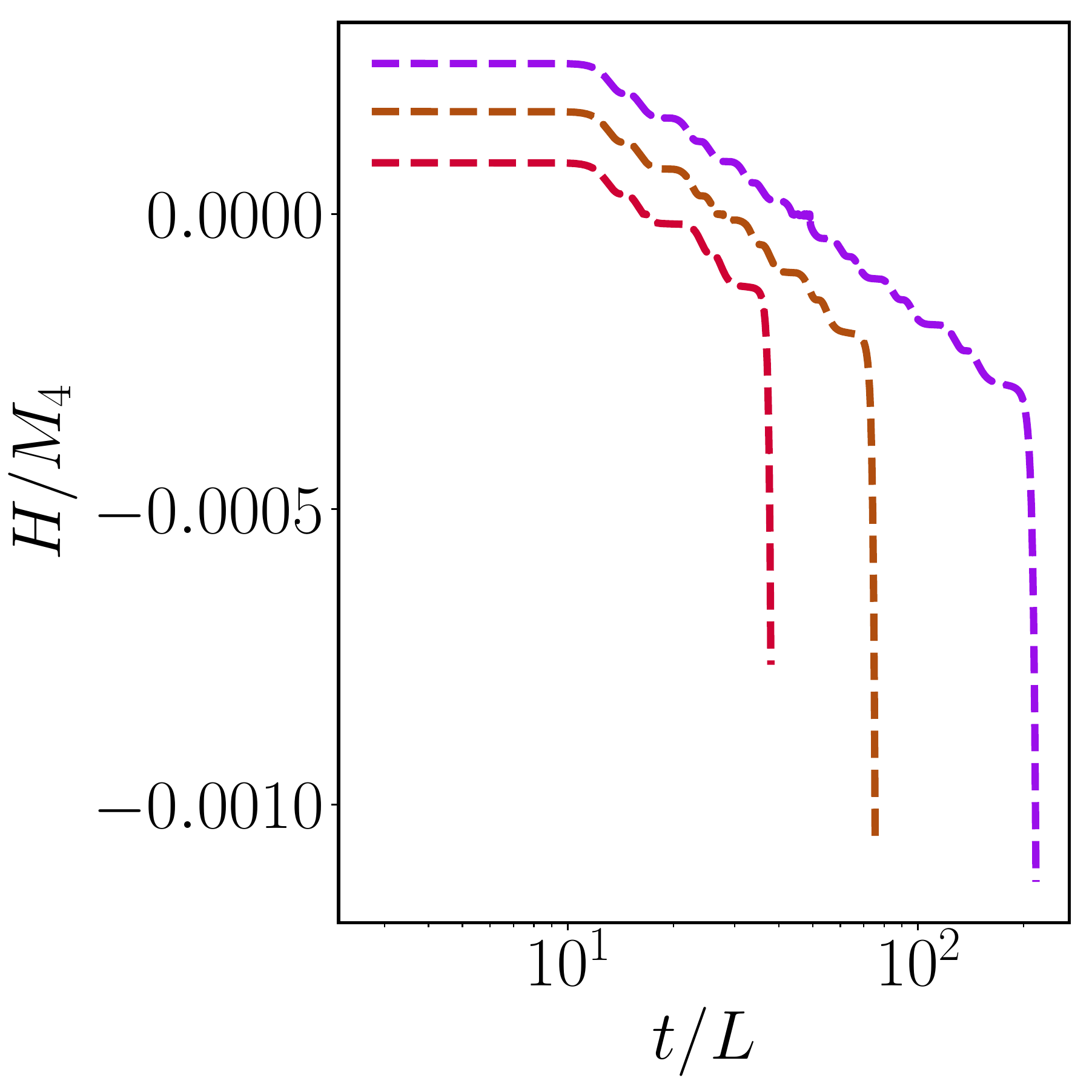}
\includegraphics[width=.327\textwidth]{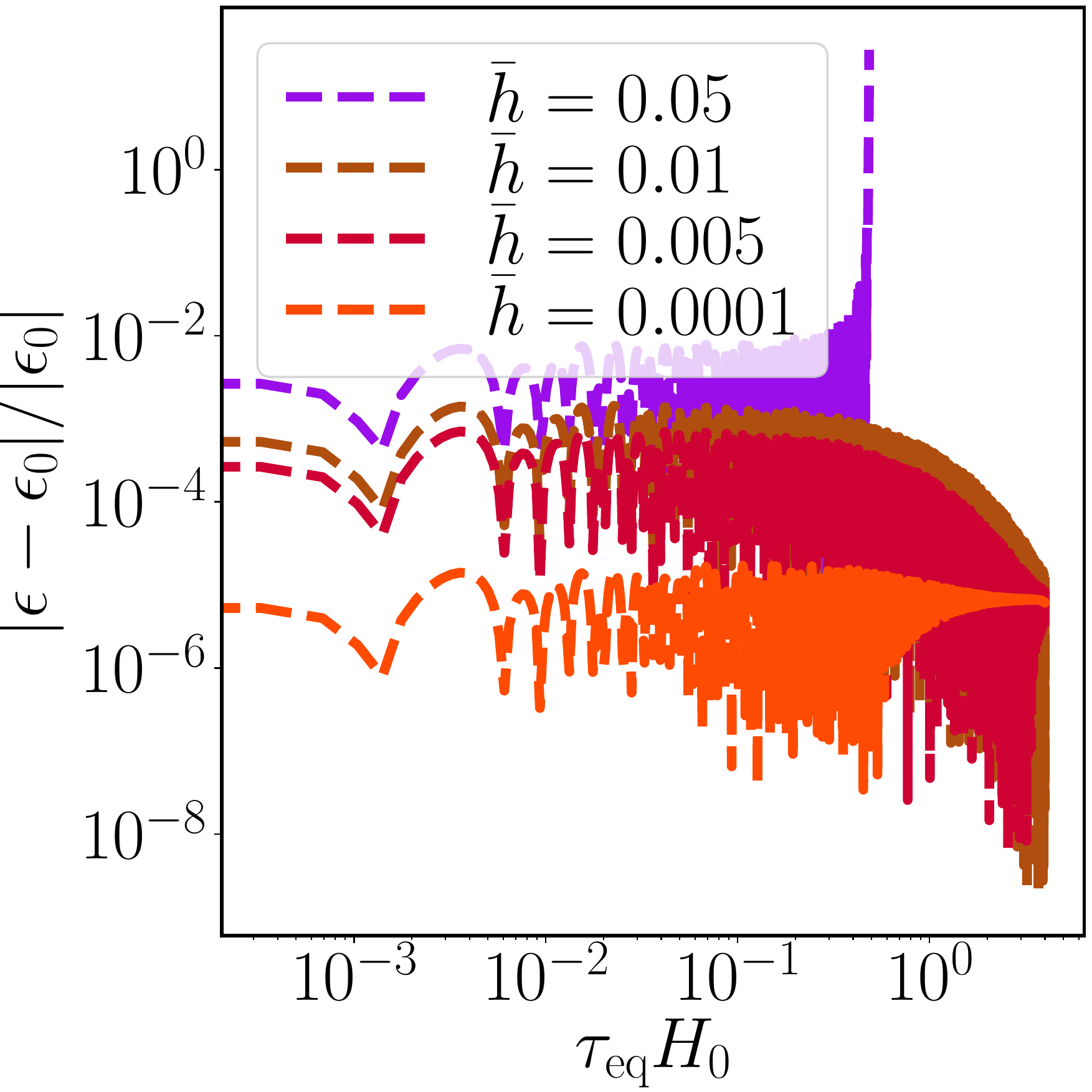}
\caption{\label{fig:WSHU} The aspect ratio (left) and effective Hubble rate (middle) for a few initially spherical
solutions with $n<n_I$. The relative difference in the aspect ratio from the background value for a sample solution on the
small Hubble warped branch with $p=q=4$, $ \Lambda_D=1$ and $H_0/M_4=1.3 \times 10^{-5}$
and successively larger $\ell=2$
perturbations (right). 
} \end{figure}

To further probe the validity of the effective potential picture, we consider perturbations
around initially oblate solutions in this flux range. As shown in
the right panel of figure{~\ref{fig:WSHU}}, small perturbations decay (consistent with the linear
stability), but modestly larger perturbations cause the solution to become
prolate, undergoing the same fate as the corresponding initially spherical
solutions. This implies the existence of a potential barrier about the oblate solutions.

\begin{figure}[h]
\centering
\includegraphics[width=.327\textwidth]{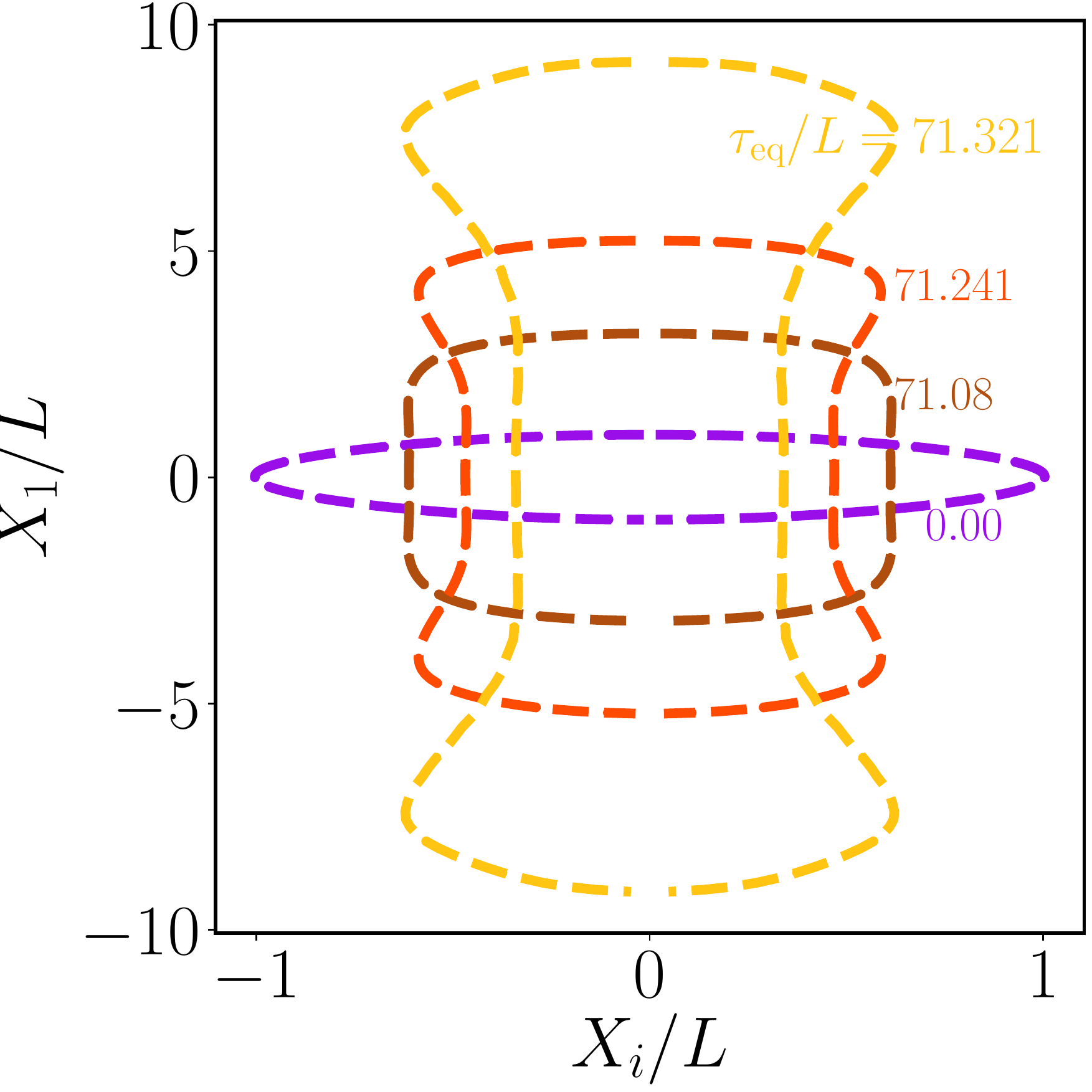}
\includegraphics[width=.327\textwidth]{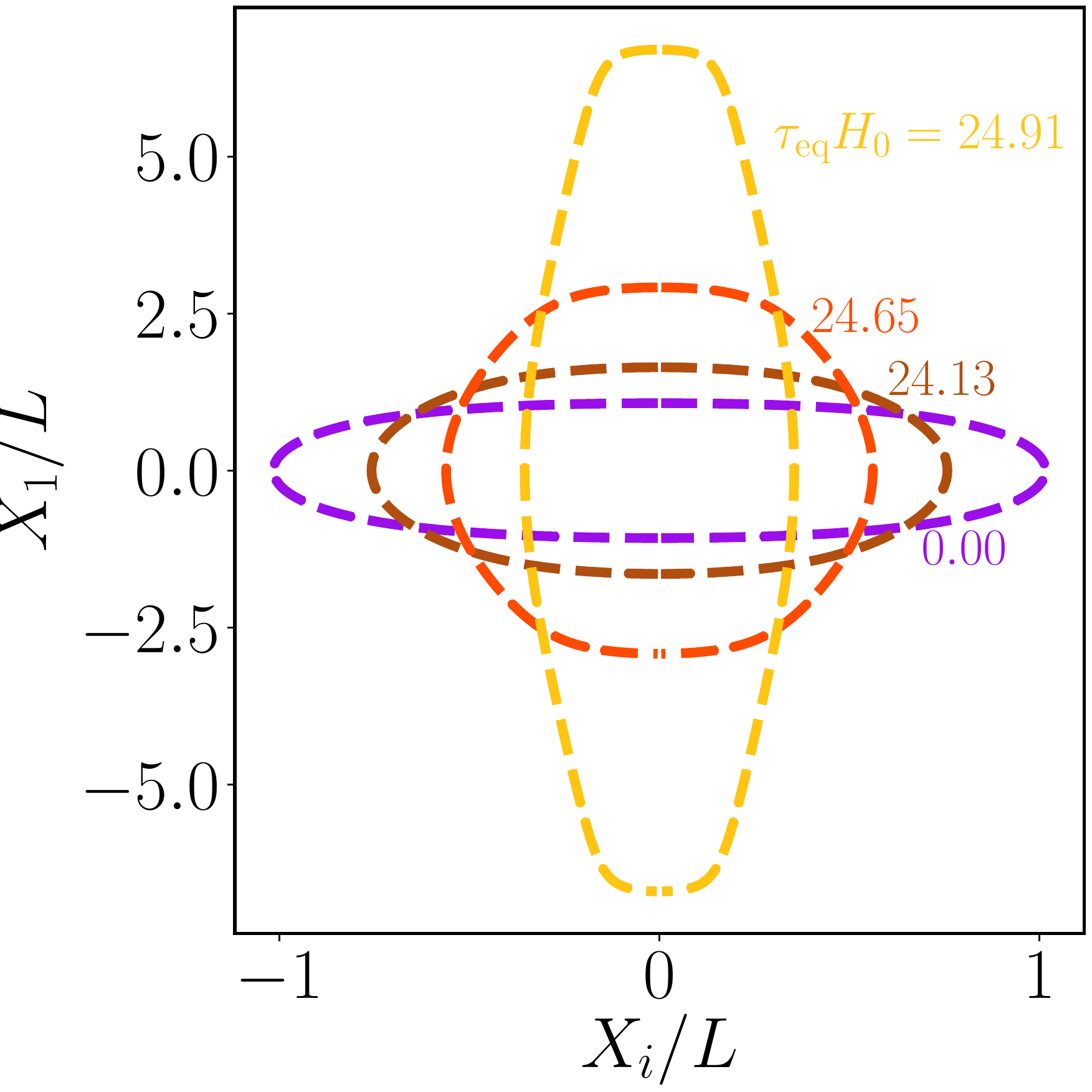}
\includegraphics[width=.327\textwidth]{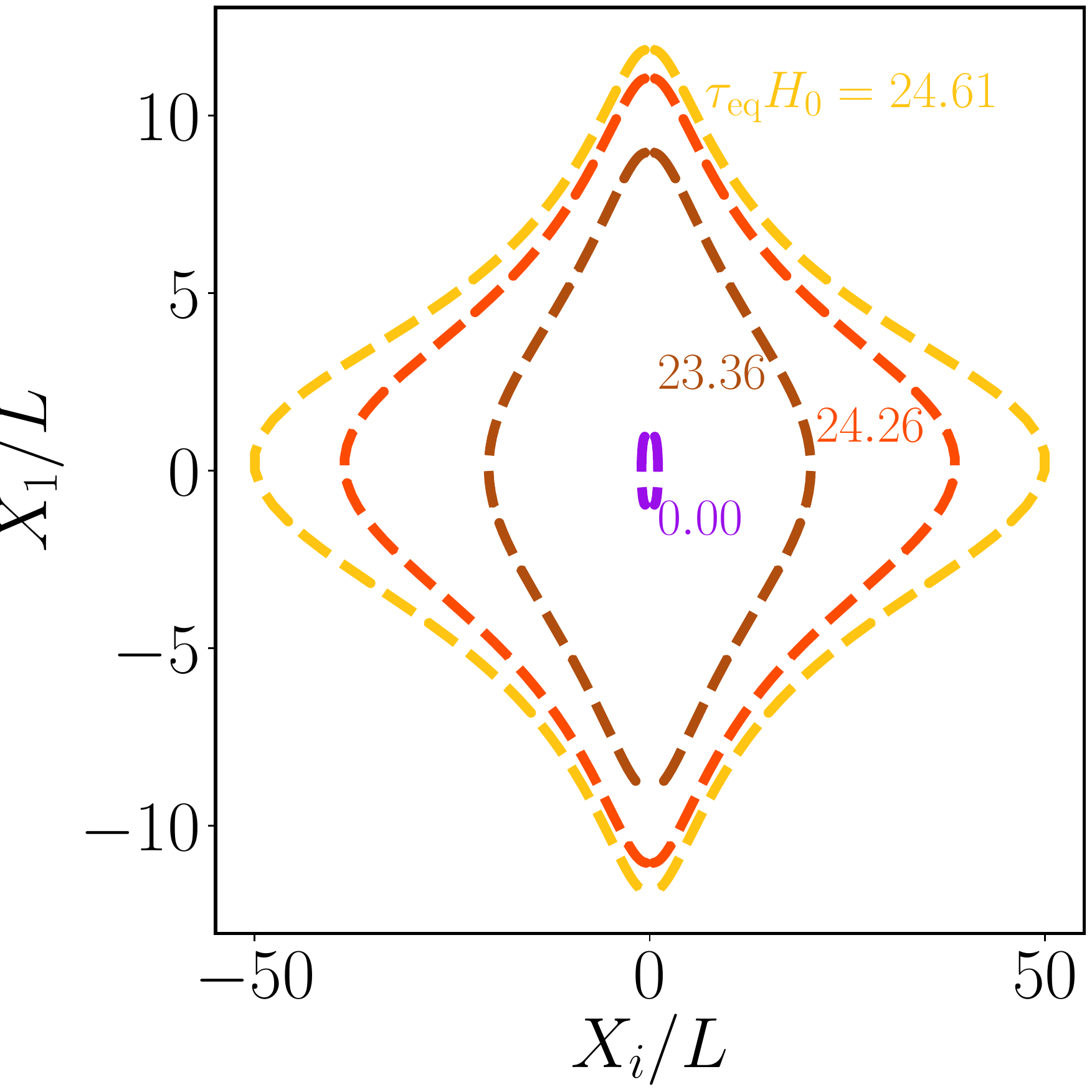}
\caption{\label{fig:Wpro}Sample solutions on the large Hubble warped branch
with $p=q=4$, $ \Lambda_D=1, \left(H_0/M_4,n\right)=(1.3 \times 10^{-5},435.23)$ (left), $\left(H_0/M_4,n\right)=(0.0077,529.29)$ (middle) and $\left(H_0/M_4,n\right)=(0.0075,402.11)$ (right) where recall $n_{\rm Mink}=435.16$.}
\end{figure}

Finally, we study initially prolate solutions to determine their fate.  We
expect that, in the range of flux where there is a corresponding solution
on the small volume branch, they will undergo the same fate as the
solutions that started out spherical or oblate but were kicked out the
potential well. The left and middle panels of figure{~\ref{fig:Wpro}} show two such
solutions, which indeed become extremely prolate as the equatorial
radius shrinks to zero.
Another interesting regime is the one with $n<n_{\mathrm{Mink}}$. As one
can see from figure{~\ref{fig:cartoon}}, these solutions do not have a
Freund-Rubin solution they could have flowed from, hence they are distinct
from the evolutions considered so far. The right panel of figure{~\ref{fig:Wpro}} shows
the embedding of the compact space for such a solution. Those solutions
become not only very oblate but also extremely large in volume. Around 
the equator, the flux density approaches zero, and the expansion rate approaches that expected from the decompactified $D$-dimensional de Sitter solution. 
However, the internal space remains very inhomogeneous.
We are unable to continue the evolution indefinitely, as tracking the distorted shape and differing expansion rates requires higher and higher numerical resolution. However, we do not find
any singular behaviour before then. This distorted shape is 
qualitatively different from any of the solutions considered above.

Finally, we briefly report on higher dimensional spaces where $q \geq 5$.
Exploring a few cases with $q=5$, 
we find similar
behaviour to the case where $q=4$. In particular, we find a range of flux
where the Freund-Rubin solution is only unstable to the total volume instability,
and another where it is unstable to the warped instability only. In the
latter case, we again find a range of the parameter space  where the
unstable solutions flow towards stable stationary oblate solutions, but
outside that range the solutions will, in a similar way to the solutions shown in figure{~\ref{fig:football}}, become increasingly prolate forming
trapped regions in the
process and eventually crunching. However, it is important to note that for $q \geq 5$, the range of
the inequalities in eqs.~\eqref{vol} and \eqref{lumpy} overlap, and hence
none of the Freund-Rubin solutions are stable.  For future work, it would
thus be interesting to see which instability dominates in this overlap
region of the parameter space.  Additionally, in contrast to $q=4$, when $q \geq
5$, higher $\ell$ modes can be unstable, which in theory would result into
further breaking of the symmetry of the compact sphere.

\section{Conclusions}
Despite extensive study of the linear stability and mass
spectrum of Freund-Rubin like solutions, very little is known about their
nonlinear evolution and dynamical formation. In this work we have explored how
such solutions might be generated and evolve. The starting point for this
flux compactification scenario was the product space of a $p$-dimensional de
Sitter space and a $q$-dimensional (in some cases warped) topological sphere. Provided we fix the higher dimensional constant $\Lambda_D$ and the dimension of the internal manifold, the
properties and stability of the $p$-dimensional vacua and the compact space,
will depend on the number
of flux units of the $q$-form field strength wrapping the sphere or the Hubble
parameter of the extended dimensions. For each value of the conserved flux
number $n < n_{\mathrm{max}}$ there are four solutions. Homogeneous solutions
can be classified into two branches, a large volume branch, unstable to the
$\ell=0$ scalar sector and a small volume branch unstable to the $\ell=2$ mode
for $n < n_c < n_{\mathrm{max}}$. On the stationary warped branch one finds the
so-called large Hubble branch, unstable to the $\ell=2$ mode for $n <
n_{\mathrm{max}}$, and the small Hubble warped branch which is stable to the
$\ell=2$ mode for $n < n_c$ but unstable for $ n_c < n < n_{\mathrm{max}}$.

To gain some understanding of the parameter space and dynamics, we studied the
evolution of initially small perturbations around those stationary solutions.
We find, in agreement with previous studies that, for any dimensionality of the
sphere, solutions on the large volume branch either decompactify to empty $D$-dimensional 
de Sitter space with cosmological constant $\Lambda_D$ or flow to the solution
on the small volume branch with the same number of flux units but a smaller
volume, and hence large enough flux density to stabilize the sphere against
collapse. 

We show that within the regime where the small volume Freund-Rubin solution is
unstable to the $\ell=2$ scalar mode, when $n_I <n <n_c $ the solutions flow to
the corresponding solution on the small Hubble warped branch.  The end point of
the instability is a stationary oblate solution where the flux is concentrated
in a band around the equator. We do not find any other instabilities and
conclude this is the final endpoint of the solution, at least under our
symmetry assumptions.  However, for $n<n_I$, but still within the regime where
the small volume Freund-Rubin branch is unstable, we find that the solution
overshoots the linearly stable oblate solution, flowing
towards an increasing prolate solution, where the flux concentrates at the
poles of the sphere. The equator of the $q$-sphere is unsupported by flux, and
the equatorial radius shrinks to zero size in finite time, forming trapped
regions around the poles in the process. The four-dimensional spacetime
undergoes a crunch. 

Finally, regarding the end state of solutions on the large Hubble branch with
$n< n_{\mathrm{Mink}}$, we find that the volume of the internal space grows,
while the shape become increasingly oblate, but with cuspy feature at the
poles. The expansion rate remains inhomogeneous, as there is no solution on the
small Hubble branch with the same number of flux units.

It follows from the above that the warping of the compact space may stabilize
initially unstable configurations. This spontaneous symmetry breaking of the
internal space to more dynamically favoured configurations is a very natural
phenomenon in cosmology. In the case of the Jeans instability, configurations
with high mass density suffer from a gravitational instability. This symmetry
breaking instability is ultimately cut-off by nonlinear terms leading to
structure formation. Other analogous examples include the Gregory-Laflamme
instability \cite{Gregory:1994bj}.

There are a number of directions in which one could expand on this study.
While here we assumed that only one of the spatial degrees of freedom in the
internal space were excited, it would be interesting to allow additional
symmetries either in the compactified, or uncompactified dimensions, to be
broken. Another possible avenue would be to study the case where the external
space is anti-de Sitter, with potential applications to the AdS/CFT
correspondence. Here, we have focused on a simple model in order to gain
insight into open questions surrounding extra dimensions and spherical
compactifications.  For example, there has been much debate and conjecture
regarding the circumstances under which it is possible to have periods of
exponential expansion in compactified
scenarios~\cite{Cicoli:2018kdo,Palti:2019pca}. The general methods presented
here could be used to explore this issue in scenarios that are dynamical and
inhomogeneous.

\acknowledgments
We thank Andrew Frey, Claire Zukowski, David Marsh, Erik Schnetter, and Leo Stein.
The authors were supported by the National Science and Engineering Research Council through a Discovery grant. This research was supported in part by Perimeter Institute for Theoretical Physics. Research at Perimeter Institute is supported by the Government of Canada through the Department of Innovation, Science and Economic Development Canada and by the Province of Ontario through the Ministry of Research, Innovation and Science. 
This research was enabled in part by support provided by SciNet
(www.scinethpc.ca) and Compute Canada (www.computecanada.ca).
Simulations were performed on the Symmetry cluster at
Perimeter Institute and the Niagara cluster at the University of Toronto.

\appendix
\section{Dimensional reduction}
\label{dimred}
In this section, we illustrate the properties of dimensional reduction with several examples. Recall that in our units $M_D\equiv (8\pi G_D)^{-1/(D-2)}=1$.

\subsection{Time-dependent Freund-Rubin Solution}
\label{dimred_fr}
We start with a simple example, where identifying scale factor and moduli fields in the four-dimensional effective theory is straightforward. Consider solutions of the form:
\begin{equation}\label{eq:tdFR}
ds^2 = -\alpha(t)^2 dt^2 +a(t)^2 d\vec{x}_{p-1}^2 + L(t)^2 d\Omega_q^2 \ .
\end{equation}
This metric ansatz encompasses the static Freund-Rubin solutions of
section~\ref{sec:FR_branch}, as well as the time-dependent solutions resulting
from total-volume ($\ell = 0$) perturbations of the static Freund-Rubin
solutions. For such solutions, we have
\begin{eqnarray}
\frac{K - {K^x}_x}{2} &=& \frac{\dot{a}}{\alpha a} + \frac{q}{2}\frac{\dot{L}}{\alpha L} \ .
\end{eqnarray}
Defining
\begin{equation}\label{eq:metric_redef_FR}
\tilde{\alpha} = \left( L / L_0 \right)^{q/2} \alpha, \ \ \ \tilde{a} = \left( L / L_0 \right)^{q/2} a \ ,
\end{equation}
where $L_0 \equiv L(t=0)$ we have that
\begin{eqnarray}
 \int d^qy \sqrt{\gamma_q} \alpha \gamma_{xx}^{3/2} \left( \frac{{K^x}_x -K}{2} \right)^2 = \left(  L_0^q \int d\Omega_q \right)  \tilde{\alpha} \tilde{a}^3 \left( \frac{\dot{\tilde{a}}}{\tilde{\alpha} \tilde{a}} \right)^2 \ .
\end{eqnarray}
From this expression we can identify $\tilde{\alpha}$ as the four dimensional lapse, $\tilde{a}$ as the four dimensional scale factor and therefore:
\begin{equation}\label{eq:FRW_iden}
M_4^2 =  L_0^q \int d\Omega_q , \ \ \ \sqrt{-g(t)} = \tilde{\alpha} \tilde{a}^3, \ \ \ H^2 = \left( \frac{\dot{\tilde{a}}}{\tilde{\alpha} \tilde{a}} \right)^2 \ .
\end{equation}
Note that the change of variables defined by eq.~\eqref{eq:metric_redef_FR} is
precisely the conformal transformation of the four-dimensional metric that
brings us to the four-dimensional Einstein frame (e.g. the conformal frame
in which the Planck mass is constant in time); for comparison, see, e.g.,
refs.~\cite{Carroll:2009dn,Dahlen:2014}.

Evaluating the area of the cosmological apparent horizon using
eq.~\eqref{eq:horizonarea}, we obtain:
\begin{eqnarray}
\mathcal{A}_H 
&=& 4 \pi \int d\Omega_q L_0^q \left( \frac{\tilde{\alpha} \tilde{a}}{\dot{\tilde{a}}} \right)^2 
= M_4^2 \frac{4\pi}{H^2} \ .
\end{eqnarray}
The entropy, eq.~\eqref{eq:entropy}, is given by $\mathcal{S}=16\pi M_4^2/H^2$,
which is the value one would have assigned based purely on the four 
dimensional effective theory.

For the time-dependent Freund-Rubin solutions, it is possible to derive
the full dimensionally reduced action. This can be found, e.g., in
refs.~\cite{Carroll:2009dn,Dahlen:2014}, which we reproduce here for
completeness. Expanding the terms in the action we obtain 
\begin{eqnarray}
S &=& \frac{1}{2 } \int d^4x d^qy \ \sqrt{-g} \left[ - 6 \left( \frac{{K^x}_x -K}{2} \right)^2 \right. \nonumber \\ 
&+& \frac{1}{2} ({K^\theta}_\theta)^2 + \frac{(q+3)(q-1)}{4} ({K^\phi}_\phi)^2 + ({K^\theta}_\theta+\frac{q-1}{2} {K^\phi}_\phi)^2 \\ 
&+&\left. {}^{(D-1)}R -2  \Lambda_D - \frac{1}{q!} \mathbf{F}_q^2 \right]  \ . \nonumber
\end{eqnarray}
Evaluating the various terms in the action for the metric ansatz eq.~\eqref{eq:tdFR} we have:
\begin{equation}
\frac{1}{2} ({K^\theta}_\theta)^2 + \frac{(q+3)(q-1)}{4} ({K^\phi}_\phi)^2 + ({K^\theta}_\theta+\frac{q-1}{2} {K^\phi}_\phi)^2 = \frac{q(q+2)}{2} \left( \frac{\dot{L}}{L}\right)^2 \ ,
\end{equation}
\begin{equation}
{}^{(D-1)}R = \frac{q(q-1)}{L^2}  
\end{equation}
and 
\begin{equation}
\frac{1}{q!} \mathbf{F}_q^2 = \frac{Q_B^2}{L^{2q}} =  \frac{1}{M_4^4} \frac{n^2}{L^{2q}} \ . 
 \end{equation}
Using the relations eqs.~\eqref{eq:metric_redef_FR} and~\eqref{eq:FRW_iden}, we have
\begin{eqnarray}
S &=&  \int d^4 x \ \tilde{\alpha} \tilde{a}^3 \left[ - \frac{M_4^2}{2} 6 H^2 + \frac{M_4^2}{2} \frac{L_0^q q(q-1)}{2 L^{q+2}} \dot{L}^2 -  V(L) \right]
\end{eqnarray}
where we have defined the effective potential
\begin{equation}
\frac{V(L)}{M^4_4} \equiv \frac{1}{2} \left(\frac{L_0}{L} \right)^q  \left( - \frac{q(q-1)}{L^2} + 2 \Lambda_D +\frac{1}{M_4^4}  \frac{n^2}{L^{2q}} \right) \ . 
\end{equation}
We see that the dimensionally-reduced theory is that of an FLRW Universe with a scalar field $L$ (with a non-canonical kinetic term) evolving in the effective potential $V(L)$ (plotted in figure~\ref{fig:Veff}).

\subsection{Factorizable warped metrics}

Another illustrative example is given by solutions of the form:
\begin{eqnarray}\label{eq:warpedmetric}
ds^2 &=& e^{2A(y,t)} \left[ - \left( \alpha(t)^2 - e^{-\frac{2(q+2)}{(q-2)}A(y,t)} \tilde{g}_{\gamma \delta} \beta^\gamma(y,t) \beta^\delta (y,t)\right) dt^2 + a(t)^2 d\vec{x}_{p-1}^2  \right] \\
&+& 2 e^{-\frac{8}{q-2} A(y,t)} \tilde{g}_{\gamma \delta} (y) \beta^\gamma(y,t) dt dy^\delta + e^{- \frac{8}{q-2} A(y,t)} \tilde{g}_{\gamma \delta}(y) dy^\gamma dy^\delta \ .
\end{eqnarray}
For $q=6$, this ansatz is characteristic of warped solutions to Type IIB string theory~\cite{Giddings:2001yu,Giddings:2005ff,Frey:2008xw}. 
We have that 
\begin{eqnarray}
\frac{K^x_x - K}{2} &=& \frac{e^{-A}}{\alpha} \left( \frac{\dot{a}}{a} - \frac{q+2}{q-2}\partial_0{A}  - \frac{1}{4} \tilde{g}^{\gamma \delta} \mathcal{L}_{\beta} \tilde{g}_{\gamma \delta} \right)\ 
\end{eqnarray}
where $\partial_0 \equiv \partial_t-\mathcal{L}_{\beta}$.
Defining
\begin{equation}\label{eq:warped_redef}
\tilde{\alpha}(y,t) = e^{-\frac{q+2}{q-2} A(y,t)} \alpha(t), \ \ \ \tilde{a}(y,t) = e^{-\frac{q+2}{q-2} A(y,t)} a(t)
\end{equation}
we have
\begin{equation}
 \int d^qy \sqrt{\gamma_q} \ \alpha \gamma_{xx}^{3/2} \left( \frac{{K^x}_x -K}{2} \right)^2 = \int d^qy \sqrt{\tilde{g}} \ \tilde{\alpha} \tilde{a}^3 \left[ \left( \frac{\dot{\tilde{a}}}{\tilde{\alpha} \tilde{a}} \right)^2 + F(\beta) \right] \ .
\end{equation}
Although the change of variables proposed above allows one to decompose the
action in a suggestive form, it is not immediately clear how to identify the
four-dimensional Planck mass, scale factor and lapse. This is due both to the
presence of terms involving the shift (written above as $F(\beta)$), as well as the average of a {\rm product} of
$y$-dependent factors over the compact space. The latter problem arises because
eq.~\eqref{eq:warped_redef} defines a conformal transformation of the four
dimensional metric that depends on both time and the coordinates on the compact
space. This can be contrasted with the standard approach to dimensional
reduction in the presence of warping, where one defines a purely time dependent
conformal transformation to the four-dimensional metric. 

Ignoring terms evolving the shift for the remainder of the calculation (i.e.
setting $\beta^\gamma=0$ and denoting the neglected terms with an ellipsis) for simplicity, we can make contact with the standard approach as follows. First,
we decompose the action as follows:
\begin{eqnarray}
\sqrt{-g_4} M_4^2 H^2 &=&  \int d^qy \sqrt{\tilde{g}} \ \alpha a^3 \left[ \frac{\dot{a}}{\alpha a} - \frac{q+2}{q-2}\frac{\dot{A}}{\alpha} + \ldots \right]^2 e^{-2 \frac{q+2}{q-2} A} \nonumber \\
&=&  
\alpha a^3 \left(\int d^qy \sqrt{\tilde{g}} \ e^{- \frac{4q}{q-2} A(y,t=0)}  \right)\times \\ && \left[
\left( \frac{\dot{a}}{\alpha a} \right)^2 e^{\phi(t)}
+\left( \frac{\dot{a}}{\alpha^2 a} \right) \frac{d e^{\phi(t)}}{dt}
+\left( \frac{1}{2\alpha}\frac{de^{\phi(t)}}{dt} \right)^2 
\right]
+ \Delta + \ldots \nonumber \\
&=& \bar{\alpha} \bar{a}^3  \left( \int d^qy \sqrt{\tilde{g}} e^{- \frac{4q}{q-2} A(y,t=0)}  \right) \left[ \frac{\dot{\bar{a}}}{\bar{\alpha} \bar{a}} \right]^2 + \Delta + \ldots \nonumber
\end{eqnarray}
where we have made the following definitions:
\begin{equation}
e^{\phi(t)} \equiv \frac{\int d^qy \sqrt{\tilde{g}} e^{-2 \frac{q+2}{q-2} A}}{\int d^qy \sqrt{\tilde{g}} \ e^{- \frac{4q}{q-2} A(y,t=0)} } \ ,
\end{equation}
and
\begin{equation}\label{eq:warped_phi_resacle}
\bar{\alpha}(t) \equiv \alpha(t) e^{\phi(t)/2} , \ \ \ \bar{a}(t) \equiv a(t) e^{\phi(t)/2}
\end{equation}
and
\begin{equation}
\Delta \equiv  \frac{(q+2)^2}{(q-2)^2} \frac{a^3}{\alpha} \int d^qy \sqrt{\tilde{g}} \ \dot{A} e^{-2 \frac{q+2}{q-2} A} \left[ \dot{A} - \langle \dot{A} \rangle \right] \ ,
\end{equation}
where
\begin{equation}
\langle \dot{A} \rangle \equiv  \frac{\int d^qy\sqrt{\tilde{g}} \ \dot{A} e^{-2 \frac{q+2}{q-2} A}}{\int d^qy \sqrt{\tilde{g}} } = -\frac{q-2}{2(q+2)}  \frac{d e^\phi}{dt} \ e^{- \frac{4q}{q-2} A(y,t=0)} \ .
\end{equation}
To the extent that $\Delta$ is small (and again, we are neglecting terms involving the shift), we can identify
\begin{equation}
M_4^2 =  \int d^qy \sqrt{\tilde{g}} \  e^{- \frac{4q}{q-2} A(y,t=0)} , \ \ \ \sqrt{-g_4} = \bar{\alpha} \bar{a}^3, \ \ \ H^2 = \left[ \frac{\dot{\bar{a}}}{\bar{\alpha} \bar{a}} \right]^2 \ .
\end{equation}
This definition for $M_4$ is not the same as in previous literature~\cite{Giddings:2001yu,Giddings:2005ff,Frey:2008xw}, but rather chosen to be consistent with our convention eq.~\eqref{eq:M4definition}.
Note that eq.~\eqref{eq:warped_phi_resacle} defines the conformal
transformation typically used in the literature to transform to the
Einstein frame. 

Evaluating the cosmological apparent horizon area
using eq.~\eqref{eq:horizonarea}
\begin{eqnarray}
\mathcal{A}_H 
&=& 4 \pi \int d^qy \sqrt{\tilde{g}} \ e^{-2 \frac{q+2}{q-2} A} \left( \frac{\dot{a}}{\alpha a} - \frac{q+2}{q-2} \frac{\dot{A}}{\alpha} + \ldots \right)^{-2} \ .
\end{eqnarray}
In the case where we 
make the approximation that $\dot{A} \simeq \langle \dot{A} \rangle$
(and neglecting terms involving the shift) we find:
\begin{eqnarray}
\mathcal{A}_H &=& 4 \pi \int d^qy \sqrt{\tilde{g}} \  e^{-2 \frac{q+2}{q-2} A} \left( \frac{\dot{\bar{a}}}{\bar{\alpha} \bar{a}} \right)^{-2} e^{-\phi} 
=  M_4^2 \frac{4 \pi }{H^2} \ .
\end{eqnarray}
Though, for general warped metrics of the form
eq.~\eqref{eq:warpedmetric}, the cosmological apparent horizon cannot
be precisely associated with the Hubble parameter as defined by
dimensional reduction in previous literature.  However, we note that
for the static warped solutions discussed in the text,
$\dot{A}=\beta^{\gamma}=0$, and the correspondence does hold.


\section{$(D-1)+1$ equations}
\label{eom}
\subsection{Maxwell equations}
Plugging in our metric and flux ansatz into the Maxwell equations \eqref{eq:ME}, the evolution equations for the electric and magnetic fluxes become
\begin{equation}\label{QE}
\begin{array}{lcl}
 \dot{Q_E}  &=&  \beta^\theta  {Q_E}  \bigg( \frac{Q_E'}{Q_E}+(q-1) \cot \theta \bigg) + \alpha K Q_E + 2\alpha  (q-1) K^{\phi_1}_{\phi_1} Q_E- \alpha \gamma_{\theta\theta}^{-1} Q_B'\vspace{0.5em} \\
 &&+\alpha \gamma_{\theta\theta}^{-1} Q_B \bigg(-(p-1)\frac{\gamma_{xx}'}{2\gamma_{xx}}+ \frac{\gamma_{\theta\theta}'}{2\gamma_{\theta\theta}} +  (q-1)\frac{\widetilde{\gamma}_{\phi_1\phi_1}'}{2\widetilde{\gamma}_{\phi_1\phi_1}} \bigg)-  \gamma_{\theta\theta}^{-1} (\partial_{\theta}\alpha)  Q_B 
\end{array}
\end{equation}
and
\begin{equation}\label{QB}
\begin{array}{lcl}
\dot{Q_B} &=& \beta^{\theta}  Q_B \bigg(\frac{Q_B'}{Q_B}+ (q-1) \cot\theta   \bigg) +  Q_B \partial_\theta \beta^\theta -\alpha \left(Q_E'+(q-1)Q_E \cot \theta\right)  \vspace{1em}\\
&&-(\partial_\theta \alpha)  Q_E 
\end{array}
\end{equation}
where the dot represents differentiation with respect to time.
\subsection{Generalized harmonic equations}
We evolve the solutions using a space-time decomposition of the generalized harmonic formulation
\cite{Garfinkle:2001ni,Pretorius:2005}. Here we write down the field equations for
completeness. In this formulation, the lapse and shift are evolution variables, in addition to  
the spatial metric and extrinsic curvature. We also introduce the
auxiliary fields $\pi$ and $\rho^{\bar{m}}$ that are directly related to the
time derivative of $\alpha$ and $\beta^{\bar{m}}$.  We fix the coordinate
degrees of freedom by specifying a so-called source vector, $H^{M}$ such
that the constraint vector 
\begin{equation}\label{eq:constraint}
C^{M} \equiv H^{M} + \left( ^{(D)}{\Gamma^{M}}_{NK}-^{(D)}{\bar{\Gamma}^{M}}_{NK}\right) g^{NK}=0
\end{equation}
vanishes. Here $^{(D)}{\bar{\Gamma}^{M}}_{NK}$ denotes a background connection.
The generalized harmonic equations are 
\begin{equation}\label{GHEFE}
\begin{array}{lcl}
^{(D)}R_{MN}-\nabla_{(M}C_{N)}=-\kappa \left[n_{(M}C_{N)} -\frac{1}{(D-2)} g_{MN}n^{L}C_{L}\right]+\left[T_{MN}-\frac{1}{D-2}g_{MN}T\right] \ . 
\end{array}
\end{equation}
These are hyperbolic, provided the source functions are specified directly as a function of the spacetime coordinates $x^M$ and the metric $g_{MN}$.

We evolve the $(D-1)+1$ form of the generalized harmonic evolution equations~\cite{Brown:2011} as follows 
\begin{subequations}\label{gamma}
\begin{equation}
\partial_t \gamma_{xx}= -2\alpha K_{xx} + \gamma_{xx}' \beta^{\theta} 
\end{equation}
\begin{equation}
\partial_t \gamma_{\theta\theta}= -2\alpha K_{\theta\theta}+ 2\gamma_{\theta\theta} \partial_{\theta}\beta^{\theta}+\beta^{\theta}\gamma_{\theta\theta}'
\end{equation}
\begin{equation}
\partial_t \widetilde{\gamma}_{\phi_1\phi_1}= -2 \alpha \widetilde{K}_{\phi_1\phi_1}+ \beta^{\theta}(\widetilde{\gamma}_{\phi_1\phi_1}'+2\cot \theta \widetilde{\gamma}_{\phi_1\phi_1})
\end{equation}
\end{subequations}
\begin{subequations}\label{Kij}
\begin{eqnarray}
\partial_t K_{xx}&=& -\frac{\gamma_{xx}'}{2\gamma_{\theta\theta}} \partial_\theta \alpha +\alpha \left( {}^{(D-1)}R_{xx}-2K^{x}_{x}K_{xx}+K K_{xx}\right)+   \alpha \left( \frac{1}{D-2}\gamma_{xx}(S-\rho)-  S_{xx} \right) \nonumber \\
&&+\beta^{\theta}\partial_{\theta}K_{xx} -\alpha C_{\perp} K_{xx} -\alpha \frac{1}{2} \gamma_{xx}' C^{\theta}-\kappa \alpha \gamma_{xx} C_{\perp}/2
\end{eqnarray}
\begin{eqnarray}
\partial_t K_{\theta\theta}&=& -\partial^2_\theta \alpha+ \frac{\gamma_{\theta\theta}'}{2\gamma_{\theta\theta}} \partial_\theta \alpha +\alpha \left( {}^{(D-1)}R_{\theta\theta}-2K^{\theta}_{\theta}K_{\theta\theta}+K K_{\theta\theta}\right)+ \alpha \left( \frac{1}{D-2}\gamma_{\theta\theta}(S-\rho)- S_{\theta\theta} \right) \nonumber \\
&&+\beta^{\theta}\partial_{\theta}K_{\theta\theta}+2K_{\theta\theta}\partial_{\theta}\beta^{\theta}-\alpha C_{\perp} K_{\theta\theta} -\alpha \frac{1}{2} \gamma_{\theta\theta}' C^\theta -\alpha  \gamma_{\theta\theta} \partial_\theta C^{\theta}-\kappa \alpha \gamma_{\theta\theta} C_{\perp}/2
\end{eqnarray}
\begin{eqnarray}
\partial_t \widetilde{K}_{\phi_1\phi_1}&=&-\left(\cot \theta \frac{\widetilde{\gamma}_{\phi_1\phi_1}}{{\gamma_{\theta\theta}}}+ \frac{\widetilde{\gamma}_{\phi_1\phi_1}'}{2\gamma_{\theta\theta}}\right)\partial_\theta \alpha  +\alpha \left( \frac{{}^{(D-1)}R_{\phi_1\phi_1}}{\sin^2 \theta} -2\widetilde{\gamma}^{\phi_1\phi_1} (\widetilde{K}_{\phi_1\phi_1})^2 +K \widetilde{K}_{\phi_1\phi_1}\right) \vspace{1em} \nonumber \\
&& +  \frac{1}{p+q-2}  \alpha \widetilde{\gamma}_{\phi_1\phi_1}\left( -2 \Lambda_D +  \widetilde{\gamma}_{\phi_1\phi_1}^{-(q-1)} (p-1) \big[   Q_E^2 -\gamma_{\theta\theta}^{-1} Q_B^2\big] \right) \vspace{1em}  \\
&& + \left(\partial_\theta \widetilde{K}_{\phi_i\phi_i} - \frac{\widetilde{\gamma}_{\phi_1\phi_1}'}{\widetilde{\gamma}_{\phi_1\phi_1}}\widetilde{K}_{\phi_i\phi_i}\right)\beta^{\theta} -\alpha C_{\perp} \widetilde{K}_{\phi_1\phi_1} -\alpha \left[\cot \theta \widetilde{\gamma}_{\phi_1\phi_1}+\frac{1}{2}\widetilde{\gamma}_{\phi_1\phi_1}'\right] C^{\theta}-\kappa \alpha \widetilde{\gamma}_{\phi_1\phi_1} C_{\perp}/2 \nonumber
\end{eqnarray}
\end{subequations}
\begin{subequations}
\begin{eqnarray}
\partial_t \alpha &=& \alpha^2 \pi-\alpha^2 H_{\perp}  +\beta^\theta \partial_\theta \alpha \vspace{1em} \\
\partial_t \beta^{\theta} &=& \beta^{\theta}\partial_\theta \beta^\theta +\alpha^2 \rho^{\theta} -\alpha \gamma^{\theta\theta} \partial_{\theta} \alpha +\alpha^2 H^{\theta}
\end{eqnarray}
\end{subequations}
and
\begin{subequations}
\begin{eqnarray}
\partial_t \pi &=& -\alpha \left((p-1) K_{xx}K^{xx} +K_{\theta\theta}K^{\theta\theta} +(q-1) K_{\phi_1\phi_1}K^{\phi_1\phi_1}\right)+D_{\bar{m}} D^{\bar{m}} \alpha  \vspace{1em} \nonumber \\
&& +C^{\theta} \partial_{\theta} \alpha -\frac{(D-3)}{(D-2)} \kappa \alpha C_{\perp} - \alpha \frac{1}{D-2}((D-3)\rho +S)+ \beta^{\theta} \partial_{\theta} \pi
\end{eqnarray}
\begin{eqnarray}
\partial_t \rho^{\theta}&=& \gamma^{\bar{n}\bar{l}} \bar{D}_{\bar{n}} \bar{D}_{\bar{l}} \beta^{\bar{\theta}}+\alpha \gamma^{\theta\theta} \partial_\theta \pi -\pi \gamma^{\theta\theta} \partial_\theta  \alpha -2K^{\theta\theta} \partial_\theta \alpha  \vspace{1em} \nonumber \\
&&+2 \alpha\left[ - (p-1)  \frac{\gamma_{xx}'}{2\gamma_{\theta\theta}}K^{xx}+ \frac{\gamma_{\theta\theta}'}{2\gamma_{\theta\theta}}K^{\theta\theta}+ (q-1) \bigg[-(\cot \theta \frac{\widetilde{\gamma}_{\phi_1\phi_1}}{\gamma_{\theta\theta}}+\frac{\widetilde{\gamma}_{\phi_1\phi_1}'}{2\gamma_{\theta\theta}})+\cot\theta\bigg] \widetilde{K}^{\phi_1\phi_1}\right] \vspace{1em} \nonumber\\
&& +\kappa \alpha C^{\theta}  -2\alpha j^{\theta}+(\beta^\theta \partial_{\theta} \rho^{\theta} -\rho^{\theta}\partial_{\theta} \beta^{\theta}  )+(q-1)\beta \widetilde{\gamma}_{\phi_1\phi_1}^{-1} \vspace{1em} \nonumber \\
\end{eqnarray}
\end{subequations}
with the non-trivial constraints
\begin{subequations}\label{CE}
\begin{equation}
C_{\perp} \equiv \pi +K =0
\end{equation}
\begin{eqnarray}
C^\theta&=& -\rho^\theta - (p-1)  \frac{\gamma_{xx}'}{2\gamma_{\theta\theta}}\gamma^{xx}+ \frac{\gamma_{\theta\theta}'}{2\gamma_{\theta\theta}}\gamma^{\theta\theta}=0 \nonumber \\
&&+ (q-1) \bigg[-(\cot \theta \frac{\widetilde{\gamma}_{\phi_1\phi_1}}{\gamma_{\theta\theta}}+\frac{\widetilde{\gamma}_{\phi_1\phi_1}'}{2\gamma_{\theta\theta}})+\cot\theta\bigg] \widetilde{\gamma}^{\phi_1\phi_1} \label{rhoa}  
\end{eqnarray}
\begin{equation}
\mathcal{H}={}^{(D-1)}R-3K_{xx}K^{xx}-K_{\theta\theta}K^{\theta\theta}-(q-1)K_{\phi_i\phi_i}K^{\phi_i\phi_i}+K^2-2 \rho=0
\end{equation}
\begin{eqnarray}\label{M}
\mathcal{M}_{\theta}&=&D_{\bar{m}}{K^{\bar{m}}}_{\theta}-D_{\theta}K- j_{\theta}=0
\end{eqnarray}
\end{subequations}
where $K \equiv \gamma^{\bar{m}\bar{n}}K_{\bar{m}\bar{n}}$,
${}^{(D-1)}R=\gamma^{\bar{m}\bar{n}}R_{\bar{m}\bar{n}}$,  $\bar{D}_{\bar{m}}$ denotes the
covariant derivative associated with the background metric $\bar{g}_{MN}$ which
we assume to have a lapse of one, shift of zero and a time-independent spatial
metric under $(D-1)+1$ splitting. We also define $H_{\perp}\equiv n^{M}H_{M}$,
$H^{\bar{m}} \equiv {\gamma^{\bar{m}}}_{N} H^{N}$ and the various projections
of stress-energy tensor $T_{MN}$ as
$S_{\bar{m}\bar{n}}=\gamma^M_{\bar{m}}\gamma^N_{\bar{n}}T_{MN}$, $\rho=n^M n^N
T_{MN}$ and $j_{\bar{m}}=-\gamma^K_{\bar{m}} T_{KN}n^N$.

We find that our solutions are more stable if we choose a gauge such
that the shift vector is driven to zero, and the lapse is constant in time for  
the stationary background solutions,
\begin{equation}
H^{\theta}=-\frac{\eta}{\alpha^2}\beta^{\theta} , \ \ \ \ H^{\perp}=-K_0
\end{equation}
where $K_0$ is the initial value of the trace of the extrinsic curvature and
$\eta$ is some constant controlling the rate at which the shift is driven to
zero. We typically set $\kappa=15$ and $\eta=10$ in units where $\Lambda_D=1$, although their exact values are not too important.

\section{Convergence tests}
\label{sec:conv}
\begin{figure}[h]
\centering
\includegraphics[width=.327\textwidth]{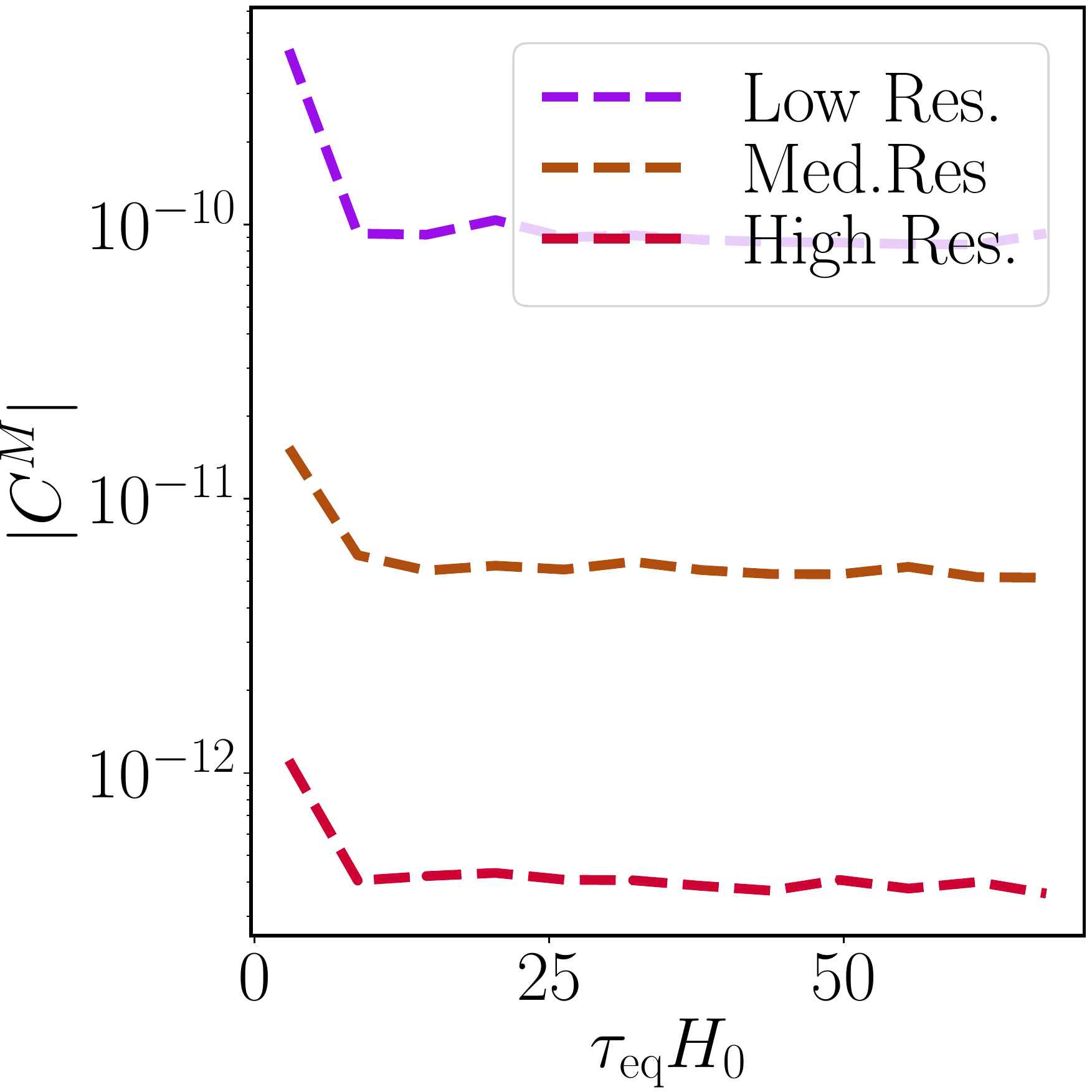}
\includegraphics[width=.327\textwidth]{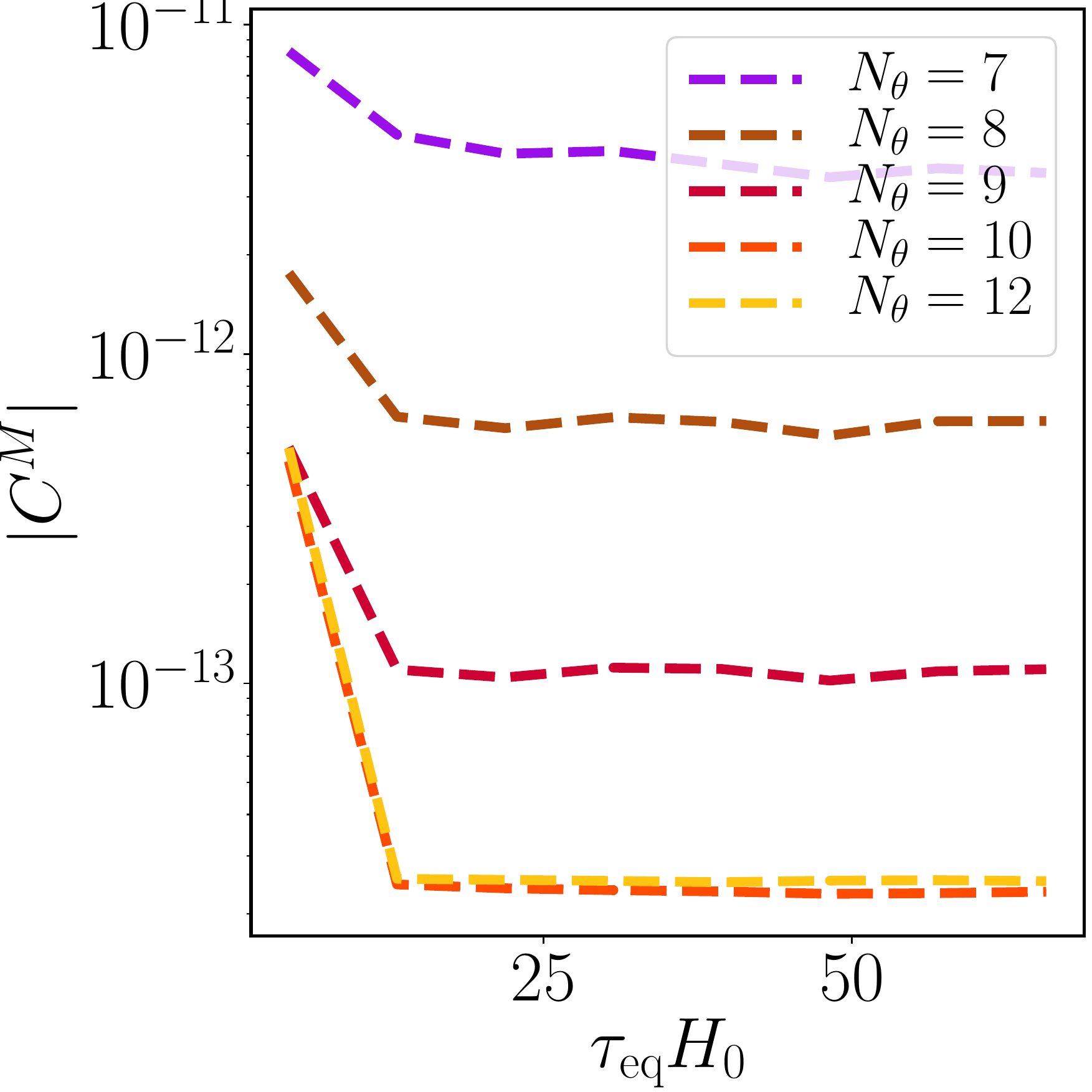}
\includegraphics[width=.327\textwidth]{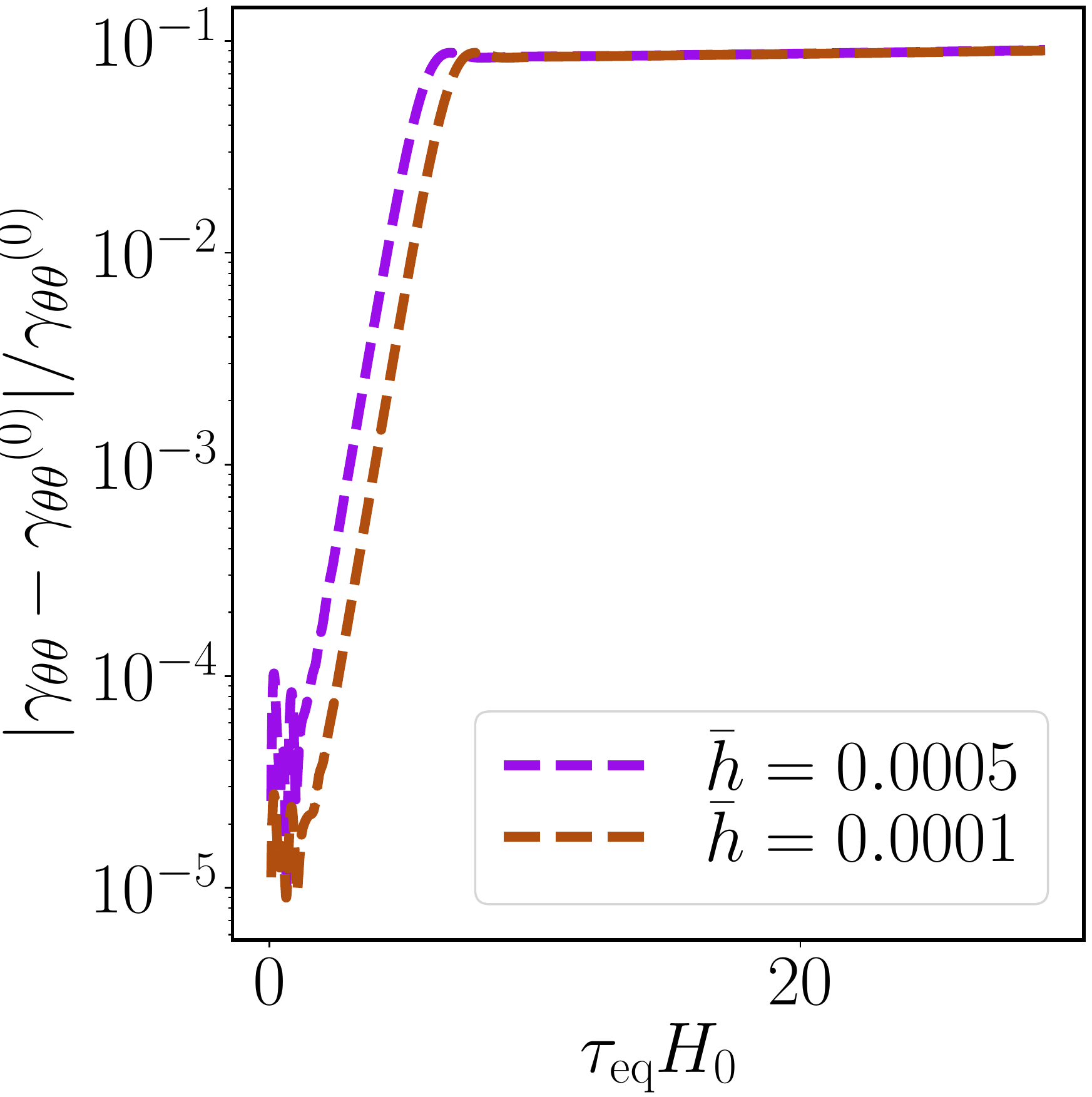}
\caption{\label{fig:conv} 
Integrated norm of the constraint violation eq.~\eqref{eq:constraint} for
different temporal (left) and spatial (middle) resolutions as a function
of proper time (in units of the background Hubble expansion) for $q=4$, $
\Lambda_D=1$, $H/M_4=0.0078$ and an initial $\ell=2$ perturbation. The
medium and high temporal resolutions have $2 \times$ and $4 \times$ the
resolution of the low resolution run. $N_\theta$ in the middle indicates
the number of collocation points used. We find that the constraint
violations converge at fourth order in time and exponentially in space.
(Right) The time evolution of spatial average of relative difference of
$\gamma_{\theta\theta}$ from its background solution for $q=4$, $H/M_4=0.0050$ and
an initial $\ell=2$ perturbation of magnitude $\bar{h}=10^{-4}$ and
$\bar{h}=5 \times 10^{-4}$. The linear warped instability is evident. }
\end{figure}

Ensuring that the constraints converge to zero with increasing numerical resolution,
and at the expected order, provides a consistency check that the numerical
solution obtained is converging to a solution of the field equations.
Our numerical scheme 
converges at fourth order
with temporal resolution and exponentially with spatial resolution.
Figure{~\eqref{fig:conv}} shows the integrated norm of the constraint
violation given by eq.~\eqref{eq:constraint} for several resolutions, demonstrating
that this quantity is converging to zero at the expected rate. The highest temporal resolution used in the resolution study is equivalent to the resolution we use for the other solutions. The spatial resolution required depends on whether the solution has inhomogeneous features that needs to be resolved or not. For homogeneous solutions we typically use $N_{\theta} \sim 20$, for stationary oblate solutions $N_{\theta} \sim 72$ and finally the prolate solutions typically require up to $N_{\theta} \sim 152$.
For
unstable solutions, we perturb the solutions with a sufficiently small
amplitude to ensure that we are in the linear regime. In
figure{~\ref{fig:conv}}, we plot the evolution of the metric
variable $\gamma_{\theta\theta}$ for a Freund-Rubin solution unstable to the warped
instability, and perturbed with an initial amplitude of $\bar{h}=10^{-5}$
and $\bar{h}=5 \times 10^{-5}$. Both solutions undergo a clear
exponential growth phase before entering the nonlinear regime, 
with the time of saturation being set by the amplitude of the initial
perturbation.

\bibliography{draft}

\providecommand{\href}[2]{#2}\begingroup\raggedright\begin{thebibliography}{10}

\bibitem{Kaluza:1921tu}
T.~Kaluza, \emph{{Zum Unit\"atsproblem der Physik}},
  \href{https://doi.org/10.1142/S0218271818700017}{\emph{Sitzungsber. Preuss.
  Akad. Wiss. Berlin (Math. Phys. )} {\bfseries 1921} (1921) 966}
  [\href{https://arxiv.org/abs/1803.08616}{{\ttfamily 1803.08616}}].

\bibitem{Klein:1926tv}
O.~Klein, \emph{{Quantum Theory and Five-Dimensional Theory of Relativity. (In
  German and English)}},
  \href{https://doi.org/10.1007/BF01397481}{\emph{Z.Phys.} {\bfseries 37}
  (1926) 895}.

\bibitem{ArkaniHamed:1998rs}
N.~Arkani-Hamed, S.~Dimopoulos and G.~Dvali, \emph{{The Hierarchy problem and
  new dimensions at a millimeter}},
  \href{https://doi.org/10.1016/S0370-2693(98)00466-3}{\emph{Phys.Lett.}
  {\bfseries B429} (1998) 263}
  [\href{https://arxiv.org/abs/hep-ph/9803315}{{\ttfamily hep-ph/9803315}}].

\bibitem{Randall:1999ee}
L.~Randall and R.~Sundrum, \emph{{A Large mass hierarchy from a small extra
  dimension}},
  \href{https://doi.org/10.1103/PhysRevLett.83.3370}{\emph{Phys.Rev.Lett.}
  {\bfseries 83} (1999) 3370}
  [\href{https://arxiv.org/abs/hep-ph/9905221}{{\ttfamily hep-ph/9905221}}].

\bibitem{Randall:1999vf}
L.~Randall and R.~Sundrum, \emph{{An Alternative to compactification}},
  \href{https://doi.org/10.1103/PhysRevLett.83.4690}{\emph{Phys.Rev.Lett.}
  {\bfseries 83} (1999) 4690}
  [\href{https://arxiv.org/abs/hep-th/9906064}{{\ttfamily hep-th/9906064}}].

\bibitem{Susskind:2003kw}
L.~Susskind, \emph{{The anthropic landscape of string theory}},  in
  \emph{Universe or Multiverse}, Cambridge University Press (2003).

\bibitem{Freund:1980}
P.G.O.~Freund and M.A.~Rubin, \emph{{Dynamics of Dimensional Reduction}},
  \href{https://doi.org/10.1016/0370-2693(80)90590-0}{\emph{Phys. Lett.}
  {\bfseries 97B} (1980) 233}.

\bibitem{Bousso:2003}
R.~Bousso, O.~DeWolfe and R.C.~Myers, \emph{{Unbounded entropy in space-times
  with positive cosmological constant}},
  \href{https://doi.org/10.1023/A:1023733106589}{\emph{Found. Phys.} {\bfseries
  33} (2003) 297} [\href{https://arxiv.org/abs/0205080}{{\ttfamily 0205080}}].

\bibitem{Maldacena:1997re}
J.M.~Maldacena, \emph{{The Large N limit of superconformal field theories and
  supergravity}}, \href{https://doi.org/10.1023/A:1026654312961}{\emph{Int. J.
  Theor. Phys.} {\bfseries 38} (1999) 1113}
  [\href{https://arxiv.org/abs/hep-th/9711200}{{\ttfamily hep-th/9711200}}].

\bibitem{Douglas:2006es}
M.R.~Douglas and S.~Kachru, \emph{{Flux compactification}},
  \href{https://doi.org/10.1103/RevModPhys.79.733}{\emph{Rev. Mod. Phys.}
  {\bfseries 79} (2007) 733}
  [\href{https://arxiv.org/abs/hep-th/0610102}{{\ttfamily hep-th/0610102}}].

\bibitem{Denef:2007pq}
F.~Denef, M.R.~Douglas and S.~Kachru, \emph{{Physics of String Flux
  Compactifications}},
  \href{https://doi.org/10.1146/annurev.nucl.57.090506.123042}{\emph{Ann. Rev.
  Nucl. Part. Sci.} {\bfseries 57} (2007) 119}
  [\href{https://arxiv.org/abs/hep-th/0701050}{{\ttfamily hep-th/0701050}}].

\bibitem{Carroll:2009dn}
S.M.~Carroll, M.C.~Johnson and L.~Randall, \emph{{Dynamical compactification
  from de Sitter space}},
  \href{https://doi.org/10.1088/1126-6708/2009/11/094}{\emph{JHEP} {\bfseries
  11} (2009) 094} [\href{https://arxiv.org/abs/0904.3115}{{\ttfamily
  0904.3115}}].

\bibitem{Brown:2013fba}
A.R.~Brown, A.~Dahlen and A.~Masoumi, \emph{{Compactifying de Sitter space
  naturally selects a small cosmological constant}},
  \href{https://doi.org/10.1103/PhysRevD.90.124048}{\emph{Phys. Rev. D}
  {\bfseries 90} (2014) 124048}
  [\href{https://arxiv.org/abs/1311.2586}{{\ttfamily 1311.2586}}].

\bibitem{Asensio:2012pg}
C.~Asensio and A.~Segui, \emph{{Exploring a simple sector of the
  Einstein-Maxwell landscape}},
  \href{https://doi.org/10.1103/PhysRevD.87.023503}{\emph{Phys. Rev. D}
  {\bfseries 87} (2013) 023503}
  [\href{https://arxiv.org/abs/1207.4662}{{\ttfamily 1207.4662}}].

\bibitem{Aguirre:2009tp}
A.~Aguirre, M.C.~Johnson and M.~Larfors, \emph{{Runaway dilatonic domain
  walls}}, \href{https://doi.org/10.1103/PhysRevD.81.043527}{\emph{Phys. Rev.
  D} {\bfseries 81} (2010) 043527}
  [\href{https://arxiv.org/abs/0911.4342}{{\ttfamily 0911.4342}}].

\bibitem{BlancoPillado:2009di}
J.J.~Blanco-Pillado, D.~Schwartz-Perlov and A.~Vilenkin, \emph{{Quantum
  Tunneling in Flux Compactifications}},
  \href{https://doi.org/10.1088/1475-7516/2009/12/006}{\emph{JCAP} {\bfseries
  0912} (2009) 006} [\href{https://arxiv.org/abs/0904.3106}{{\ttfamily
  0904.3106}}].

\bibitem{Brown:2010bc}
A.R.~Brown and A.~Dahlen, \emph{{Small Steps and Giant Leaps in the
  Landscape}},
  \href{https://doi.org/10.1103/PhysRevD.82.083519}{\emph{Phys.Rev.} {\bfseries
  D82} (2010) 083519} [\href{https://arxiv.org/abs/1004.3994}{{\ttfamily
  1004.3994}}].

\bibitem{BlancoPillado:2010et}
J.J.~Blanco-Pillado, H.S.~Ramadhan and B.~Shlaer, \emph{{Decay of flux vacua to
  nothing}}, \href{https://doi.org/10.1088/1475-7516/2010/10/029}{\emph{JCAP}
  {\bfseries 1010} (2010) 029}
  [\href{https://arxiv.org/abs/1009.0753}{{\ttfamily 1009.0753}}].

\bibitem{Contaldi:2004hr}
C.R.~Contaldi, L.~Kofman and M.~Peloso, \emph{{Gravitational instability of de
  Sitter compactifications}},
  \href{https://doi.org/10.1088/1475-7516/2004/08/007}{\emph{JCAP} {\bfseries
  0408} (2004) 007} [\href{https://arxiv.org/abs/hep-th/0403270}{{\ttfamily
  hep-th/0403270}}].

\bibitem{Krishnan:2005}
C.~Krishnan, S.~Paban and M.~Zanic, \emph{{Evolution of gravitationally
  unstable de Sitter compactifications}},
  \href{https://doi.org/10.1088/1126-6708/2005/05/045}{\emph{JHEP} {\bfseries
  05} (2005) 045} [\href{https://arxiv.org/abs/hep-th/0503025}{{\ttfamily
  hep-th/0503025}}].

\bibitem{BlancoPillado:2009mi}
J.J.~Blanco-Pillado, D.~Schwartz-Perlov and A.~Vilenkin,
  \emph{{Transdimensional Tunneling in the Multiverse}},
  \href{https://doi.org/10.1088/1475-7516/2010/05/005}{\emph{JCAP} {\bfseries
  1005} (2010) 005} [\href{https://arxiv.org/abs/0912.4082}{{\ttfamily
  0912.4082}}].

\bibitem{Kinoshita:2007}
S.~Kinoshita, \emph{{New branch of Kaluza-Klein compactification}},
  \href{https://doi.org/10.1103/PhysRevD.76.124003}{\emph{Phys. Rev. D}
  {\bfseries 76} (2007) 124003}
  [\href{https://arxiv.org/abs/0710.0707}{{\ttfamily 0710.0707}}].

\bibitem{Kinoshita:2009hh}
S.~Kinoshita and S.~Mukohyama, \emph{{Thermodynamic and dynamical stability of
  Freund-Rubin compactification}},
  \href{https://doi.org/10.1088/1475-7516/2009/06/020}{\emph{JCAP} {\bfseries
  06} (2009) 020} [\href{https://arxiv.org/abs/0903.4782}{{\ttfamily
  0903.4782}}].

\bibitem{Lim:2012}
Y.-K.~Lim, \emph{{Warped branches of flux compactifications}},
  \href{https://doi.org/10.1103/PhysRevD.85.064027}{\emph{Phys. Rev.}
  {\bfseries D85} (2012) 064027}
  [\href{https://arxiv.org/abs/1202.3525}{{\ttfamily 1202.3525}}].

\bibitem{Dahlen:2014}
A.~Dahlen and C.~Zukowski, \emph{{Flux Compactifications Grow Lumps}},
  \href{https://doi.org/10.1103/PhysRevD.90.125013}{\emph{Phys. Rev.}
  {\bfseries D90} (2014) 125013}
  [\href{https://arxiv.org/abs/1404.5979}{{\ttfamily 1404.5979}}].

\bibitem{Giddings:2001yu}
S.B.~Giddings, S.~Kachru and J.~Polchinski, \emph{{Hierarchies from fluxes in
  string compactifications}},
  \href{https://doi.org/10.1103/PhysRevD.66.106006}{\emph{Phys. Rev. D}
  {\bfseries 66} (2002) 106006}
  [\href{https://arxiv.org/abs/hep-th/0105097}{{\ttfamily hep-th/0105097}}].

\bibitem{Kachru:2003aw}
S.~Kachru, R.~Kallosh, A.D.~Linde and S.P.~Trivedi, \emph{{De Sitter vacua in
  string theory}},
  \href{https://doi.org/10.1103/PhysRevD.68.046005}{\emph{Phys. Rev. D}
  {\bfseries 68} (2003) 046005}
  [\href{https://arxiv.org/abs/hep-th/0301240}{{\ttfamily hep-th/0301240}}].

\bibitem{Kachru:2003sx}
S.~Kachru, R.~Kallosh, A.D.~Linde, J.M.~Maldacena, L.P.~McAllister et~al.,
  \emph{{Towards inflation in string theory}},
  \href{https://doi.org/10.1088/1475-7516/2003/10/013}{\emph{JCAP} {\bfseries
  0310} (2003) 013} [\href{https://arxiv.org/abs/hep-th/0308055}{{\ttfamily
  hep-th/0308055}}].

\bibitem{DeWolfe:2001nz}
O.~DeWolfe, D.Z.~Freedman, S.S.~Gubser, G.T.~Horowitz and I.~Mitra,
  \emph{{Stability of AdS(p) x M(q) compactifications without supersymmetry}},
  \href{https://doi.org/10.1103/PhysRevD.65.064033}{\emph{Phys.Rev.} {\bfseries
  D65} (2002) 064033} [\href{https://arxiv.org/abs/hep-th/0105047}{{\ttfamily
  hep-th/0105047}}].

\bibitem{Brown:2013}
A.R.~Brown and A.~Dahlen, \emph{{Spectrum and stability of compactifications on
  product manifolds}},
  \href{https://doi.org/10.1103/PhysRevD.90.044047}{\emph{Phys. Rev. D}
  {\bfseries 90} (2014) 044047}
  [\href{https://arxiv.org/abs/1310.6360}{{\ttfamily 1310.6360}}].

\bibitem{Hinterbichler:2014}
K.~Hinterbichler, J.~Levin and C.~Zukowski, \emph{{Kaluza-Klein Towers on
  General Manifolds}},
  \href{https://doi.org/10.1103/PhysRevD.89.086007}{\emph{Phys. Rev. D}
  {\bfseries 89} (2014) 086007}
  [\href{https://arxiv.org/abs/1310.6353}{{\ttfamily 1310.6353}}].

\bibitem{Garfinkle:2008ei}
D.~Garfinkle, W.C.~Lim, F.~Pretorius and P.J.~Steinhardt, \emph{{Evolution to a
  smooth universe in an ekpyrotic contracting phase with w \ensuremath{>} 1}},
  \href{https://doi.org/10.1103/PhysRevD.78.083537}{\emph{Phys. Rev. D}
  {\bfseries 78} (2008) 083537}
  [\href{https://arxiv.org/abs/0808.0542}{{\ttfamily 0808.0542}}].

\bibitem{Wainwright:2013lea}
C.L.~Wainwright, M.C.~Johnson, H.V.~Peiris, A.~Aguirre, L.~Lehner and
  S.L.~Liebling, \emph{{Simulating the universe(s): from cosmic bubble
  collisions to cosmological observables with numerical relativity}},
  \href{https://doi.org/10.1088/1475-7516/2014/03/030}{\emph{JCAP} {\bfseries
  03} (2014) 030} [\href{https://arxiv.org/abs/1312.1357}{{\ttfamily
  1312.1357}}].

\bibitem{East:2015ggf}
W.E.~East, M.~Kleban, A.~Linde and L.~Senatore, \emph{{Beginning inflation in
  an inhomogeneous universe}},
  \href{https://doi.org/10.1088/1475-7516/2016/09/010}{\emph{JCAP} {\bfseries
  09} (2016) 010} [\href{https://arxiv.org/abs/1511.05143}{{\ttfamily
  1511.05143}}].

\bibitem{East:2016anr}
W.E.~East, J.~Kearney, B.~Shakya, H.~Yoo and K.M.~Zurek, \emph{{Spacetime
  Dynamics of a Higgs Vacuum Instability During Inflation}},
  \href{https://doi.org/10.1103/PhysRevD.95.023526}{\emph{Phys. Rev. D}
  {\bfseries 95} (2017) 023526}
  [\href{https://arxiv.org/abs/1607.00381}{{\ttfamily 1607.00381}}].

\bibitem{Clough:2016ymm}
K.~Clough, E.A.~Lim, B.S.~DiNunno, W.~Fischler, R.~Flauger and S.~Paban,
  \emph{{Robustness of Inflation to Inhomogeneous Initial Conditions}},
  \href{https://doi.org/10.1088/1475-7516/2017/09/025}{\emph{JCAP} {\bfseries
  09} (2017) 025} [\href{https://arxiv.org/abs/1608.04408}{{\ttfamily
  1608.04408}}].

\bibitem{DeWolfe:2002nn}
O.~DeWolfe and S.B.~Giddings, \emph{{Scales and hierarchies in warped
  compactifications and brane worlds}},
  \href{https://doi.org/10.1103/PhysRevD.67.066008}{\emph{Phys.Rev.} {\bfseries
  D67} (2003) 066008} [\href{https://arxiv.org/abs/hep-th/0208123}{{\ttfamily
  hep-th/0208123}}].

\bibitem{Giddings:2005ff}
S.B.~Giddings and A.~Maharana, \emph{{Dynamics of warped compactifications and
  the shape of the warped landscape}},
  \href{https://doi.org/10.1103/PhysRevD.73.126003}{\emph{Phys.Rev.} {\bfseries
  D73} (2006) 126003} [\href{https://arxiv.org/abs/hep-th/0507158}{{\ttfamily
  hep-th/0507158}}].

\bibitem{Frey:2008xw}
A.R.~Frey, G.~Torroba, B.~Underwood and M.R.~Douglas, \emph{{The Universal
  Kahler Modulus in Warped Compactifications}},
  \href{https://doi.org/10.1088/1126-6708/2009/01/036}{\emph{JHEP} {\bfseries
  0901} (2009) 036} [\href{https://arxiv.org/abs/0810.5768}{{\ttfamily
  0810.5768}}].

\bibitem{Jacobson:1995ab}
T.~Jacobson, \emph{{Thermodynamics of space-time: The Einstein equation of
  state}}, \href{https://doi.org/10.1103/PhysRevLett.75.1260}{\emph{Phys. Rev.
  Lett.} {\bfseries 75} (1995) 1260}
  [\href{https://arxiv.org/abs/gr-qc/9504004}{{\ttfamily gr-qc/9504004}}].

\bibitem{Padmanabhan:2002sha}
T.~Padmanabhan, \emph{{Classical and quantum thermodynamics of horizons in
  spherically symmetric space-times}},
  \href{https://doi.org/10.1088/0264-9381/19/21/306}{\emph{Class. Quant. Grav.}
  {\bfseries 19} (2002) 5387}
  [\href{https://arxiv.org/abs/gr-qc/0204019}{{\ttfamily gr-qc/0204019}}].

\bibitem{Hayward:1998ee}
S.A.~Hayward, S.~Mukohyama and M.C.~Ashworth, \emph{{Dynamic black hole
  entropy}}, \href{https://doi.org/10.1016/S0375-9601(99)00225-X}{\emph{Phys.
  Lett. A} {\bfseries 256} (1999) 347}
  [\href{https://arxiv.org/abs/gr-qc/9810006}{{\ttfamily gr-qc/9810006}}].

\bibitem{Frolov:2002va}
A.V.~Frolov and L.~Kofman, \emph{{Inflation and de Sitter thermodynamics}},
  \href{https://doi.org/10.1088/1475-7516/2003/05/009}{\emph{JCAP} {\bfseries
  05} (2003) 009} [\href{https://arxiv.org/abs/hep-th/0212327}{{\ttfamily
  hep-th/0212327}}].

\bibitem{Cai:2005ra}
R.-G.~Cai and S.P.~Kim, \emph{{First law of thermodynamics and Friedmann
  equations of Friedmann-Robertson-Walker universe}},
  \href{https://doi.org/10.1088/1126-6708/2005/02/050}{\emph{JHEP} {\bfseries
  02} (2005) 050} [\href{https://arxiv.org/abs/hep-th/0501055}{{\ttfamily
  hep-th/0501055}}].

\bibitem{GalvezGhersi:2011tx}
J.T.~Galvez~Ghersi, G.~Geshnizjani, F.~Piazza and S.~Shandera, \emph{{Eternal
  inflation and a thermodynamic treatment of Einstein's equations}},
  \href{https://doi.org/10.1088/1475-7516/2011/06/005}{\emph{JCAP} {\bfseries
  06} (2011) 005} [\href{https://arxiv.org/abs/1103.0783}{{\ttfamily
  1103.0783}}].

\bibitem{Garfinkle:2001ni}
D.~Garfinkle, \emph{{Harmonic coordinate method for simulating generic
  singularities}},
  \href{https://doi.org/10.1103/PhysRevD.65.044029}{\emph{Phys. Rev. D}
  {\bfseries 65} (2002) 044029}
  [\href{https://arxiv.org/abs/gr-qc/0110013}{{\ttfamily gr-qc/0110013}}].

\bibitem{Pretorius:2005}
F.~Pretorius, \emph{Numerical relativity using a generalized harmonic
  decomposition},
  \href{https://doi.org/10.1088/0264-9381/22/2/014}{\emph{Classical and Quantum
  Gravity} {\bfseries 22} (2005) 425}.

\bibitem{Majda7511}
A.J.~Majda, D.~Qi and T.P.~Sapsis, \emph{Blended particle filters for
  large-dimensional chaotic dynamical systems},
  \href{https://doi.org/10.1073/pnas.1405675111}{\emph{Proceedings of the
  National Academy of Sciences} {\bfseries 111} (2014) 7511}
  [\href{https://arxiv.org/abs/https://www.pnas.org/content/111/21/7511.full.pdf}{{\ttfamily
  https://www.pnas.org/content/111/21/7511.full.pdf}}].

\bibitem{Kreiss}
H.-O.~Kreiss and J.~Oliger, \emph{Comparison of accurate methods for the
  integration of hyperbolic equations},
  \href{https://doi.org/10.3402/tellusa.v24i3.10634}{\emph{Tellus} {\bfseries
  24} (1972) 199}
  [\href{https://arxiv.org/abs/https://doi.org/10.3402/tellusa.v24i3.10634}{{\ttfamily
  https://doi.org/10.3402/tellusa.v24i3.10634}}].

\bibitem{Gregory:1994bj}
R.~Gregory and R.~Laflamme, \emph{{The Instability of charged black strings and
  p-branes}},
  \href{https://doi.org/10.1016/0550-3213(94)90206-2}{\emph{Nucl.Phys.}
  {\bfseries B428} (1994) 399}
  [\href{https://arxiv.org/abs/hep-th/9404071}{{\ttfamily hep-th/9404071}}].

\bibitem{Cicoli:2018kdo}
M.~Cicoli, S.~De~Alwis, A.~Maharana, F.~Muia and F.~Quevedo, \emph{{De Sitter
  vs Quintessence in String Theory}},
  \href{https://doi.org/10.1002/prop.201800079}{\emph{Fortsch. Phys.}
  {\bfseries 67} (2019) 1800079}
  [\href{https://arxiv.org/abs/1808.08967}{{\ttfamily 1808.08967}}].

\bibitem{Palti:2019pca}
E.~Palti, \emph{{The Swampland: Introduction and Review}},
  \href{https://doi.org/10.1002/prop.201900037}{\emph{Fortsch. Phys.}
  {\bfseries 67} (2019) 1900037}
  [\href{https://arxiv.org/abs/1903.06239}{{\ttfamily 1903.06239}}].

\bibitem{Brown:2011}
J.~Brown, \emph{{Generalized Harmonic Equations in 3+1 Form}},
  \href{https://doi.org/10.1103/PhysRevD.84.124012}{\emph{Phys. Rev. D}
  {\bfseries 84} (2011) 124012}
  [\href{https://arxiv.org/abs/1109.1707}{{\ttfamily 1109.1707}}].

\end{thebibliography}\endgroup
\bibliographystyle{JHEP}

\end{document}